\documentclass[]{jfm}

\usepackage{graphicx}
\usepackage{newtxtext}
\usepackage{newtxmath}
\usepackage{natbib}
\usepackage{hyperref}
\usepackage{graphicx}
\usepackage{epstopdf, epsfig}
\usepackage{subfigure}
\usepackage{amsmath}
\usepackage{gensymb}
\usepackage{array}
\usepackage{amsfonts}
\usepackage{textcomp}
\usepackage{array}
\newcolumntype{C}[1]{>{\centering\arraybackslash}m{#1}}

\usepackage[bbgreekl]{mathbbol}

\usepackage{bm}
\hypersetup{
    colorlinks = true,
    urlcolor   = blue,
    citecolor  = black,
}

\newcommand{\RomanNumeralCaps}[1]
\title{Transport and orientation of anisotropic particles settling in surface gravity waves}

\author{Himanshu Mishra
   \and
  Anubhab Roy \corresp{\email{anubhab@iitm.ac.in}}}
\affiliation{Department of Applied Mechanics, Indian Institute of Technology Madras, Chennai, India, 600036}
\begin{document}
\maketitle

\begin{abstract}

We study the translation and orientation dynamics of an anisotropic particle settling in monochromatic linear surface gravity waves. Recent work has shown that a neutrally buoyant spheroid attains a preferred mean orientation in such wave fields, independent of its initial state and determined solely by its aspect ratio. Comparing the settling parameter $\mathrm{Sv}$, the ratio of settling speed to wave speed, with the asymptotically small wave steepness $\epsilon$, we investigate the long time dynamics of a negatively buoyant particle. We examine the transition from aspect ratio-dependent equilibrium orientation in the weak settling regime ($\mathrm{Sv} \ll \epsilon^2$) to initial-condition-dependent alignment in the strong settling limit ($\mathrm{Sv} \gg 1$). Since translation and orientation are coupled for anisotropic particles, we use orientation dynamics to predict net horizontal transport. Fluid inertia induces an inertial torque that breaks the Stokesian degeneracy and drives broadside alignment. We analyze the influence of this torque on drift and alignment rate as functions of settling and wave parameters. Finally, we evaluate finite-size effects through the parameter $\sigma$, showing that a neutrally buoyant finite-size spheroid exhibits $\sigma$-dependent drift, validating the finite-size approximation when the spheroid size approaches the wavelength.

\end{abstract}

% \begin{keywords}

% \end{keywords}

% {\bf MSC Codes }  {\it(Optional)} Please enter your MSC Codes here

\section{Introduction}
\label{sec:intro}

Environmental flows are abundant with dispersed particulate matter, which are often non-spherical in shape. Some of them include the dynamics of ice crystals in clouds, dispersion of pollen grains in atmospheric flows, and transport of plankton, microplastics, and marine snow in oceanic flows. Anisotropic particulate matter, through their shape and the ambient flow field coupling, can exhibit intriguing dynamical behavior even in the dilute regime that is otherwise absent in their oft-studied isotropic counterparts. Flows in the atmosphere and the oceans are turbulent. What would be the dispersion of particles in such flows? This question has been addressed extensively in the literature when one ignores the orientation dynamics of the particles. However, considering the rotational motion of a particle in a turbulent flow, the scenario becomes significantly more complex \citep{vothsoldati2017}.   

The particles of interest often inhabit a dissipation-dominated regime due to their small sizes compared to the characteristic flow length scales. In such viscous scenarios, the translational dynamics is determined by the particle's orientation, which could arise due to an interplay of external forces (for example, gravity) and the background flow. Shape- and size-influenced dispersion becomes particularly important when considered in the context of marine particulate matter. There are various mechanisms by which particles are transported in oceanic flows. These include, but are not limited to, windage, Stokes drift, Langmuir circulations, internal tides, and Ekman drift \citep{van2020physical}. Therefore, analyzing how different flow and particle parameters influence the particle transport dynamics in an oceanic environment is essential; whether one is interested in quantifying the dispersion of microplastics and its contamination of the marine environment \citep{law2017plastics}, investigating the dynamics of planktonic and crustacean matter and their debatable role in biogenic mixing \citep{kunze2019biologically} or the role of marine snow in the biological carbon pump \citep{turner2015zooplankton}.

 In a surface gravity wave, a tracer drifts along the wave propagation direction in an oscillatory manner. These oscillations are induced by the waves, where a tracer spends most of its time going forward (on crest) rather than backward (on trough), resulting in the net drift in a horizontal direction, called Stokes drift. A good comprehensive review has been done by \cite{van2018stokes} on the Stokes drift. The analytical model was developed \citep{stokes1847theory} where the Lagrangian description of mean horizontal drift is explained by imposing an irrotational assumption on a flow field. The theoretical model can capture the mean drift of a neutrally buoyant particle in a horizontal direction, which is valid for waves with a low amplitude. Despite the constraint on the height of the amplitude and the nature of the flow field, the model has been studied well in the past to understand the tracer transport. A tracer particle remains mainly at the surface, either swept away by wind or translated by Stokes drift. For a larger particle, particle inertia starts playing a role in the transport dynamics.  With particle inertia, many studies have modified the model to analyze the horizontal drift of the spherical particle. Also, the particulate matter in the oceanic flows are of varying density. The density of a particle in the oceanic flows changes with fragmentation, bio-fouling, and erosion \citep{ter2016understanding}; therefore, it is crucial to understand the dynamics of a heavy particle as it settles. The general transport equation used for analyzing weak inertial spherical particles in the background flow field is given by \cite{mr1983}. \cite{eames2008} has conducted a detailed investigation on the settling of an inertial spherical particle. The maximum horizontal distance a spherical particle can transport is obtained before it settles vertically in the flow field. Using perturbation techniques, \cite{Santamaria_2013} have shown that a weakly inertial spherical particle can also drift in the vertical direction. Also, particle inertia can enhance the horizontal drift of a positively buoyant particle. \cite{stocchino2019sea} have performed the experiments to analyze the transport of a spherical particle when interacting with sea waves via Stokes drift. Below the interface, an inertial effect is shown to dominate the transport dynamics of a negatively buoyant particle. Moreover, they have concluded that the Stokes drift is an effective way to analyze the transportation of a positive buoyant particle since it lies at the surface, where the oscillatory wavy effects are dominant. Recently, \cite{dib_clark_pujara_2022} have studied the enhanced settling of a spherical particle induced due to the non-linear drag force. Along with \cite{eames2008}, they have examined the difference in the drift velocities of the heavy particles and flow. 

%3. anisotropic modelling

  \begin{table}
  \begin{center}
\def~{\hphantom{0}}
  \begin{tabular}{C{12em} C{5em} C{3em} C{3em} C{5cm}}
      Citations  & Nature   &   Shape & Density & Remarks  \\ [ 3pt]
      \hline
       \cite{eames2008}  & T  & \includegraphics[scale=0.04]{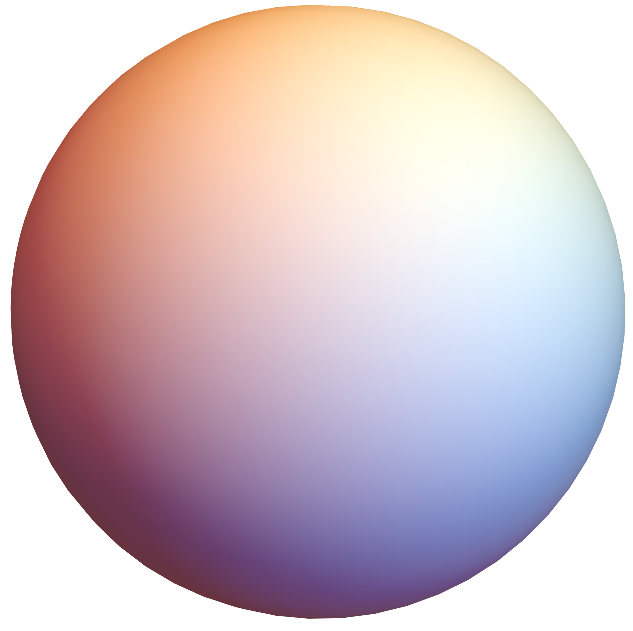} & B  & Examine the dynamics of inertial spherical particle in surface gravity waves\\ [ 8pt]
        \cite{Santamaria_2013} & T  & \includegraphics[scale=0.04]{sp.eps} & B  & Effect of weak particle inertia on particle's drift\\ [ 8pt]
        \cite{dibenedetto_2018} & T  & \includegraphics[scale=0.04]{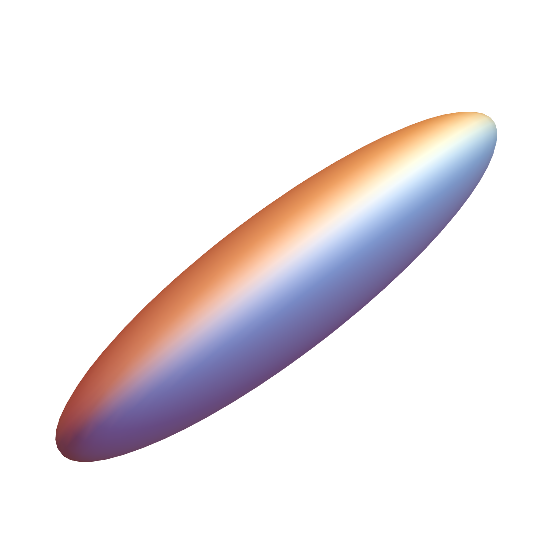} \includegraphics[scale=0.04]{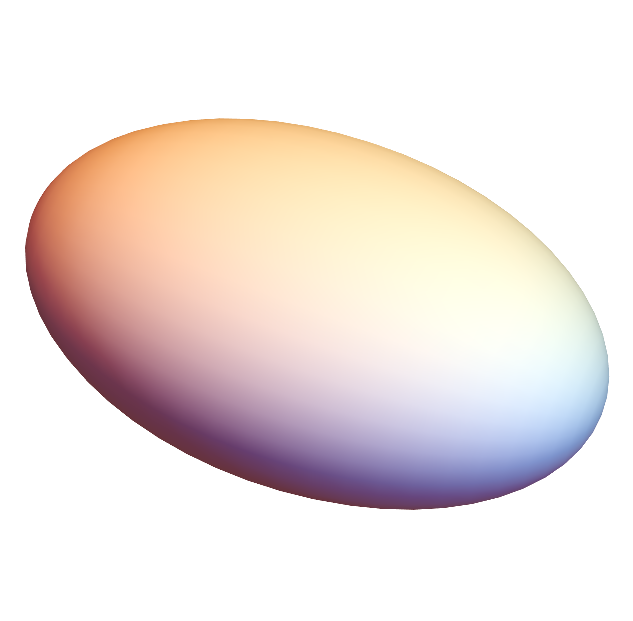}  & B  & Dynamics of inertial spheroidal particle \\ [ 8pt]
        \cite{dibenedetto_ouellette_2018} & T & \includegraphics[scale=0.04]{pro.eps} \includegraphics[scale=0.04]{ob.eps} & NB &  Orientation dynamics of neutrally buoyant spheroid \\ [ 8pt]
        \cite{dib2019oriexp} & E & \includegraphics[scale=0.04]{pro.eps} \includegraphics[scale=0.04]{ob.eps} & B  & Experiments on orientation dynamics of a buoyant particle\\ [ 8pt]
        \cite{stocchino2019sea} & T  & \includegraphics[scale=0.04]{sp.eps} & B   & Role of Particle Inertia in the Dynamics of Buoyant Particles\\ [ 8pt]
          \cite{clarkesett2020} & E & \includegraphics[scale=0.04]{pro.eps} \includegraphics[scale=0.04]{ob.eps} & B  & Experiment on settling of a buoyant particle\\ [ 8pt]
           \cite{leo21} & E and T & \includegraphics[scale=0.04]{sp.eps} & B  & Quantify inertial effect from different wave regimes \\ [ 8pt]
        \cite{leo2021} & T & \includegraphics[scale=0.04]{sp.eps} & B & Dispersion of a very small particle\\ [ 8pt]
        \cite{ma22} & T & \includegraphics[scale=0.04]{pro.eps} \includegraphics[scale=0.04]{ob.eps} & B & Transport of microswimmers in the wavy field \\ [ 8pt]
         \cite{dib_clark_pujara_2022} & T & \includegraphics[scale=0.04]{sp.eps} & B & Nonlinear effects on inertial spherical particle transport \\ [ 8pt]        \cite{clark_dibenedetto_ouellette_koseff_2023} & E & \includegraphics[scale=0.04]{pro.eps} \includegraphics[scale=0.04]{ob.eps} & B & Experiments on dispersion of particles \\ [ 8pt]
         \cite{pujara2023wave} & T & \includegraphics[scale=0.04]{pro.eps} \includegraphics[scale=0.04]{ob.eps} & B & Dynamics of a negatively buoyant spheroid in wavy flows \\ [ 8pt]
         \cite{sunberg2024parametric} & T & \includegraphics[scale=0.04]{pro.eps} \includegraphics[scale=0.04]{ob.eps} & B &  Parametric study of a settling spheroid in wave-current flows \\ [ 8pt]
           \cite{de2025rigid} & E & \includegraphics[scale=0.04]{pro.eps}  & B &  Tumbling of a rod-shaped particle in non-homogeneous turbulent flow\\ [ 8pt]
       \end{tabular}
  \caption{Literature summary of recent studies on particle transport in oceanic flows. Here, B denotes positively or negatively buoyant particles, NB indicates neutrally buoyant particles, T refers to theoretical studies (including numerical simulations), and E denotes experimental investigations. Particle shapes are classified as spherical or non-spherical based on the images presented in the table.}
  \label{tab:kd}
  \end{center}
\end{table}

Most of the studies mentioned above have approximated the shape of a particle as spherical. The assumption of the isotropic geometry of a particle while analyzing the transport dynamics is very restrictive. A diverse range of particulate matter found in environmental flows, such as plankton, textile particles, pollen grains, volcanic ash, and cellulose fibre, is rather non-spherical in shape. Many complex approximations of shape have been adopted in the past to model the realistic dynamics of a particle in a variety of flow fields. The most widely used approximation of shape is anisotropic geometry. The dynamics of anisotropic particles is complex because they exhibit orientation-dependent dispersion (or settling) in the flow field. Therefore, several authors have started attempting to explore the rotational dynamics of anisotropic particles. The rotational dynamics of an anisotropic particle in the Stokesian regime is well studied in the literature. In a seminal work of \cite{jef1922} has studied the orientation dynamics of a small ellipsoidal particle in a simple shear flow. The rotational trajectories of a particle are shown to follow one of an infinite family of initial condition-dependent elliptical orbits. The rotational dynamics is indeterminate since the choice of the rotational trajectory depends on the initial orientation. The rotational dynamics obtained from the Jeffery solution results in a tumbling motion of a spheroid, where a symmetry axis of a spheroid rotates along the flow gradient plane, and log-rolling motion, where a symmetry axis coincides with a vorticity axis while rotating.

Recent studies have extensively explored the transport dynamics of anisotropic particles in surface gravity waves. When an anisotropic particle moves through a linear wave field, it adopts a preferred orientation about which it oscillates \citep{dibenedetto_2018}. The particle’s trajectory is strongly influenced by this wave-induced preferential alignment. \cite{dibenedetto_ouellette_2018} further examined the translation dynamics of an inertial anisotropic particle in a wavy flow field, demonstrating that the particle reaches a wave-preferred orientation regardless of its initial orientation. Additionally, they found that the particle dispersion depends on its initial depth, as different depths expose the particle to varying shear forces within the flow. Beyond orientation, the rate at which a particle aligns also impacts its translation. While particles with the same aspect ratio eventually reach the same steady-state orientation, differences in the transient of orientation can introduce the variations in their transport. More recently, \cite{pujara2023wave} have investigated the dynamics of negatively buoyant spheroids, showing that such particles settle into a shape-dependent wave-preferred orientation. Using multiple-scale analysis, they have derived the mean equilibrium orientation and demonstrated its crucial role in determining the mean horizontal translation of the particle. In contrast, our work reveals that a heavier particle aligns according to a more complex interplay of factors, which includes non-dimensional settling velocity, density, and initial orientation of a particle. The steady-state orientation described in previous studies applies only to neutrally buoyant particles with negligible settling velocity relative to the wave velocity. For negatively buoyant particles, we demonstrate that the steady orientation undergoes multiple transitions depending on the dimensionless settling velocity. This buoyancy-induced alignment significantly influences the particle trajectories, leading to a transient orientation dynamics that differ from the dynamics of a neutrally buoyant particle. Furthermore, we use the asymptotic analysis to derive expressions for the long-time orientation and the maximum horizontal transport in the high-settling regime.

% inertial torque

Experimental and field observations in atmospheric contexts consistently report that settling rod-shaped particles adopt a broadside alignment across diverse flow conditions \citep{jayaweera1965behaviour,zikmunda1972fall,platt1978lidar,sassen1980remote,breon2004horizontally}, a finding that long lacked theoretical underpinning. At particle scales, inertial corrections to Stokes flow become significant, with contributions arising either from ambient shear or from translational motion, the latter generating a fore-aft asymmetric wake structure that produces a restoring torque driving the particle toward the broadside state. \citet{khayat_cox_1989} have captured this mechanism using matched asymptotic expansions for a slender body, matching an inner Stokes solution to an outer Oseen approximation to obtain an inertial torque of the form $T_{I}\sim \rho_f a^3 W^{2} \mathcal{F}\sin2\theta$, here $\rho_f$ is a fluid density, $W$ is a slip velocity based on a settling, $a$ is a characteristic length of a particle, $\mathcal{F}$ is an aspect ratio dependent quantity, and $\theta$ is an orientation angle of a spheroid with the veritical axis. Extending this framework, \citet{dabade2015effects} have derived the corresponding expression for spheroids, likewise predicting broadside-on alignment. Analytical studies have further explored the role of inertial torque in particle orientation dynamics. \citet{roy_2019} have investigated the preferred alignment of an asymmetric fibre resulting from the interplay between gravitational and inertial torques. More recently, \citep{menon2017theoretical, roy2023orientation} have considered the symmetric fibre and demonstrated that inertial torque induces a particle's rotation toward the broadside-on face alignment in homogeneous isotropic turbulence at low turbulent intensities, although this tendency can be disrupted under strong turbulence. While laboratory experiments with rods in cellular flows \citep{lopez2017inertial} and spheroids of varying aspect ratio \citep{cabrera2022experimental} demonstrated quantitative agreement with these theories, the slender-body model was accurate at high aspect ratios, and the spheroidal model better captured moderate aspect ratios. Subsequent work further showed that the asymptotic form of $\boldsymbol{T}_I$ qualitatively reproduces orientation trends even beyond its formal regime of validity. Related studies in surface gravity waves \citep{dib2019oriexp} and recent parametric analyses \citep{sunberg2024parametric} confirmed that inertial torque robustly enforces broadside alignment and can strongly influence particle dispersion relative to other effects, though its precise role in long-time dispersion and clustering in turbulence remains an open problem. More recently, \citet{de2025rigid} have conducted experiments to investigate the tumbling dynamics of long fibres in non-homogeneous and non-stationary turbulent flows. They have demonstrated that long fibres can serve as effective probes of the statistical properties of upper-ocean turbulence. Furthermore, their findings provide valuable insights into the transport mechanisms of mesoscale and macroscale floating litter in coastal environments.

\begin{figure}
    \centering
    \includegraphics[width=0.95\linewidth]{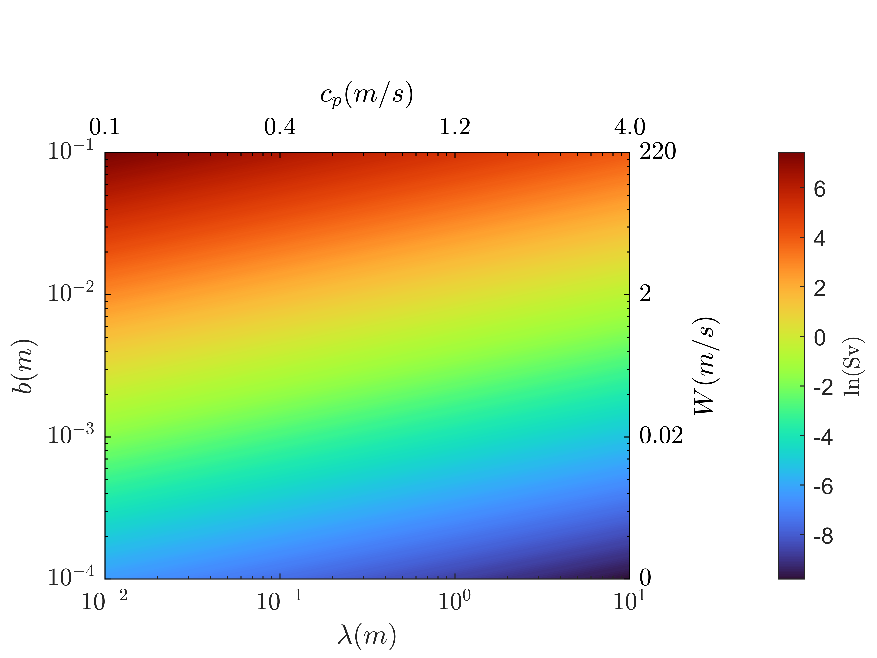}
    \caption{Contour plot of $\mathrm{Sv}$ as a function of semi-minor length $b$ of a particle and wavelength $\lambda$. Here, $ c_p$  represents the phase speed of the wave, and $W$  denotes the settling velocity of a particle.}
    \label{fig:count1}
\end{figure}

This paper investigates the translational and orientational dynamics of a heavy, finite-size anisotropic particle in surface gravity waves. In a non-dimensional framework, the relative importance of particle settling is characterised by the settling parameter, $\mathrm{Sv}$, which also plays a central role in the presence of inertia-induced torque. In figure, \ref{fig:count1}, we show a contour plot of $\mathrm{Sv}$ at wavelength $\lambda$ and $b$ plane. Here, $b$ is the length of the semi-minor axis of a particle, and we have defined the Settling parameter as $\mathrm{Sv}=W/c_{p}$, where $c_{p}$ is the phase speed of the wave. The results show that $\mathrm{Sv}$ increases with particle size, indicating enhanced settling for larger particles. Consequently, inertia-induced torques become increasingly important for the orientational dynamics in this regime. We first analyse the motion of a point particle subject to inertial torque, focusing on the rate of change of orientation and the resulting particle transport for different values of $\mathrm{Sv}$. We then extend the analysis to finite-size particles and compare their dynamics with those of point particles.

\section{Problem Formulation}

\begin{figure}
    \centering
    \includegraphics[scale=0.65]{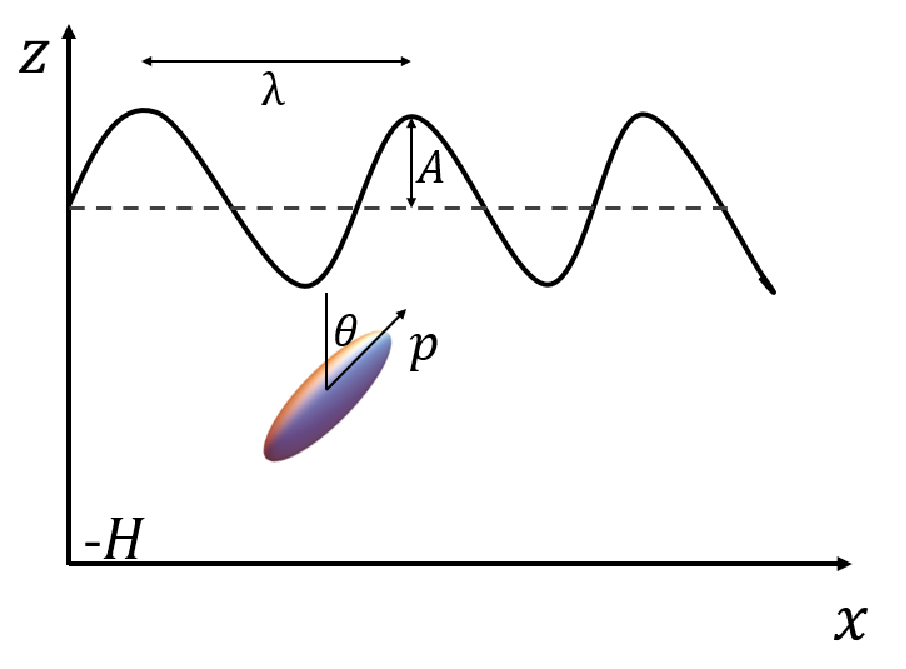}
    \caption{Schematic diagram of an anisotropic particle subjected to the wavy flow field. The orientation vector along the symmetry axis of the particle is denoted by $\boldsymbol{p}$, making an angle $\theta$ with a vertical axis. $\lambda$ and $A$ are the wavelength and amplitude of the wave. Here, $H$ represents the total depth of the flow field. The direction of gravity is taken parallel to the $z$ axis.}
    \label{fig:diag}
\end{figure}

We consider the velocity field generated by a surface gravity wave with horizontal and vertical components $u_x$ and $u_z$ in the $x$ and $z$ directions, respectively, where $x$ is aligned with the propagation direction and $z$ with the vertical direction. The flow is assumed two-dimensional. Within the framework of linear wave theory, the velocity components are given by
\begin{equation}\label{eq:ux}
 u_{x}=A\omega \dfrac{\cosh[k(z+H)]\cos(kx-\omega t)}{\sinh(kH)}, \qquad
 u_{z}=A\omega \dfrac{\sinh[k(z+H)]\sin(kx-\omega t)}{\sinh(kH)},
\end{equation}
where $A$ is the wave amplitude, $\omega$ is the frequency, $k$ is the wavenumber, and $H$ is the water depth (see figure \ref{fig:diag}). The wave field is irrotational and inviscid. The associated dispersion relation is 
\begin{equation}
\omega^{2}=gk\tanh(kH),
\end{equation}
and in the deep-water limit ($kH\to \infty$) one may take $k>0$. Surface tension modifies this relation through
\begin{equation}
\omega^{2}=k\left(g+\dfrac{\Gamma k^{2}}{\rho_{f}}\right)\tanh(kH),
\end{equation}
where $\Gamma$ is the surface tension coefficient and $\rho_{f}$ is the fluid density. For wavelengths exceeding $\mathcal{O}(1)$ cm, surface tension is negligible and gravity dominates, therefore when $k=O(10)$ cm$^{-1}$ the surface tension correction is $\Gamma k^{2}/\rho_{f}\sim O(10^{-3})$ and can be safely ignored. 

In natural oceanic flows, a wide distribution of particulate matter exists: plankton span millimetres to $\mathcal{O}(10)$ cm, while microplastics range from microns to millimetres in size \citep{hidalgo2012microplastics,enders2015abundance,isobe2016percentage,pabortsava2020high}. Densities vary widely between $800$ and $1500\,\mathrm{kg/m^{3}}$ \citep{kooi2019simplifying}, with processes such as biofouling and erosion altering effective density over time \citep{ter2016understanding}. Particles heavier than water settle, while lighter particles remain near the surface. Here we focus on a negatively buoyant spheroid in the deep-water limit, for which the depth dependence in (\ref{eq:ux}) vanishes. 

The motion of a rigid body in a fluid is described by the Newton-Euler equations. For a spheroid of volume $V=(4/3)\pi ab^{2}$, with $a$ and $b$ are defined as the  length of semi-major and -minor axes of a particle, respectively. The density of a particle is denoted by $\rho_p$, the particle mass is $m=\rho_p V$ and the displaced fluid mass is $m_f=\rho_f V$. The governing equations are given as

\begin{subequations}
\begin{equation}\label{eq:odfor}
 m\dfrac{d\boldsymbol{v}}{dt}=\int_{\mathcal{S}_{b}}\mathbb{\Sigma}\cdot \boldsymbol{n}\,ds+(m-m_{f})\boldsymbol{g},
\end{equation} 
\begin{equation}\label{eq:odrot1}
 \mathbb{J}\cdot\dfrac{d\boldsymbol{\Omega}}{dt}+\boldsymbol{\Omega}\times (\mathbb{J}\cdot\boldsymbol{\Omega})=\int_{\mathcal{S}_{b}}\boldsymbol{r}\times (\mathbb{\Sigma}\cdot\boldsymbol{n})\,ds,
\end{equation}
\end{subequations}

where $\mathbb{J}$ is the moment of inertia tensor of the particle, $\mathcal{S}_{b}$ is the particle surface, $\boldsymbol{n}$ is the position vector with respect to the center of gravity of the particle, and $\mathbb{\Sigma}$ is the stress tensor. Additionally, \(\boldsymbol{v}\) represents the velocity of a particle, while \(\boldsymbol{\Omega}\) denotes the angular velocity. In the Stokesian regime, the Faxén laws relate the disturbance velocity field to the hydrodynamic forces and torques, $\boldsymbol{F}$ and $\boldsymbol{T}$ \citep{KIM1985}. For a prolate spheroid, these yield

\begin{subequations}\label{eq:fsge}
\begin{multline}\label{eq:for}
F_{i}=3\pi\mu  \left\{X_{A}p_{i}p_{j}+Y_{A}(\delta_{ij}-p_{i}p_{j})\right\}\dfrac{1}{e}\int_{-\alpha}^{\alpha}\left\{1+(\alpha^{2}-\xi^{2})\dfrac{1-e^{2}}{4e^{2}}\nabla^{2}\right\}(u_{j}-v_{j})\,d\xi
\end{multline}
\begin{multline}\label{eq:tor}
T_{i}=3\pi \mu  \left\{X_{C}p_{i}p_{j}+Y_{C}(\delta_{ij}-p_{i}p_{j})\right\}\dfrac{1}{e^{3}}\int_{-\alpha}^{\alpha}\left(\alpha^{2}-\xi^{2}\right)(\left(\nabla \times u\right)_{j}-2\Omega_{j})\,d\xi\\
-6\pi \mu Y_{H}\varepsilon_{ijl} p_{l}p_{k}\dfrac{1}{e^{3}}\int_{-\alpha}^{\alpha}\left[(\alpha^{2}-\xi^{2})(1+(\alpha^{2}-\xi^{2})\dfrac{1-e^{2}}{8e^{2}}\nabla^{2})\right]S_{jk}\,d\xi.
\end{multline}
\end{subequations}

Here $a$ is the semi-major axis, $r$ is the aspect ratio, $e=\sqrt{1-(\min(r,1/r))^{2}}$ the eccentricity, $\alpha=ae$, $\xi$ the spheroidal coordinate, and $X_{A},Y_{A},X_{C},Y_{C},Y_{H}$ are resistance functions depending only on aspect ratio \citep[see also Appendix~\ref{appB}]{kim2013microhydrodynamics}. Also, $\boldsymbol{\varepsilon}$ is the Levi-Civita tensor.  For an oblate spheroid, the above expressions are changed by substituting $\alpha\rightarrow i |\alpha|$ \citep{hasimoto1983extension}. The orientation vector $\boldsymbol{p}$ describes the particle symmetry axis (see figure \ref{fig:diag}), with components $p_{x}=\sin\theta\sin\phi$, $p_{y}=\cos\phi$, $p_{z}=\cos\theta\sin\phi$. In the deep-water limit, the background strain-rate tensor simplifies to
\begin{subequations}\label{eq:str}
\begin{equation}
S_{xx}=-kA\omega e^{kz}\sin(kx-\omega t), \qquad 
S_{xz}=kA\omega e^{kz}\cos(kx-\omega t),
\end{equation}
\end{subequations}

with $S_{xx}=-S_{zz}$ and $S_{xz}=S_{zx}$. In this work, we use analytical expressions for the hydrodynamic force and torque on a spheroidal particle in a surface gravity wave by explicitly evaluating the integrals in (\ref{eq:fsge}). A detailed analysis of finite-size corrections is presented in \S\ref{sec:fssec}; until then, we approximate the background flow in the neighborhood of the spheroid by its local linearisation about the particle centroid. In this approximation, the force reduces to a drag proportional to the slip velocity between the particle and the surrounding flow, while the torque depends solely on the local velocity gradients. 

At particle scales, the flow is accurately described by Stokesian dynamics; however, at larger scales fluid inertia becomes significant. For a particle settling in a background wavy flow, inertia arises from multiple mechanisms.  
\begin{figure}
    \centering
    \includegraphics[scale=0.35]{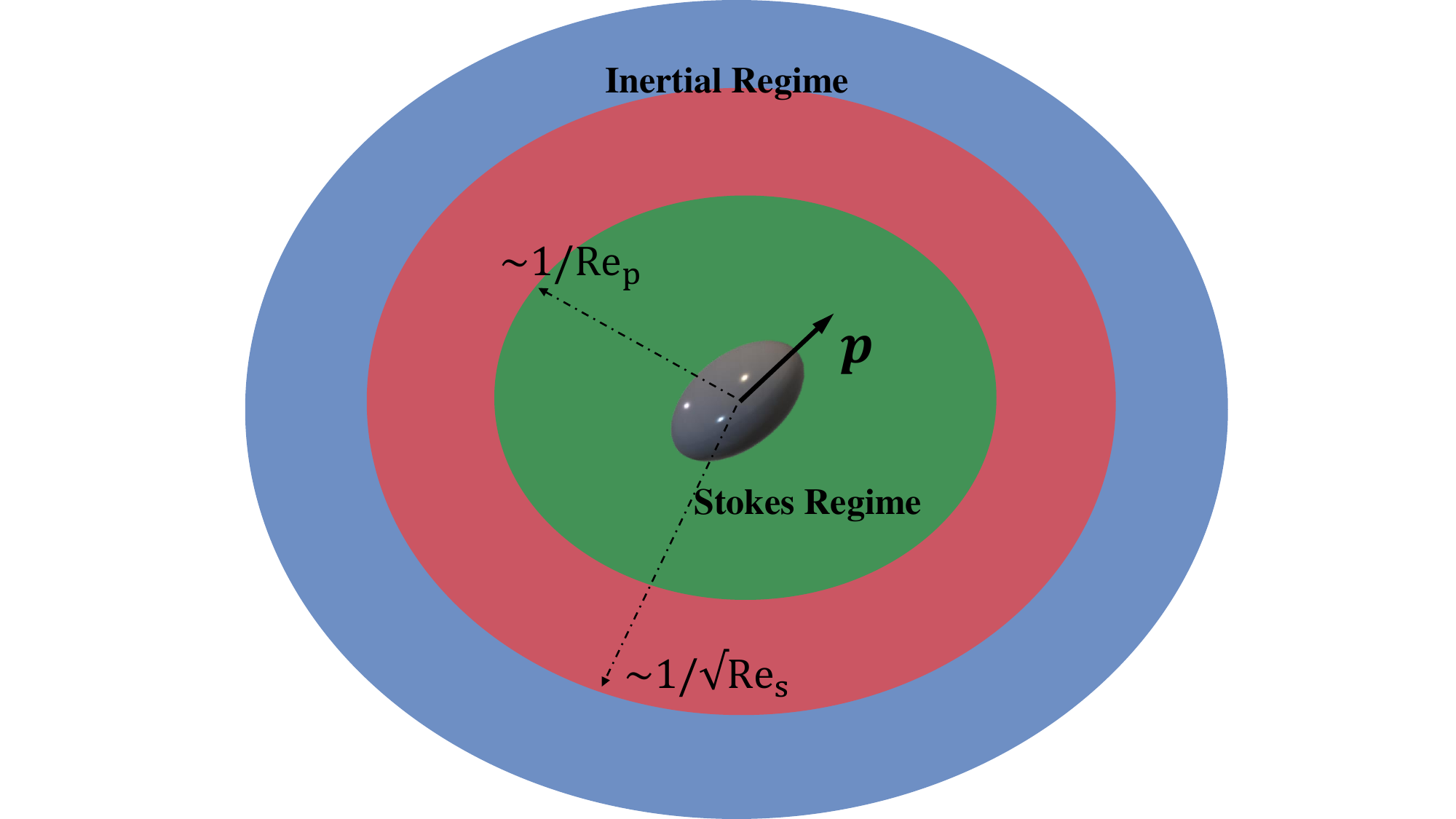}
    \caption{Schematic diagram illustrating various inertial regimes based on particle and shear Reynolds numbers. The green region denotes the viscosity-dominated Stokes regime, while fluid inertia becomes increasingly important outside this region.}
    \label{fig:IR}
\end{figure}
Upon non-dimensionalisation with $x^{*}=kx$, $t^{*}=\omega t$, and $u^{*}=u/c_{p}$, we identify four Reynolds numbers,  
\begin{equation}\label{eq:ndim}
\mathrm{Re}_{p}=\dfrac{Wa}{\nu}, \qquad 
\mathrm{Re}_{s}=\dfrac{\dot{\gamma}a^{2}}{\nu}, \qquad 
\mathrm{Re}_{w}=\dfrac{c_{p}}{\nu k}, \qquad 
\mathrm{Re}_{r}=\dfrac{\omega a^{2}}{\nu},
\end{equation}
which represent particle-, shear-, wave-, and rotation-based inertia, respectively. Their relative magnitudes determine the dominant physical mechanism (figure \ref{fig:IR}). For example, $\mathrm{Re}_{p}/\mathrm{Re}_{w}\sim(W/c_{p})(a/\lambda)\ll 1$ for small particles, though large slip velocities may still generate appreciable inertial corrections. Similarly, $\mathrm{Re}_{p}\gg \mathrm{Re}_{s}^{1/2}$ for long waves and small $a$, while $\mathrm{Re}_{w}\gg \mathrm{Re}_{s}$ for realistic amplitudes. In the regime of interest, highlighted in figure \ref{fig:IR}, the settling Reynolds number dominates over shear-based inertia, so translation-induced effects are expected to be the leading inertial contribution.  

In the present study, the settling velocity $U_{s}$ of the spheroid typically exceeds the characteristic shear-induced velocity scale $A\omega k$, where $A$ is the wave amplitude and $\omega=(gk)^{1/2}$. This suggests that inertial effects are primarily governed by translation across the inhomogeneous wave field, while the direct contribution of the ambient shear may be subdominant at leading order. For propagating waves, the relevant unsteadiness arises from the relative velocity $U_{r}=U_{s}-c_{p}$, such that the effective frequency is $\omega\sim kU_{r}$. By contrast, for standing waves or weak-amplitude long waves, the corresponding estimates of $\Omega a^{2}/\nu$ (with $\Omega$ a characteristic rate of slip-velocity variation) remain small, reinforcing the view that ambient shear contributes only indirectly.  

This motivates a quasi-steady treatment of the inertial torque: the $O(\mathrm{Re}_{p})$ correction can be retained while higher-order unsteady drag and torque corrections are neglected. In this limit, the memory dependence of the disturbance flow is weak, and the outer-region dynamics relax on a timescale long compared with the wave period. The quasi-steady assumption is therefore reasonable for the present parameter regime, but would become restrictive if $O(\mathrm{Re}_{p})$ drag or higher-order torques were included, since in that case the development of the steady outer response proceeds more slowly.  

Particle inertia, quantified by the Stokes number $\mathrm{St}=\omega\tau_{p}$, is vanishingly small here ($\mathrm{St}\to 0$). Here, $\tau_{p}$ is a relaxation time of a particle defined as $\tau_{p}=m/6\pi \mu a$. We therefore adopt the overdamped limit of (\ref{eq:odfor}, \ref{eq:odrot1}) and restrict attention to the effects of fluid inertia arising solely from sedimentation. In what follows, we retain only the translation-induced $O(\mathrm{Re}_{p})$ torque as the leading inertial correction to Jeffery dynamics.  

Including both viscous and inertial contributions, the torque on a spheroid can be written as
\begin{multline}\label{eq:initor1}
T_{i}=8\pi\mu a^{3}\left[\{X_{C}p_{i}p_{j}-Y_{C}(\delta_{ij}-p_{i}p_{j})\}(\Omega^{\infty}_{j}-\Omega_{j})
-Y_{H}\varepsilon_{ijl}p_{l}p_{k}S_{jk}\right] \\
-\rho_{f}a^{3}\mathcal{F}\,(W_{l}p_{l})(\varepsilon_{imn}W_{m}p_{n}),
\end{multline}
where the first bracketed term represents the Jeffery torque and the second term is the inertial correction, with $\mathcal{F}$ a shape factor depending only on particle aspect ratio \citep[see also Appendix~\ref{appB}]{dabade2015effects}.

Finally, under force- and torque-free conditions, the coupled equations of motion reduce to
\begin{subequations}\label{eq:3dem}
\begin{equation}
\dot{x}=kA e^{z}\cos(x-t)-\mathrm{Sv}(\mathcal{X}_{A}-\mathcal{Y}_{A})\sin\theta\cos\theta\sin^{2}\phi,
\end{equation}
\begin{equation}
\dot{y}=-\mathrm{Sv}(\mathcal{X}_{A}-\mathcal{Y}_{A})\cos\theta\sin\phi\cos\phi,
\end{equation}
\begin{equation}
\dot{z}=kA e^{z}\sin(x-t)-\mathrm{Sv}[\{\mathcal{X}_{A}-\mathcal{Y}_{A}\}\cos^{2}\theta\sin^{2}\phi+\mathcal{Y}_{A}],
\end{equation}
\begin{equation}
\dot{\theta}=\mathcal{B}kA e^{z}\cos(x-t+2\theta)-\mathrm{Sv}^{2}\mathrm{Re}_{w}\mathcal{F}_{p}\sin 2\theta,
\end{equation}
\begin{equation}
\dot{\phi}=\tfrac{1}{2}\left[\mathcal{B}kA e^{z}\sin(x-t+2\theta)+\mathrm{Sv}^{2}\mathrm{Re}_{w}\mathcal{F}_{p}(1+\cos 2\theta)\right]\sin 2\phi,
\end{equation}
\end{subequations}
where $\mathrm{Sv}=W/c_p=g\tau_{p}k(1-\gamma)/\omega$ is the settling parameter, $\mathcal{B}=Y_{H}/Y_{C}=(r^2-1)/(r^2+1)$ the Bretherton constant, and $\mathcal{F}_{p}=\mathcal{F}/(16\pi Y_{C}X_{A}Y_{A})$ the modified shape factor. Also, $\gamma$ denotes the density ratio, defined as $\gamma=\rho_{f}/\rho_p$. For prolate spheroids $0<\mathcal{B}<1$ ($\mathcal{F}_{p}<0$), and for oblates $-1<\mathcal{B}<0$ ($\mathcal{F}_{p}>0$). Also, $\mathcal{X}_{A}$ and $\mathcal{Y}_{A}$ are the mobility functions, and are the inverse of the resistance functions. These equations govern the coupled translation–rotation dynamics of spheroids in surface gravity waves, capturing the competition between Jeffery and inertial torques. 

The equation for orientation dynamics [see (\ref{eq:3dem}d--e)] contains an inertial torque term of $\mathcal{O}(\mathrm{Sv}^{2}\mathrm{Re}_{w})$. In earlier studies on anisotropic particles settling in homogeneous isotropic turbulence \citep{menon2017theoretical,anand2020orientation,gustavsson2021effect, roy2023orientation}, the inertial torque was found to be $\mathcal{O}(\mathrm{Sv}^{2})$ because the characteristic Reynolds number associated with a Kolmogorov eddy is $\mathcal{O}(1)$. In contrast, the present work considers wave-driven flows where $\mathrm{Re}_{w}$ is large. As we will elaborate in Section~\ref{sec:weak_settling}, this makes the inertial torque effectively $\mathcal{O}(\mathrm{Sv})$, thereby allowing it to significantly influence the orientation dynamics even when inertial effects are considered weak.

\section{Role of settling in translational and orientation dynamics}

Earlier studies have shown that a neutrally buoyant spheroid in surface gravity waves attains a shape-dependent, wave-preferred orientation while translating horizontally \citep{dibenedetto_2018}. We examine how negative buoyancy modifies this behaviour. For a heavy particle, alignment arises from the balance between buoyancy and quasi-steady drag, coupling orientation and translation. \citet{pujara2023wave} have shown via multiple-scale analysis that the steady alignment of a heavy spheroid depends only on aspect ratio and not on size or density. In contrast, we find that a settling spheroid deviates from this tracer-like behaviour: the preferred orientation depends on the initial state and settling parameters. Since orientation transients directly affect particle trajectories, the long-time alignment and its approach are central to the transport problem. At depth, when wave effects decay, the motion reduces to quiescent settling, but the final orientation retains memory of the wave–settling interplay and initial conditions.  

To isolate the role of settling, we first neglect both inertial torque and orientation-dependent gravitational settling. The resulting equations for a heavy point particle are  
\begin{eqnarray} \label{eq:isoem}
    \dot{x}=\epsilon e^{z}\cos(x-t),\\
    \dot{z}=\epsilon e^{z}\sin(x-t)-\mathrm{Sv},\\
\dot{\theta}=\mathcal{B}\epsilon e^{z}\cos(x-t+2\theta).
\end{eqnarray}
Here anisotropy appears only in the orientation dynamics. Also, $\epsilon=kA$ is a small parameter. For nearly neutrally buoyant particles, if we assume $\mathrm{Sv}=\mathcal{O}(\epsilon^{2})$ the analysis reduces to that of the neutrally buoyant limit. By contrast, for particles of size $\mathcal{O}(10^{-3})$ m transported by waves of order tens of centimetres, we obtain $\mathrm{Sv}\sim\mathcal{O}(10^{-1})$ and $\epsilon\sim\mathcal{O}(10^{-2})$ (figure \ref{fig:count1}). Moreover, wave and settling effects are comparable near the surface, but buoyancy dominates with increasing depth. Accordingly, we focus on regimes where settling remains non-negligible by considering $\mathrm{Sv} = \mathcal{O}(1)$.  

On expanding in small parameter $\epsilon$,
\begin{subequations}\label{eq:serrps}
\begin{equation}
    x=x_{0}+\epsilon x_{1}+\epsilon^{2}x_{2}+\cdots,\qquad
    z=z_{0}+\epsilon z_{1}+\epsilon^{2}z_{2}+\cdots,\qquad
    \theta=\theta_{0}+\epsilon\theta_{1}+\epsilon^{2}\theta_{2}+\cdots,
\end{equation}
\end{subequations}
we obtain at leading order $x_{0}=X$, $z_{0}=Z-\mathrm{Sv}\,t$, $\theta_{0}=\Theta$, where $(X, Z)\, \& \,\Theta$ specify the initial position and orientation respectively. In the long-time limit $t\to\infty$, the horizontal displacement (``spreading'') $x_{\infty}$ and long-time orientation $\theta_{\infty}$ are  
\begin{eqnarray}
x_{\infty}&=& X+\epsilon\dfrac{e^{Z}(\mathrm{Sv}\cos X+\sin X)}{1+\mathrm{Sv}^{2}}
+\epsilon^{2}\dfrac{e^{2Z}}{2(\mathrm{Sv}+\mathrm{Sv}^{3})}+\mathcal{O}(\epsilon^3),\label{eq:rpsvxth}\\
\theta_{\infty}&=& \Theta+\epsilon\dfrac{\mathcal{B} e^{Z}\{\mathrm{Sv}\cos(X+2\Theta)+\sin(X+2\Theta)\}}{1+\mathrm{Sv}^{2}} \nonumber\\
&&+\dfrac{\epsilon^{2}}{2\mathrm{Sv}\left(1+\mathrm{Sv}^{2}\right)^{2}}
\Big[\mathcal{B}e^{2Z}\{(\mathrm{Sv}^{2}+1)\cos 2\Theta+2\mathcal{B}\mathrm{Sv}^{2}\cos(2(X+2\Theta)) \nonumber\\
&&+(\mathrm{Sv}^{2}+1)(\mathcal{B}-\mathrm{Sv}\sin 2\Theta)-\mathcal{B}\mathrm{Sv}(\mathrm{Sv}^{2}-1)\sin(2(X+2\Theta))\}\Big]+\mathcal{O}(\epsilon^3).  
\end{eqnarray}

\begin{figure}
\centering  
\subfigure[]{\includegraphics[width=0.475\linewidth]{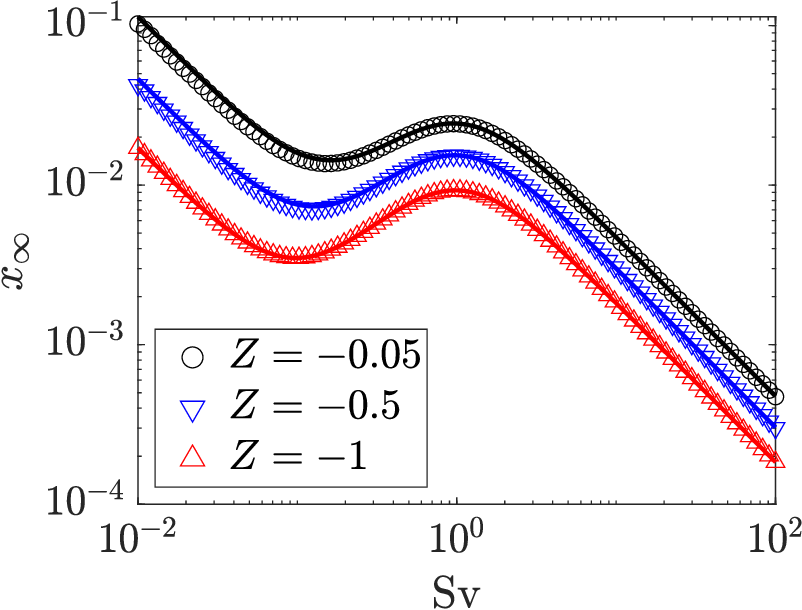}}
\subfigure[]{\includegraphics[width=0.475\linewidth]{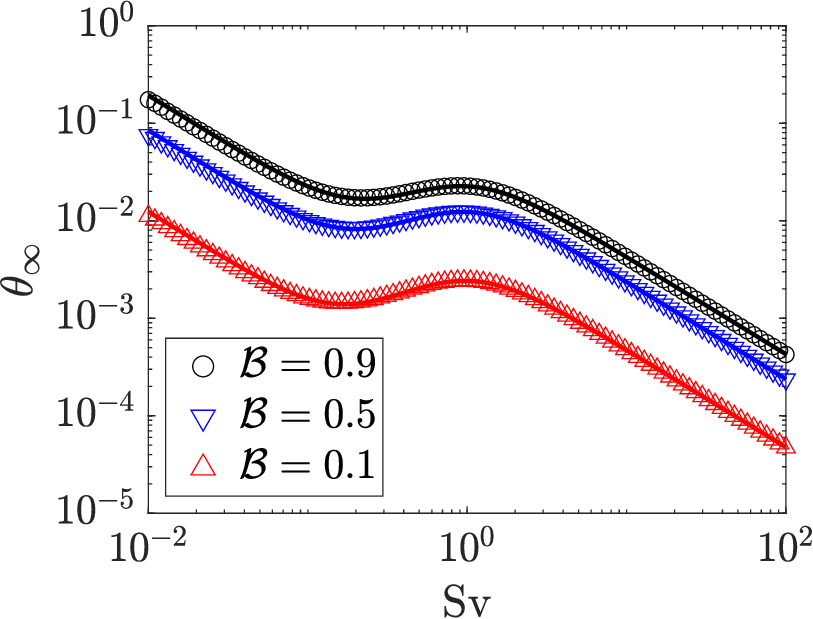}}
\caption{ Comparison of asymptotic predictions of $x_{\infty}$ and $\theta_{\infty}$ with numerical results. Scatter points denote numerical simulations, and solid lines show the corresponding asymptotic approximations. (a) Horizontal spreading $x_{\infty}$ versus $\mathrm{Sv}$, and (b) Preferred orientation $\theta_{\infty}$ versus $\mathrm{Sv}$. Here, $\mathcal{B}=0.9$, $\epsilon=0.05$.} 
\label{fig:asymisotrop}
\end{figure}

\begin{figure}
\centering  
\subfigure[]{\includegraphics[width=0.51\linewidth]{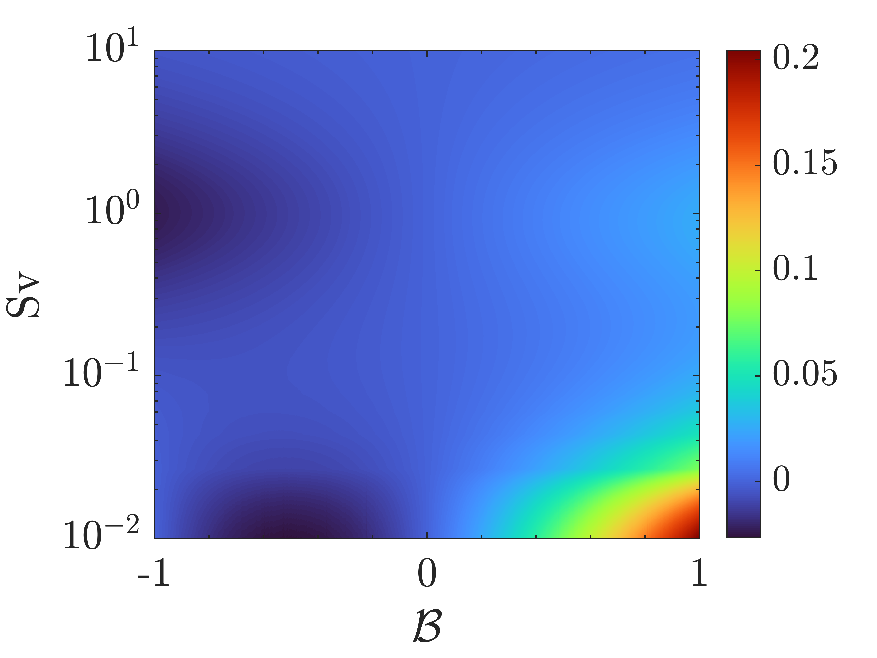}}
\subfigure[]{\includegraphics[width=0.47\linewidth]{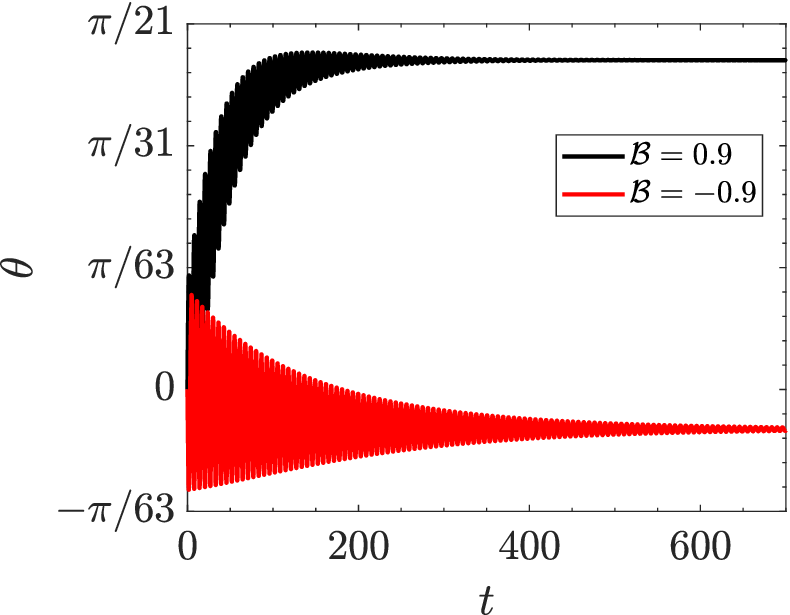}}
\caption{ (a) Contour plot of long-time orientation $\theta_{\infty}$ in a $\mathrm{Sv}-\mathcal{B}$ plane obtained from (\ref{eq:isoem}), and (b) temporal evolution of $\theta$ for prolate and oblate spheroids, computed from (\ref{eq:velff}). For this figure, parameters are: $\gamma=0.99$, $\epsilon=0.05$ and $\mathrm{Sv}=5\times10^{-3}$.} 
\label{fig:theiso}
\end{figure}

Figure \ref{fig:asymisotrop} compare the asymptotic predictions with the numerical results. The spreading length decays as $x_{\infty}\sim \mathrm{Sv}^{-1}$ for large $\mathrm{Sv}$, consistent with the spherical-particle scaling of \citet{eames2008}, while intermediate $\mathrm{Sv}$ exhibits a non-monotonic dependence. Particles released closer to the interface exhibit larger spreading, consistent with earlier findings by \citet{dib_clark_pujara_2022} for spherical particles. Moreover, a particle exhibits higher spreading $x_{\infty}$ at lower values of $\mathrm{Sv}$.  The long-time orientation increases with aspect ratio for prolates, while for oblates it peaks at intermediate $\mathrm{Sv}$ and small $\mathcal{B}$. Thin disks achieve maximum orientation at intermediate $\mathrm{Sv}$, whereas nearly spherical particles remain close to their initial orientation (see figure \ref{fig:theiso}).

Next, we consider the more realistic case where buoyancy modifies both translation and orientation of a spheroid with an arbitrary aspect ratio. Unlike the isotropic limit, the horizontal velocity now depends on the mobility functions and thus on particle orientation. The coupled equations of motion for a negatively buoyant spheroid are  
\begin{subequations}\label{eq:velff}
\begin{equation}
    \Dot{x}=\epsilon e^{z}\cos(x-t)-\mathrm{Sv}(\mathcal{X}_{A}-\mathcal{Y}_{A})\sin\theta\cos\theta,
\end{equation}
\begin{equation}
    \Dot{z}=\epsilon e^{z}\sin(x-t)-\mathrm{Sv}[(\mathcal{X}_{A}-\mathcal{Y}_{A})\cos^{2}\theta+\mathcal{Y}_{A}],
\end{equation}
\begin{equation}
    \Dot{\theta}=\mathcal{B}\epsilon e^{z}\cos(x-t+2\theta).
\end{equation}
\end{subequations}

 With only quasi-steady drag, the mean translation in the wave direction is linear in time, and the particle oscillates indefinitely about the wave-preferred orientation. With buoyancy included, translation can occur in either direction, and the long-time behaviour is controlled by $\mathrm{Sv}$ through its effect on the preferred orientation. The rotational equation remains unchanged, and inertial torque is neglected. As shown in figure \ref{fig:theiso}b, numerical solutions of (\ref{eq:velff}) for prolate and oblate spheroids ($X=0$, $Z=-0.05$, $\Theta=0$, $\epsilon=0.05$, $\gamma=0.99$) reveal that a heavy particle achieves a steady buoyancy-induced alignment during sedimentation, with oscillations in $\theta$ gradually damped as wave effects decay exponentially with depth.

We solve the above system using a regular perturbation expansion. When the anisotropic mobility functions are retained in the translational dynamics, the long-time horizontal displacement does not saturate, and no finite spreading length, $x_\infty$, exists; the particle therefore continues to drift horizontally with a non-zero mean velocity. The corresponding long-time preferred orientation is

\begin{eqnarray}\label{eq:anisothetal1} 
  \theta_{\infty}&=&\Theta  +\epsilon \,\dfrac{\mathcal{B}e^{Z}\{\mathrm{Sv}(\mathcal{X}_{A}-\mathcal{Y}_{A})\cos X+\mathrm{Sv}(\mathcal{X}_{A}+\mathcal{Y}_{A})\cos{(X+2\Theta)+2\sin{(X+2\Theta)}}\}}{2\{(1+\mathrm{Sv}^{2}\mathcal{X}_{A}^{2})\cos^{2}\Theta+(1+\mathrm{Sv}^{2}\mathcal{Y}_{A}^{2})\sin^{2}\Theta+\mathrm{Sv}(\mathcal{X}_{A}-\mathcal{Y}_{A})\sin{2\Theta}\}}+\nonumber\\
  &&\epsilon^2 \,\mathcal{J}(\mathcal{B},\mathrm{Sv};X,Z,\Theta)+\mathcal{O}(\epsilon^3),
\end{eqnarray}
where $\mathcal{J}(\mathcal{B},\mathrm{Sv};X,Z,\Theta)$ denotes an algebraically lengthy expression. Its explicit forms for specific initial conditions are provided in the Appendix \ref{app_asym_orientation}.

\begin{figure}
\centering  
\subfigure[]{\includegraphics[width=0.49\linewidth]{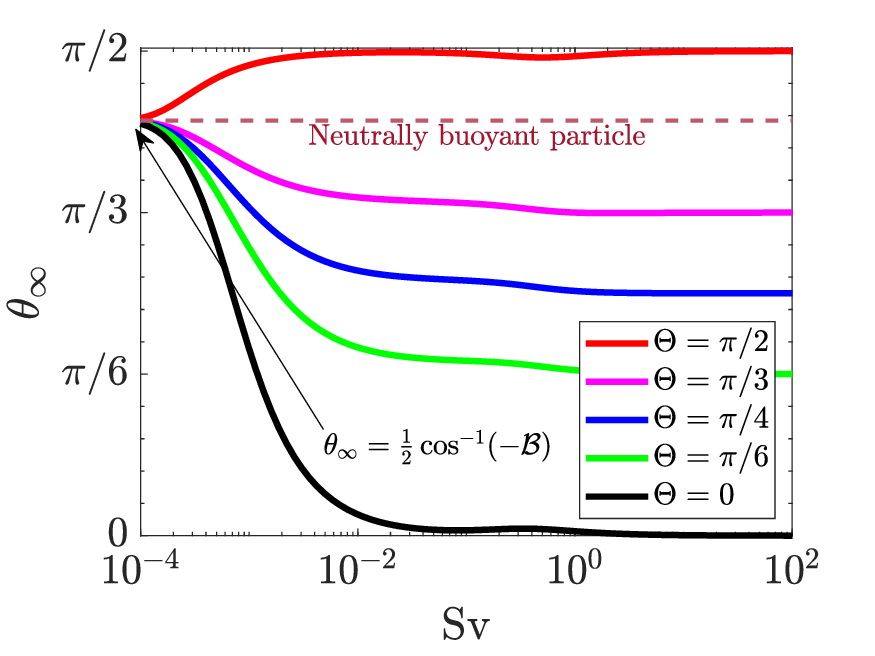}}
\subfigure[]{\includegraphics[width=0.5\linewidth]{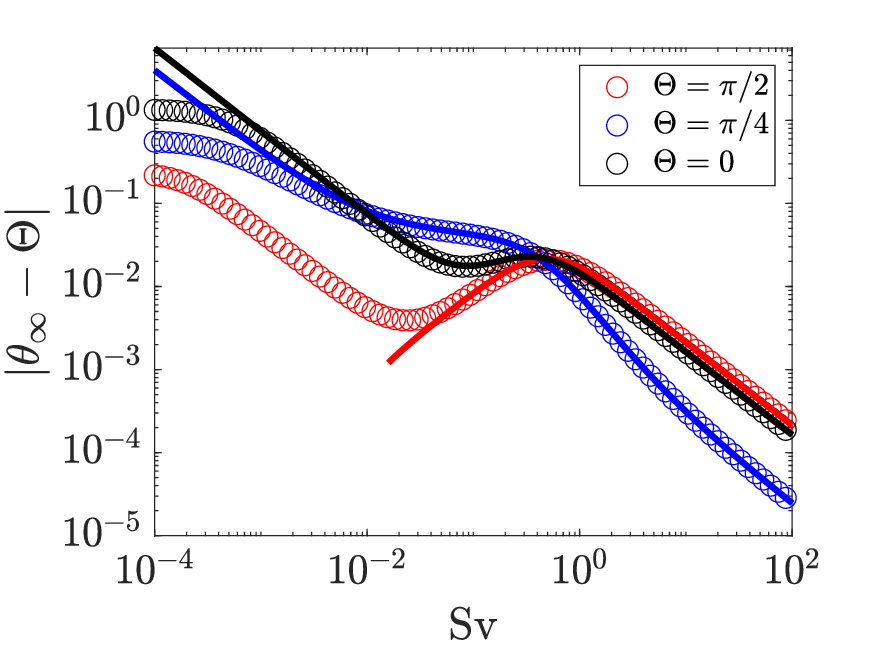}}
\caption{ (a) Variation of $\theta_{\infty}$ as a function of $\mathrm{Sv}$, and (b) Difference of a long-time orientation from the initial orientation $|\theta_{\infty}-\Theta|$ versus $\mathrm{Sv}$. In the right panel, the curves indicate the asymptotic calculations, and the circles correspond to the numerical results. The results are shown for $\mathcal{B}=0.9$.} 
\label{fig:thinfysv}
\end{figure} 

Lateral dispersion controls how anisotropic particles such as marine snow, plankton and microplastics spread relative to surface wave fields, which in turn affects horizontal transport, clustering, and vertical export. It is therefore useful to relate long-time lateral drift to the preferred orientation and settling characteristics of a spheroid. In figure \ref{fig:thinfysv}, we show the long-time orientation $\theta_{\infty}$ attained by a negatively buoyant particle as a function of $\mathrm{Sv}$. The dashed lines denotes the long-time orientation of a neutrally buoyant particle and the solid lines are corresponds to negatively buoyant spheroid, for different initial orientations. For large $\mathrm{Sv}$, $\theta_{\infty}$ coincides with the initial angle $\Theta$, as captured by the leading-order term in (\ref{eq:anisothetal1}). As $\mathrm{Sv}$ decreases, $\theta_{\infty}$ departs from $\Theta$, with the strongest variation at small $\mathrm{Sv}$. In the limit $\mathrm{Sv}\rightarrow 0$, the preferred orientation approaches the anisotropic tracer result $\theta_{\infty}=(1/2)\cos^{-1}(-\mathcal{B})$ \citep{dibenedetto_ouellette_2018,pujara2023wave}, becoming independent of $\Theta$ and depending only on particle shape. In the next panel, we compare the asymptotic predictions, shown by curves, with the numerical results, indicated by circles. The asymptotic solution agrees well with numerics down to intermediate $\mathrm{Sv}$.

\begin{figure}
\centering  
\subfigure[]{\includegraphics[width=0.49\linewidth]{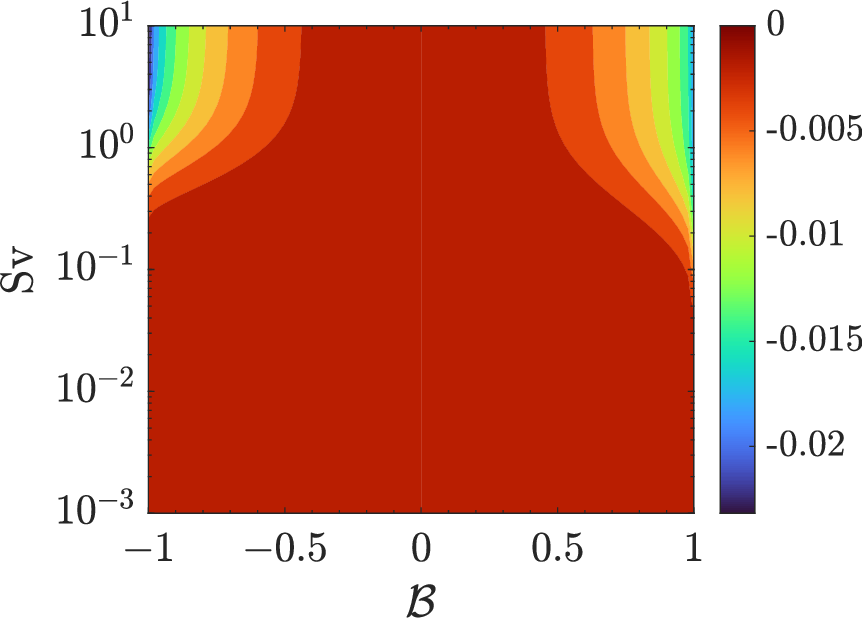}}
\subfigure[]{\includegraphics[width=0.49\linewidth]{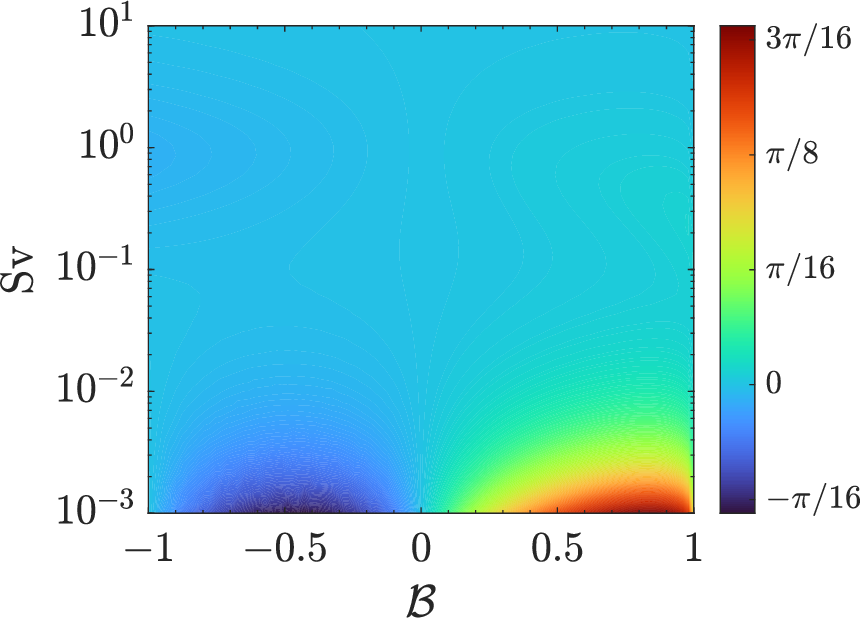}}
\caption{ (a) Contour plot of horizontal drift velocity $\mathrm{v}_{x}$ and (b) orientation in a long-time limit $\theta_{\infty}$ in the $\mathrm{Sv}-\mathcal{B}$ plane. The initial conditions are $X=0$, $Z=-0.05$ and $\Theta=0$.} 
\label{fig:anisomobcont}
\end{figure}

In figure \ref{fig:anisomobcont}, we present the long-time dynamics in the $\mathrm{Sv}$--$\mathcal{B}$ plane. The left panel shows the numerically computed horizontal velocity component $\mathrm{v}_{x}$, for both prolate and oblate spheroids, with $X=0$,  $Z=-0.05$ and $\Theta=0$. Rod- and disk-like particles exhibit larger drift velocities than nearly spherical ones. In the presence of buoyancy and anisotropic mobility, the particle experiences a persistent horizontal translation; once it reaches its preferred orientation, the horizontal displacement increases linearly in time, and the drift velocity approaches a constant value far from the interface. Consequently, determining the preferential alignment arising from buoyancy-induced settling is essential, since this alignment governs the direction of particle motion. In that regime, the drift is controlled solely by settling. Close to the interface, however, wave-induced oscillations produce an orbital motion (OM) before the trajectory relaxes to a wave-absent regime (WOM).

In the Stokesian regime, the particle relaxes to a buoyancy-induced preferred alignment that minimises drag while settling. The right panel of figure \ref{fig:anisomobcont} shows $\theta_{\infty}$ in the $\mathrm{Sv}$--$\mathcal{B}$ plane. For prolates, $\theta_{\infty}$ is largest at low $\mathrm{Sv}$, whereas oblates attain larger magnitudes of orientations at low and intermediate $\mathrm{Sv}$. The yellow region at intermediate $\mathrm{Sv}$ indicates a band of large $\theta_{\infty}$ for prolates. For $\mathrm{Sv}=\mathcal{O}(10)$, $\theta_{\infty}$ is small for all $\mathcal{B}$, so the orientation changes little from its initial value. Nearly spherical particles remain close to $\theta_{\infty}\approx 0$. The asymptotic expansion (\ref{eq:anisothetal1}) captures these trends: at large $\mathrm{Sv}$, the steady orientation $\theta_{\infty}$ scales as $\mathrm{Sv}^{-1}$, whereas at smaller $\mathrm{Sv}$ the higher-order terms become necessary to capture the dynamics. Comparing with the nearly isotropic limit shows qualitatively similar contour shapes, although the magnitude of $\theta_{\infty}$ is reduced across all $\mathrm{Sv}$ and $\mathcal{B}$ when anisotropic mobility is included. This reflects the coupling between translation and rotation: the modified translation equation and its appearance in the orientation dynamics change the quantitative alignment behaviour.

\begin{figure}
\centering  
\subfigure[]{\includegraphics[width=0.49\linewidth]{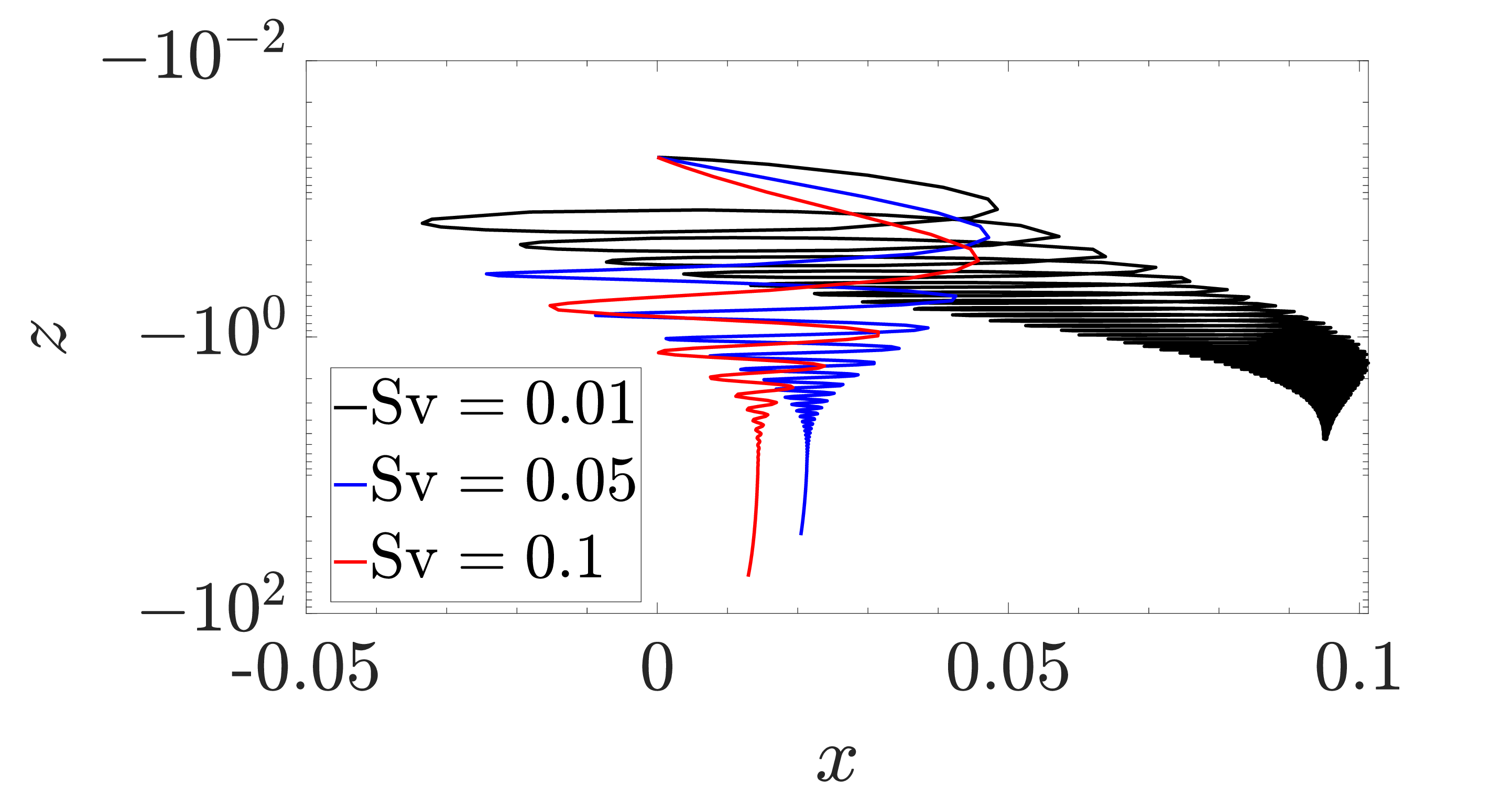}}
\subfigure[]{\includegraphics[width=0.49\linewidth]{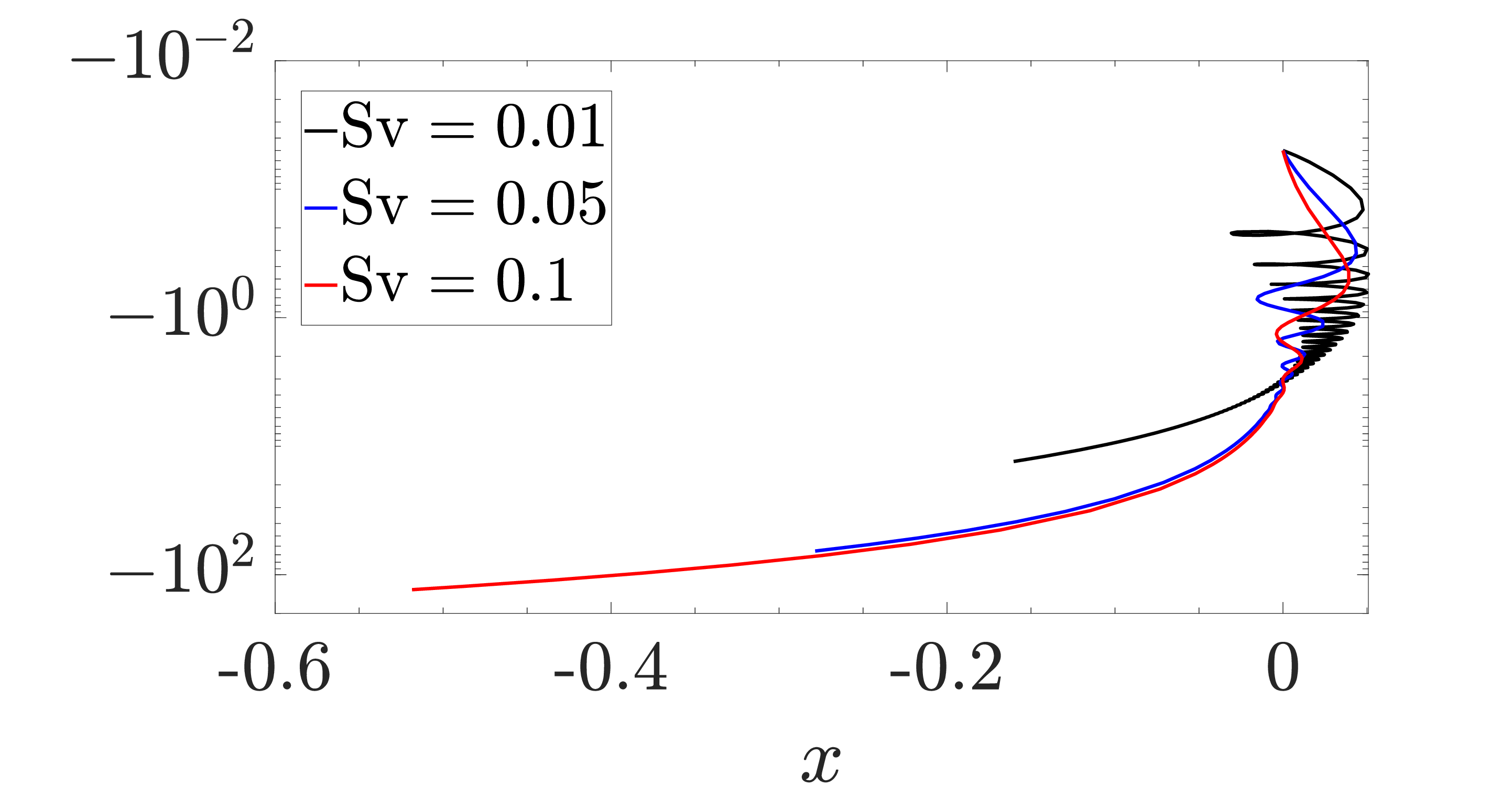}}
\caption{ Trajectories of a prolate spheroid shown till time $t=500$ for (a) $\mathcal{B}=0.1$ and (b) $\mathcal{B}=0.9$. The initial conditions are: $X=0$, $\Theta=0$, and $Z=-0.05$. The calculations are done with anisotropic mobility without considering inertial torque. } 
\label{fig:trjanisomobpro}
\end{figure}

In figure \ref{fig:trjanisomobpro}, we plot the trajectories of a prolate spheroid for different $\mathrm{Sv}$. As discussed above, the particle drifts against the wave propagation direction. Near the surface, the particle undergoes orbital motion (OM) driven by the wave; this decays as the particle settles away from the interface. The damping of the orbital loops is controlled by the approach to a steady orientation, which in turn generates an oscillatory drag during settling. Further, the frequency of OM decreases with increasing $\mathrm{Sv}$. For a nearly spherical particle at low $\mathrm{Sv}$, the oscillation frequency is high, and the particle spends most of its trajectory in OM at a given instant. As $\mathrm{Sv}$ increases, OM weakens, and the particle spends more time in a WOM, with larger net horizontal drift. For fixed $\mathrm{Sv}$, a prolate spheroid with $\mathcal{B}=0.9$ settles more rapidly and exhibits greater horizontal drift than the $\mathcal{B}=0.1$ case. 

\begin{figure}
\centering  
\subfigure[]{\includegraphics[width=0.49\linewidth]{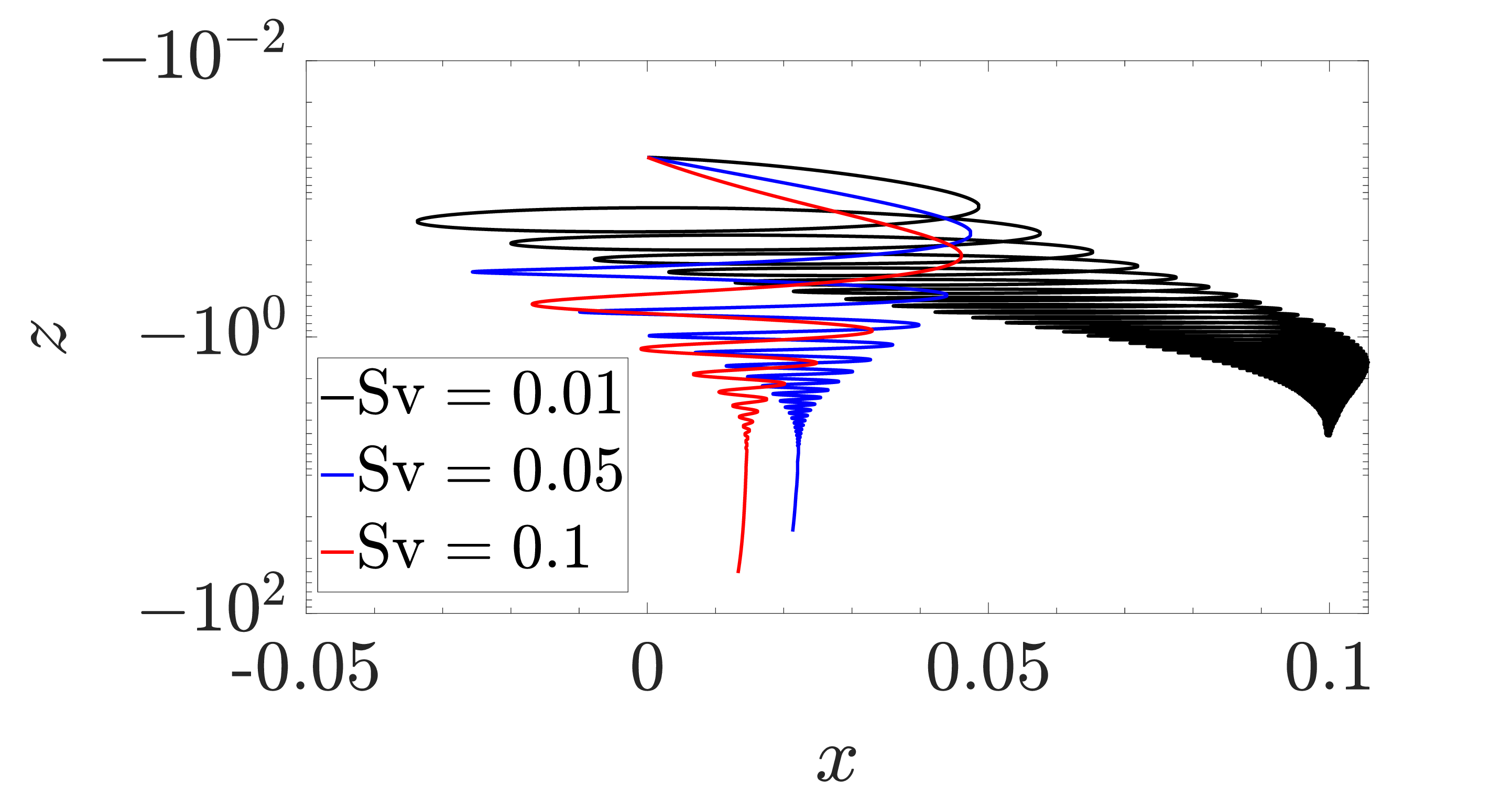}}
\subfigure[]{\includegraphics[width=0.49\linewidth]{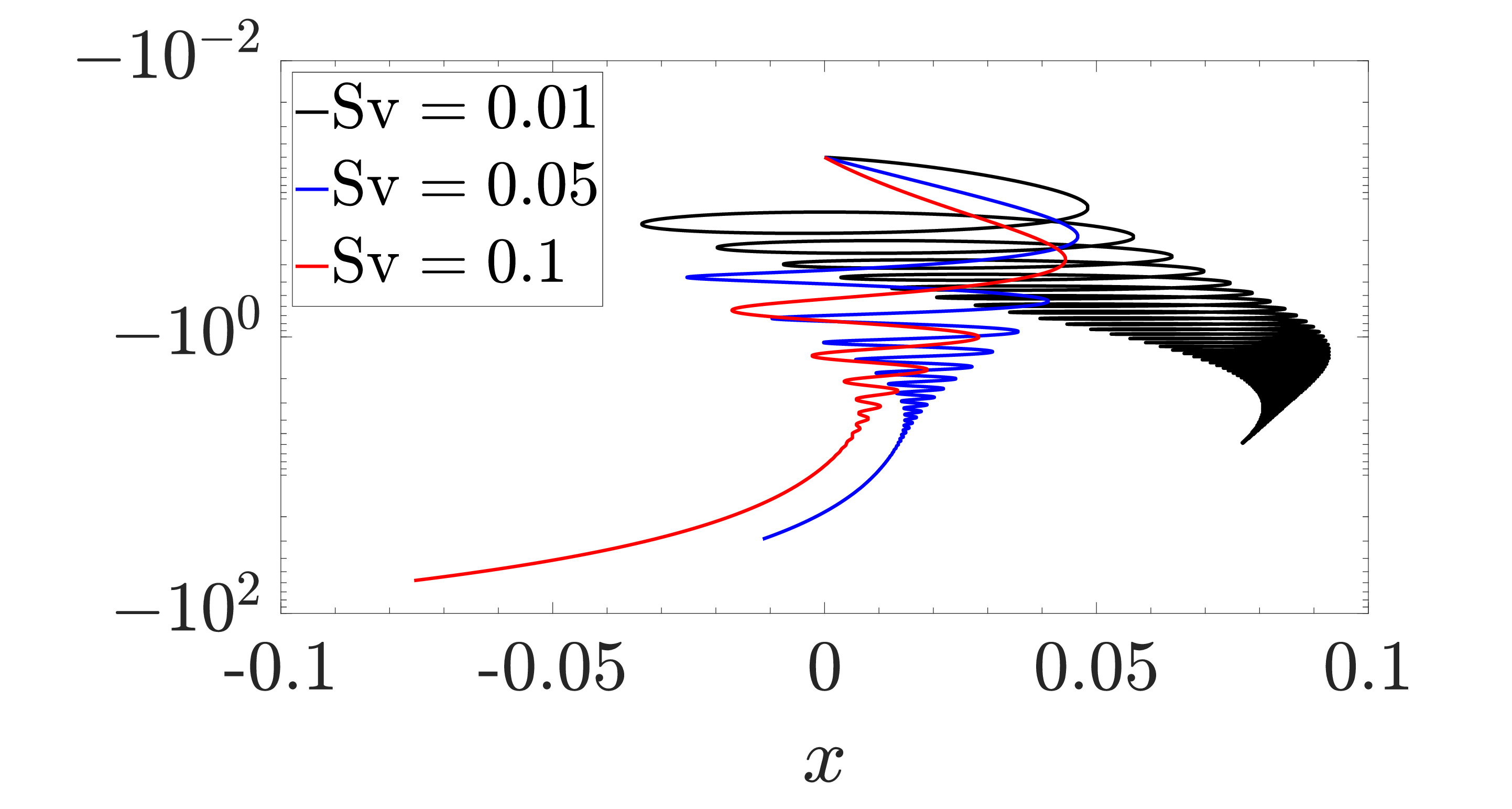}}
\caption{ Trajectories of an oblate spheroid shown till time $t=500$ for (a) $\mathcal{B}=-0.1$ and (b) $\mathcal{B}=-0.9$. The initial conditions are: $X=0$, $\Theta=0$, and $Z=-0.05$. The calculations are done with anisotropic mobility without considering inertial torque.} 
\label{fig:trjanisomobob}
\end{figure}

In figure \ref{fig:trjanisomobob}, we show the trajectories of an oblate spheroid up to $t=500$. Compared with prolate spheroids, oblates settle and drift less horizontally at large $\mathrm{Sv}$ and $|\mathcal{B}|$. Oblates also display stronger and more persistent OM at higher aspect ratios. For small $|\mathcal{B}|$, the transport is similar to the prolate case. As discussed earlier, oblates reorient more slowly than prolates, taking longer to reach their preferred alignment. The timing of this alignment controls the transition from OM to WOM and therefore the net horizontal displacement. A delayed alignment for oblates leads to longer residence in OM and hence smaller horizontal dispersion than prolates at comparable parameters. These trajectories highlight that particles spending more time in OM translate less horizontally than those that quickly enter WOM.

\begin{figure}
\centering  
\subfigure[]{\includegraphics[width=0.495\linewidth]{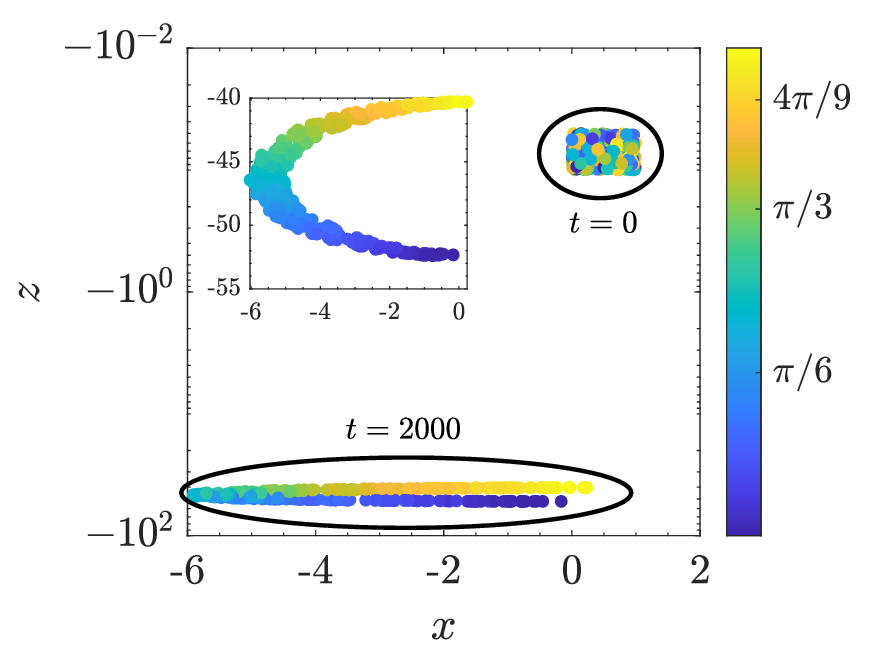}}
\subfigure[]{\includegraphics[width=0.49\linewidth]{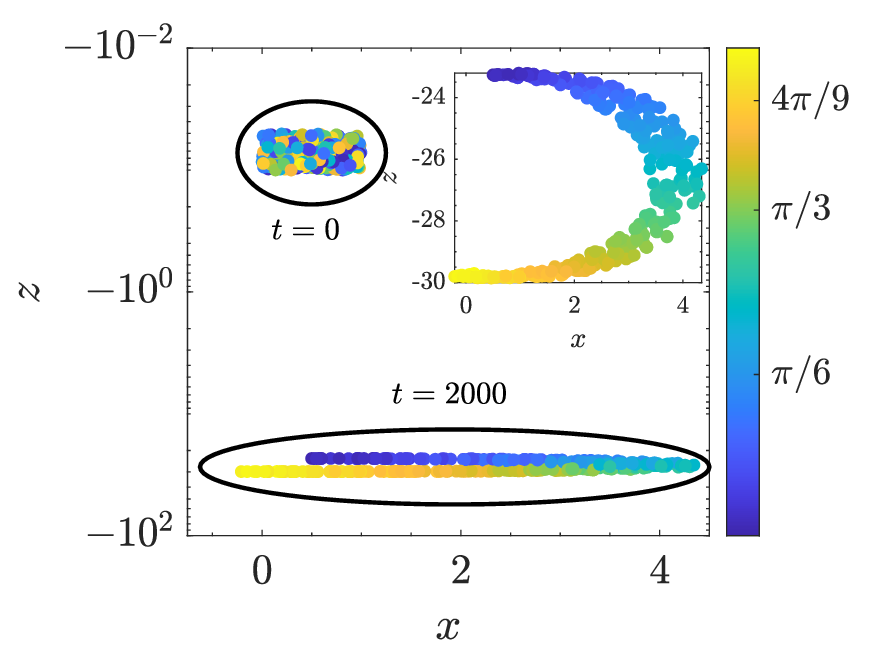}}
\caption{Dispersion of anisotropic particles in the $x-z$ plane. (a) Prolate spheroid and (b) Oblate spheroid. The colorbar represents the magnitude of initial orientation $\Theta$. Parameters: $\mathrm{Sv}=10^{-2}$ and $|\mathcal{B}|=0.9$. Inset figures provides a magnified view of the particle dispersion at $t=2000$.} 
\label{fig:dispsett}
\end{figure} 

Figure \ref{fig:dispsett} shows the dispersion of anisotropic particles in the $x$--$z$ plane, starting from random initial positions and orientations at $t=0$ and sampled again at $t=2000$. The colorbar indicates the initial orientation $\Theta$. Particles organize into orientation-dependent bands, with less dispersion for particles initially aligned either vertically or horizontally with the wave direction. Consistent with the settling part of (\ref{eq:velff}), the largest horizontal spread occurs for orientations near $\Theta\approx \pi/4$. Prolate spheroids drift predominantly against wave propagation, whereas oblates typically drift in the direction of the waves. However, because the net translation depends on both the initial orientation and the equilibrium angle after OM has decayed, oblates can also exhibit backward drift despite $\mathcal{Y}_{A}-\mathcal{X}_{A}>0$. Vertically aligned particles settle more than horizontally aligned ones, since vertical alignment minimises drag during settling. Consequently, prolate spheroids with $\Theta \approx 0$ and oblate spheroids with $\Theta \approx \pi/2$ settle the fastest. Overall, in the Stokesian regime, a negatively buoyant spheroid evolves toward a buoyancy-induced equilibrium orientation that optimizes settling while modulating lateral transport.

\begin{figure}
    \centering
    \includegraphics[width=0.65\linewidth]{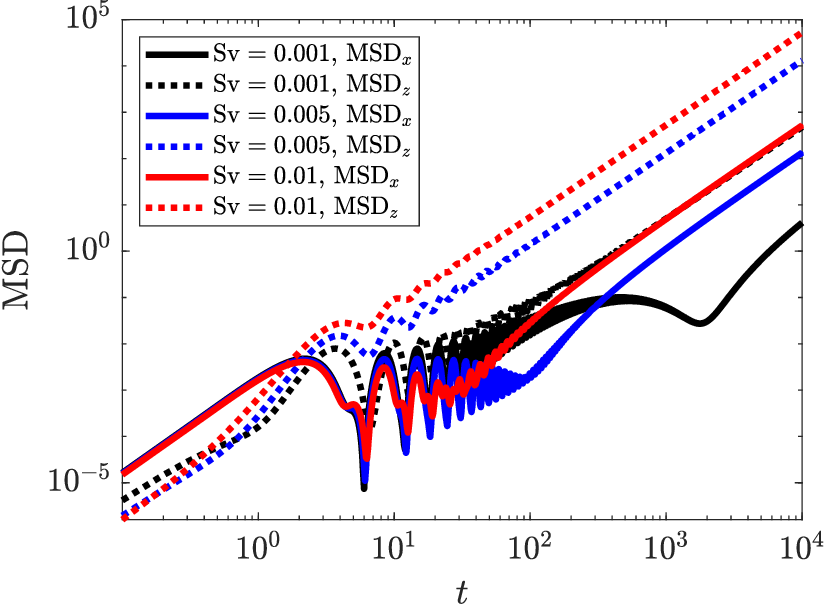}
    \caption{Mean-square displacement (MSD) curve of a prolate spheroid versus time for different values of $\mathrm{Sv}$. Solid lines indicate the MSD of a horizontal displacement, and dashed lines correspond to the vertical displacement. The result is shown for $\mathcal{B}=0.9$.}
    \label{fig:msd_aniso}
\end{figure}

The dispersion can be quantified via the mean-square displacement (MSD),
\begin{equation}
    \mathrm{MSD}=\left\langle \frac{1}{N} \sum^{N}_{i=1} \bigl\|x_{i}(t)-x_{i}(0)\bigr\|^{2}\right\rangle,
\end{equation}
where the angle brackets denote an ensemble average over particles with random initial positions and orientations. Figures \ref{fig:trjanisomobpro}--\ref{fig:trjanisomobob} highlight the contrast between prolate and oblate trajectories, with OM near the surface relaxing to steady drift at depth. The corresponding dispersion patterns in figure \ref{fig:dispsett} show orientation-dependent clustering. Figure \ref{fig:msd_aniso} displays the MSD of a prolate spheroid for different $\mathrm{Sv}$, with solid and dashed lines indicating horizontal and vertical components, respectively, for $\mathcal{B}=0.9$. We found that MSD grows approximately as $t^{2}$, and its magnitude decreases as $\mathrm{Sv}$ is reduced. As anticipated from the drift analysis, particles with smaller $\mathrm{Sv}$ display reduced horizontal and vertical excursions, confirming that settling strength strongly controls lateral dispersion in surface gravity waves.

\section{Role of inertial torque on particle's dynamics}

In the previous section, the particle orientation evolved solely under the quasi–steady Stokes torque generated by the wave–induced velocity gradients. We now incorporate the effects of weak fluid inertia on the coupled translation–orientation dynamics of a settling spheroid. Although $\mathrm{Re}\ll1$, inertial corrections become appreciable at distances $\sim a\,\mathrm{Re}^{-\alpha}$ from the particle, where the value of $\alpha(>0)$ depends on whether translational or shear–induced inertia dominates. Here, we retain only the leading contribution arising from translation. A negatively buoyant spheroid generates a fore–aft asymmetric wake as it settles, producing a hydrodynamic torque that drives the particle toward a preferred alignment.

In the absence of inertia, the particle preserves its initial orientation. With inertial torque, however, the broadside–on configuration becomes the stable equilibrium, as documented in laboratory and atmospheric observations of millimetre-scale rods \citep{jayaweera1965behaviour,zikmunda1972fall,platt1978lidar,sassen1980remote,breon2004horizontally,noel2010global, roy2023orientation}. The torque magnitude scales with particle size and settling speed \citep{dabade2015effects}, so large or fast–settling particles exhibit pronounced alignment. Experiments in wave-driven environments further show that negatively buoyant spheroids ultimately adopt this broadside orientation regardless of their initial state \citep{dib2019oriexp}.

\cite{sunberg2024parametric} have analyzed the influence of different non-dimensional parameters on the dispersion of particles, accounting for the inertial torque and wave currents in a horizontal direction. Based on their interest, the domain of their study is limited to the upper part of the flow field where wavy effects from the flow dominate, and particles are carried away due to currents. Although their parametric study reveals that the inertial torque is necessary to capture the realistic dynamics of a settling particle beyond the Stokesian regime, the effect of the inertial torque on the particle's translation and orientation dynamics remains to be understood. In contrast to purely buoyancy-driven motion—where horizontal translation can remain unbounded—we find that the inclusion of inertial torque leads to confined horizontal motion, a feature not reported in earlier studies. Specifically, the horizontal translation in the long-time limit correlates with the critical time at which the particle reaches its equilibrium orientation.

At low $\mathrm{Re}_{p}$, symmetry arguments imply that the inertial torque is proportional to $(\hat{\boldsymbol{W}}\cdot\boldsymbol{p})(\hat{\boldsymbol{W}}\times\boldsymbol{p})$ \citep{khayat_cox_1989,roy_2019}, where $\hat{\boldsymbol{W}}$ is the direction of settling. The torque vanishes when $\boldsymbol{p}$ is parallel or normal to $\hat{\boldsymbol{W}}$. The former is an unstable longitudinal alignment; the latter is the stable broadside–on state. In shear or wave-driven flows, the orientation dynamics arise from the competition between this inertial torque and the Jeffery torque. For the present problem, the appropriate inertial parameter is the wave Reynolds number $\mathrm{Re}_{w}$, defined earlier via the wavenumber and phase speed. The alignment timescale is set by $\mathrm{Sv}$, $\mathrm{Re}_{w}$, and the particle shape factor $\mathcal{F}_{p}$.

Including inertial torque in (\ref{eq:3dem}) yields the below equations of motion 
\begin{subequations}\label{eq:fiem}
    \begin{equation}
    \dot{x}=\epsilon e^{z}\cos(x-t)-\mathrm{Sv}\left(\mathcal{X}_{A}-\mathcal{Y}_{A}\right)\sin\theta\cos\theta,
\end{equation}
\begin{equation}
    \dot{z}=\epsilon e^{z}\sin(x-t)-\mathrm{Sv}\left[\left(\mathcal{X}_{A}-\mathcal{Y}_{A}\right)\cos^{2}\theta+\mathcal{Y}_{A}\right],
\end{equation}
\begin{equation}
    \dot{\theta}=\mathcal{B}\epsilon e^{z}\cos(x-t+2\theta)-\mathrm{Sv}^{2}\mathrm{Re}_{w}\mathcal{F}_{p}\sin2\theta.
\end{equation}
\end{subequations}

To determine the leading-order contribution to long-term horizontal motion considering inertial torque, it is essential to identify the permissible range of the settling parameter $\mathrm{Sv}$ over which the expression for spreading length remains valid. The expression for inertial torque applies in the regime $\mathrm{Re}_{p} < 1$, where the particle Reynolds number is given by the non-dimensional relation (see (\ref{eq:ndim})),
\[
\mathrm{Re}_{p} = \left(\frac{\sigma}{e}\right) \left(\dfrac{\mathrm{Sv}}{1 - \gamma}\right) \mathrm{Re}_{w}.
\]
Details regarding the finite-size parameter $\sigma$ are provided in \S\ref{sec:fssec}. For representative values, $\mathrm{Re}_{w} \sim \mathcal{O}(10^3)$, $\sigma \sim \mathcal{O}(10^{-2})$, and $\mathrm{Sv} \sim \mathcal{O}(10^{-3})$, we obtain $\mathrm{Re}_{p} \sim \mathcal{O}(10^{-1} - 10^{0})$, ensuring the validity of the inertial torque approximation. This sets an upper bound on $\mathrm{Sv}$ for which the asymptotic expansion remains applicable.

A second constraint arises in the regime where inertial torque dominates, requiring $\mathrm{Sv} \mathrm{Re}_{p} \mathcal{F}_{p} \gg \epsilon \mathcal{B} \sigma (1 - \gamma)$, which ensures that the asymptotic solution is physically meaningful. To capture settling-dominated dynamics, we further impose $\mathrm{Sv} > 10^{-3}$. Together, these criteria guide the range of $\mathrm{Sv}$ used for asymptotic analysis, which we restrict to $10^{-2} \leq \mathrm{Sv} \leq 10^{1}$.

\begin{figure}
    \centering
    \includegraphics[width=0.65\linewidth]{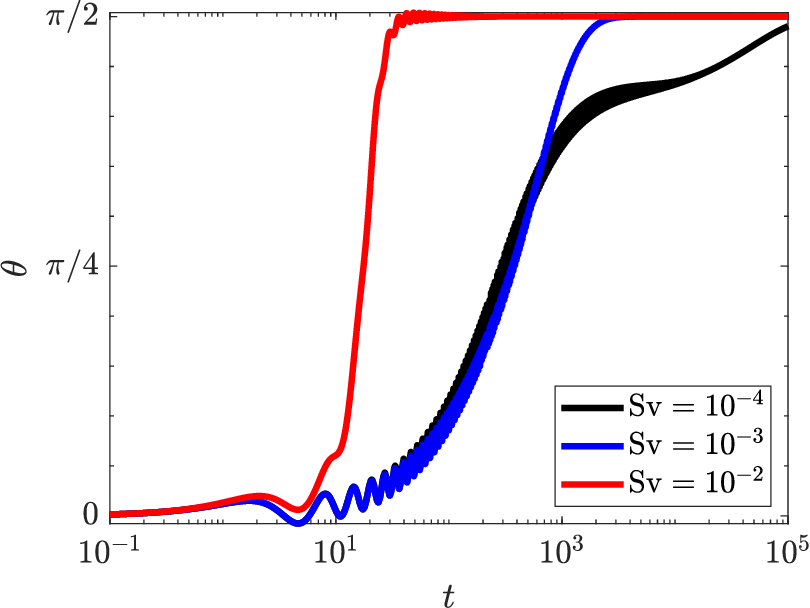}
    \caption{Time evolution of orientation angle $\theta$ for different $\mathrm{Sv}$ values. Parameters: $\mathcal{B} = 0.9$, $\mathrm{Re}_{w} = 3000$ and $\Theta=0$.}
    \label{fig:tht_fi}
\end{figure}

Figure~\ref{fig:tht_fi} shows the evolution of orientation angle $\theta$ with time for varying $\mathrm{Sv}$. At low $\mathrm{Sv}$, the particle exhibits larger orientation oscillations that are gradually damped out. These damped oscillations resemble behavior seen in the purely buoyant case. Notably, the evolution of orientation is non-monotonic before converging to a steady-state angle, indicating that this transient behavior may imprint on the horizontal trajectory. This observation is consistent with the experimental findings of \citet{dib2019oriexp}, in which a slightly negatively buoyant particle attains a broadside-on alignment. For the size and wave parameters considered in their study, the settling parameter is of order $\mathrm{Sv}=10^{-3}$.

As $\mathrm{Sv}$ increases, particles tend to align broadside-on more rapidly, with oscillations suppressed earlier. Similar orientation dynamics are observed for different values of $\mathcal{B}$. This behavior contrasts with the neutrally buoyant case studied by \citet{dibenedetto_2018}, where persistent oscillations about the preferential angle occur and the equilibrium depends solely on $\mathcal{B}$. In the settling regime, by contrast, the terminal orientation depends on $\mathrm{Sv}$, $\mathcal{B}$, and the initial condition. 

We now proceed to analyze the particle dynamics in two distinguished limits: weak settling ($\mathrm{Sv} = \mathcal{O}(\epsilon^2)$) and rapid settling ($\mathrm{Sv} \gg 1$).

\subsection{Weak settling limit ($\mathrm{Sv}=\mathcal{O}(\epsilon^{2})$)} \label{sec:weak_settling}

First, we have considered a weak settling case where $\mathrm{Sv}=\mathcal{O}(\epsilon^{2})$ and wave term is $\mathcal{O}(\epsilon)$. For this case, we have used the multiple scale analysis to obtain the wave-averaged quantities, following the approach outlined by \cite{pujara2023wave}, and additionally incorporating the inertial torque term at $\mathcal{O}(\epsilon^{2})$.
 The fast and slow time scales, representing an oscillatory and mean motion of a particle, are denoted by $t$ and $T=\epsilon^{2} t$. We employ a two-time-scale expansion of the particle position and orientation to derive the wave-averaged motion:
\begin{subequations}
    \begin{equation}
        \boldsymbol{x}(t)=\boldsymbol{x}_{0}(T)+\epsilon\,\boldsymbol{x}_{1}(t,T)+\epsilon^{2}\boldsymbol{x}_{2}(t,T)+\cdots,
    \end{equation}
    \begin{equation}
        \theta(t)=\theta_{0}(t,T)+\epsilon\,\theta_{1}(t,T)+\epsilon^{2}\theta_{2}(t,T)+\cdots.
    \end{equation}
\end{subequations}

 At $\mathcal{O}(\epsilon^{2})$, the evolution of wave-averaged quantities with respect to the slow time scale becomes,

\begin{subequations}\label{eq:ms_fieq}
       \begin{equation}
        \dfrac{d\bar{X}}{dT}=e^{2\bar{Z}}+\mathrm{Sv}^{*}\left(\mathcal{Y}_{A}-\mathcal{X}_{A}\right)\cos\vartheta\sin\vartheta,
    \end{equation} \begin{equation}
        \dfrac{d\bar{Z}}{dT}=\mathrm{Sv}^{*}\left[-\mathcal{Y}_{A}+(\mathcal{Y}_{A}-\mathcal{X}_{A})\cos^{2}\vartheta\right],
    \end{equation}
    \begin{equation}
        \dfrac{d\vartheta}{dT}= \mathcal{B}e^{2\bar{Z}}\left(\mathcal{B}+\cos{2\vartheta}\right)-\mathrm{Sv}^{*}\tau\mathcal{F}_{p}\sin{2\vartheta}.
    \end{equation}
\end{subequations}

\begin{figure}
\centering  
\subfigure[]{\includegraphics[width=0.49\linewidth]{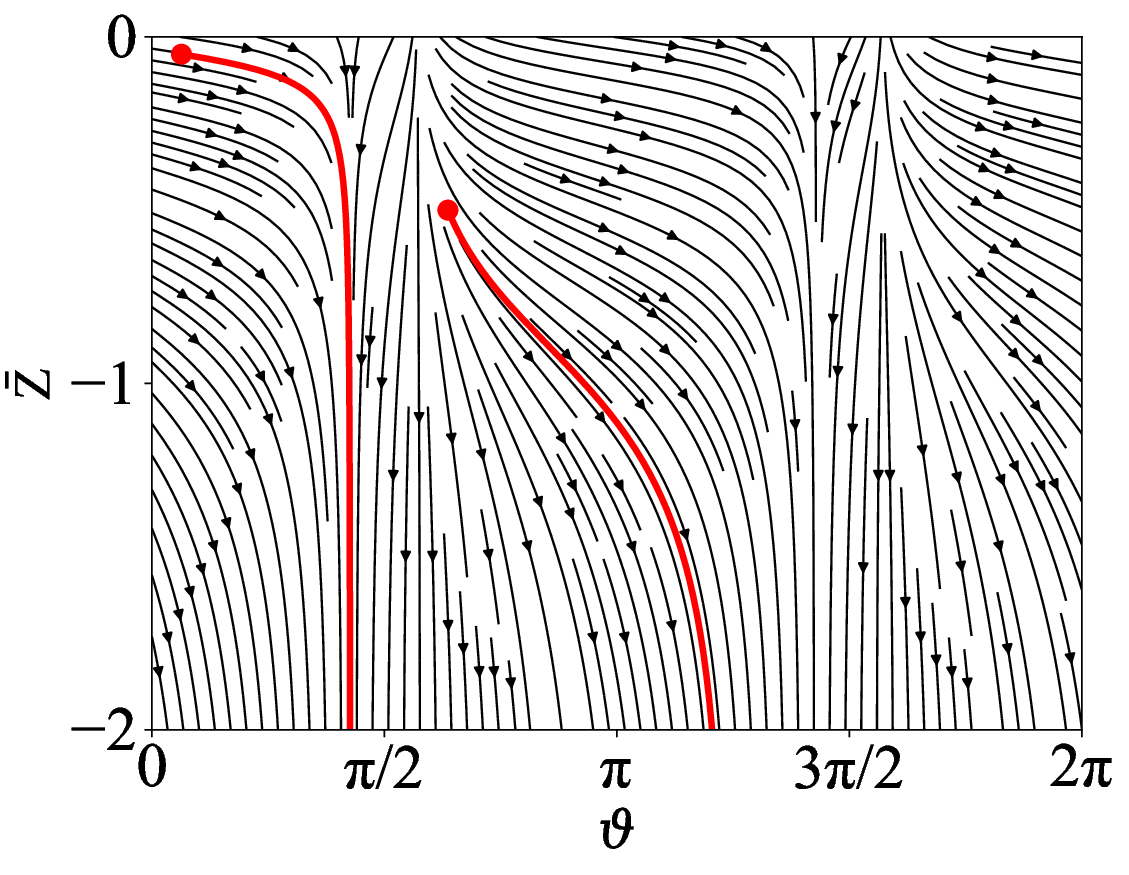}}
\subfigure[]{\includegraphics[width=0.49\linewidth]{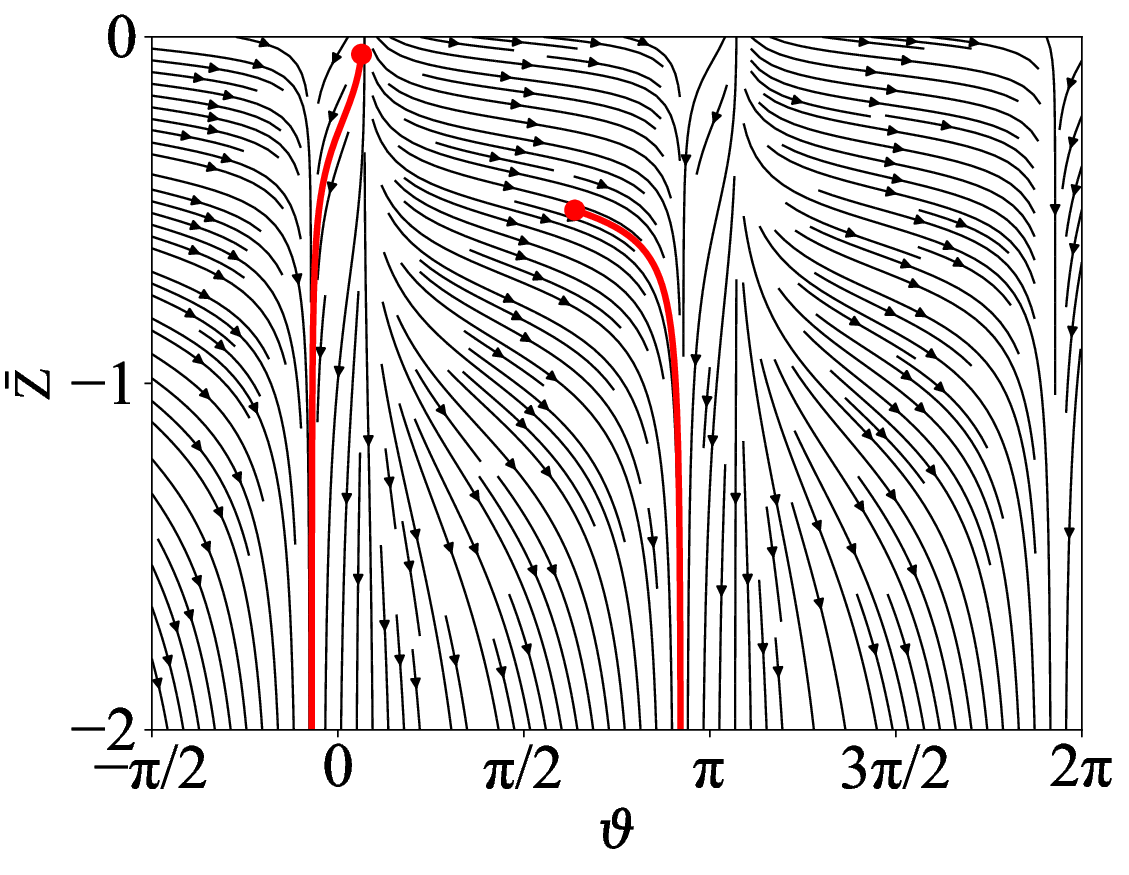}}
\subfigure[]{\includegraphics[width=0.49\linewidth]{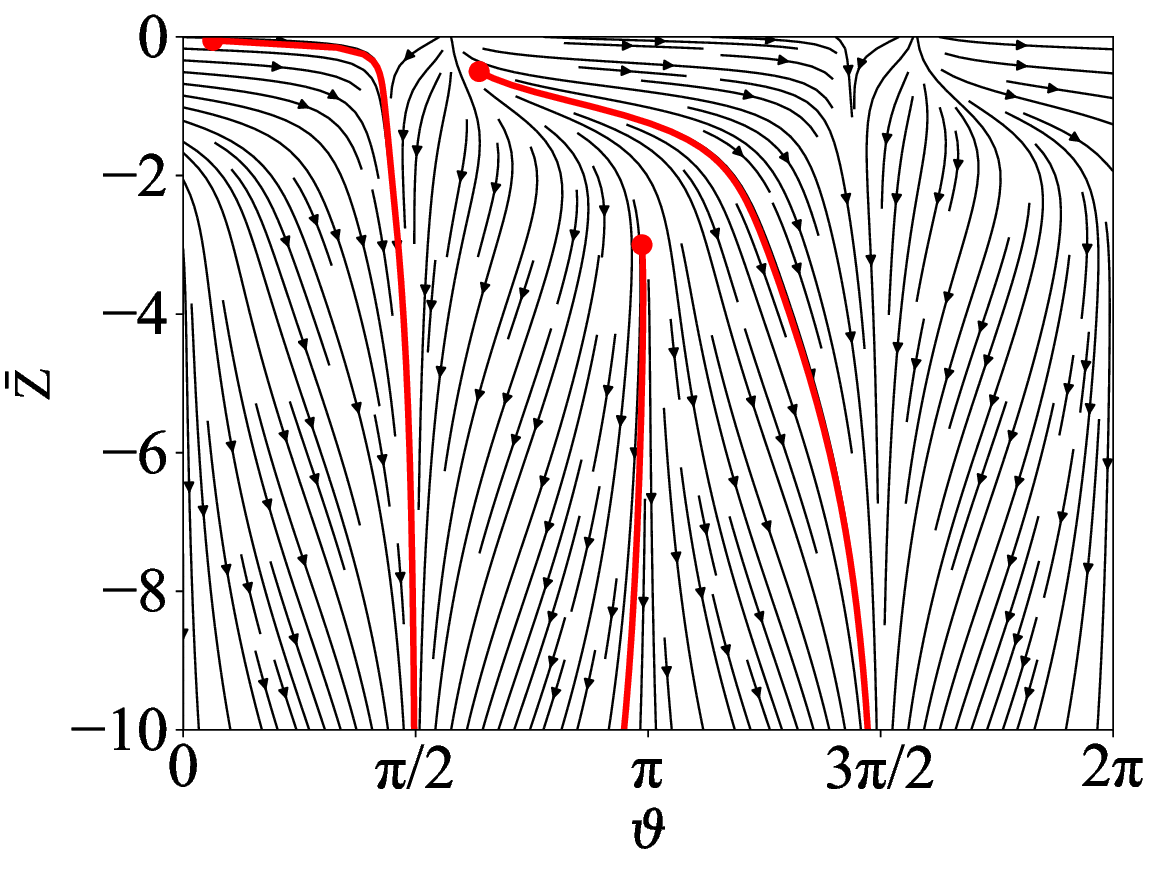}}
\subfigure[]{\includegraphics[width=0.49\linewidth]{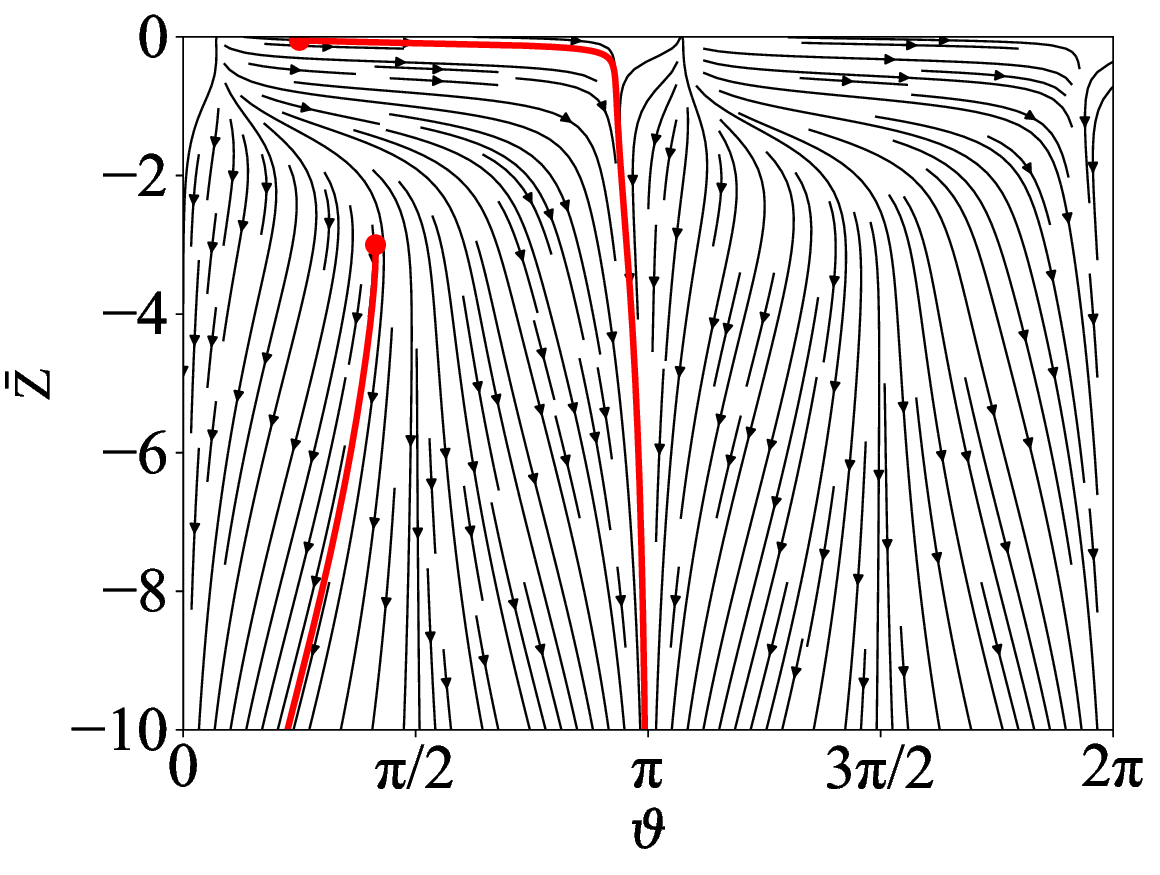}}
\caption{Panels (a) and (b) show phase-portrait diagrams in the $\bar{Z}-\vartheta$ plane for both prolate and oblate spheroids, respectively, with inertial torque neglected. Panels (c) and (d) show the corresponding phase portraits when inertial torque is included. Here, the timescale ratio is $\tau=1$. Red curves denote trajectories in phase space, with filled dots indicating the initial conditions. Parameters are, $|\mathcal{B}|=0.9$ and $\mathrm{Sv}^{*}=4\times10^{-2}$.} 
\label{fig:phase_sv}
\end{figure} 

In the above equations, $\tau$ denotes the ratio of the fall timescale, $t_f=(kW)^{-1}$, which represents the time required for a particle to fall a distance equal to one wavelength, to the diffusion timescale, $t_{d}=(\nu k^{2})^{-1}$. Also, $\mathrm{Sv}^{*}=\mathrm{Sv}/\epsilon^{2}$.  In figure~\ref{fig:phase_sv} we present the phase space diagrams in the $\bar{Z}-\vartheta$ plane for both prolate and oblate spheroids. In the top row, we first neglect the inertial torque. Parameters for these figures are, $\mathrm{Sv}^{*}=4\times 10^{-2}$ and $|\mathcal{B}|=0.9$. The red lines correspond to the trajectories with dots representing the initial conditions. As discussed in the previous section, the long-time orientation of the particle depends on the initial orientation, especially at large values of $\mathrm{Sv}^{*}$. For sufficiently small $\mathrm{Sv}^{*}$, the asymptotic orientation approaches the tracer limit (see figure~\ref{fig:thinfysv}). The phase portraits further reveal that, for certain initial orientations, a spheroidal particle undergoes a complete rotation before reaching its equilibrium orientation, thereby influencing its horizontal motion.

In the bottom row, we present the phase portrait diagrams accounting for the inertial torque at $\tau=1$. Even in the weak-settling limit, a spheroidal particle ultimately attains a broadside-on alignment. However, at low $\mathrm{Sv}^{*}$, the relaxation toward the asymptotic orientation $\theta_{\infty}$ is slow, resulting in larger horizontal and vertical displacements prior to alignment. As in the case without inertial torque, there exist initial orientations for which the particle undergoes a complete rotation before reaching equilibrium, attaining $\vartheta=3\pi/2$ for a prolate spheroid and $\pi$ for an oblate spheroid. 

We now turn our attention to the opposite regime, examining the particle dynamics in the rapid settling limit $\mathrm{Sv} \gg 1$.

\subsection{Rapid settling limit ($\mathrm{Sv}\gg 1$)}

When both background shear and inertial torques act on a settling spheroid, analytical insight becomes possible in the so-called rapid-settling limit. Originally introduced by \citep{menon2017theoretical, roy2023orientation} for anisotropic particles settling through homogeneous isotropic turbulence, this limit describes the regime in which a particle descends through a characteristic flow length scale, such as the Kolmogorov scale, faster than the surrounding eddies can turn over. In this setting, the particle aligns in a broadside orientation with weak oscillations, whose amplitude is governed by the settling parameter $\mathrm{Sv}$, which quantifies the ratio of gravitational to flow time scales.

While earlier studies focused primarily on orientation dynamics, the implications of this `flutter' on translational motion have remained less clear. Before turning to the specific case of surface gravity waves, we highlight how even a weak background flow is essential to induce lateral drift. Consider a prolate spheroid with orientation vector $\boldsymbol{p}$. In the rapid-settling regime, the particle predominantly maintains a broadside orientation ($|p_3| \ll 1$), and the vertical component of the orientation vector can be approximated by
\begin{equation}
p_3 \sim \frac{(R_{3j}+\mathcal{B}S_{3j})p^t_j}{\mathrm{Sv}^{2}\mathrm{Re}_{w}\mathcal{F}_{p}},
\end{equation}
where $p^t_i$ denotes the transverse (horizontal) components ($i=1,2$) of the orientation vector, governed by the Jeffery equation in the horizontal plane, and $\mathbf{R}, \mathbf{S}$ are the rotation and strain-rate tensors, respectively. This expression shows that the flutter amplitude decreases with increasing $\mathrm{Sv}$. The horizontal velocity in this limit becomes
\begin{equation}
\dot{x}^t_i \sim -\mathrm{Sv}(\mathcal{X}_{A}-\mathcal{Y}_{A})p^t_i p_3 \sim -\frac{(\mathcal{X}_{A}-\mathcal{Y}_{A})(R_{3j}+\mathcal{B}S_{3j})p^t_j p^t_i}{\mathrm{Sv}\mathrm{Re}_{w}\mathcal{F}_{p}}.
\end{equation}
This confirms that the presence of a weak background flow is essential for enabling horizontal drift, which scales as $\mathrm{Sv}^{-1}$. In the absence of any flow field, the particle would rapidly align with its broadside orientation and cease to drift laterally. In the present study, the background turbulence generated by the monochromatic wave field continuously perturbs the orientation, sustaining finite horizontal drift. We now return our focus to the two-dimensional monochromatic deep-water wave field to investigate drift in the rapid-settling regime.

In a deep-water wave field, the dynamics simplify further. As the particle settles, the local velocity-gradient field decays rapidly with depth, causing the orientation dynamics to naturally reduce to the rapid-settling regime. In this case, inertial torques dominate, and Jeffery torques become negligible. The resulting orientation dynamics become independent of the initial conditions or wave parameters, and the spheroid invariably approaches
\begin{equation}\label{eq:firpsth}
  \theta_{\infty}= \left\{
    \begin{array}{cl}
      \pi/2, & \text{Prolate}, \\[6pt]
      0,     & \text{Oblate},
    \end{array} \right.
\end{equation}
corresponding to the classical broadside-on alignment of rods and disks under gravity.

In the absence of wave effects ($\epsilon = 0$), the governing equations~(\ref{eq:fiem}) admit a closed-form expression for the long-time horizontal displacement:
\begin{equation}\label{eq:firpsx}
    x_{\infty}(\epsilon=0)
    = X+\dfrac{\mathcal{X}_{A}-\mathcal{Y}_{A}}{2\,\mathrm{Sv}\,\mathrm{Re}_{w}\,\mathcal{F}_{p}}
      \left(\theta_{\infty}-\Theta\right),
\end{equation}
which depends solely on the net change in orientation between the initial condition and the asymptotic equilibrium value. Wave-induced corrections first appear at $O(\epsilon)$ and depend on the depth, Reynolds number $\mathrm{Re}_{w}$, and the particle’s geometry. Considering the particle to be initially perturbed from its broadside orientation, i.e., $\Theta \sim \theta_{\infty} + \epsilon \Theta_1$, we obtain the following expressions for the wave-induced spreading length:

\begin{subequations}
    \begin{align}
x^{\text{prolate}}_{\infty} &\sim X -\frac{(\mathcal{X}_A - \mathcal{Y}_A)\,\epsilon\Theta_1}{2\mathrm{Sv}\,\mathrm{Re}_w\,\mathcal{F}_p} + \frac{\epsilon e^{Z}}{1 + \mathrm{Sv}^2\,\mathcal{Y}_A^2} \left(1 +  \frac{\mathcal{B}(\mathcal{X}_A - \mathcal{Y}_A)}{2\,\mathrm{Re}_w\,\mathcal{F}_p\,\mathrm{Sv}}\right) \left(\mathrm{Sv}\,\mathcal{Y}_A\,\cos{X} + \sin{X} \right), \\
x^{\text{oblate}}_{\infty} &\sim X -\frac{(\mathcal{X}_A - \mathcal{Y}_A)\,\epsilon \Theta_1}{2\mathrm{Sv}\,\mathrm{Re}_w\,\mathcal{F}_p} + \frac{\epsilon e^{Z}}{1 + \mathrm{Sv}^2\,\mathcal{X}_A^2} \left(1 -  \frac{\mathcal{B}(\mathcal{X}_A - \mathcal{Y}_A)}{2\,\mathrm{Re}_w\,\mathcal{F}_p\,\mathrm{Sv}}\right) \left(\mathrm{Sv}\,\mathcal{X}_A\,\cos{X} + \sin{X} \right).
\end{align}
\end{subequations}

\begin{figure}
\centering  
\subfigure[]{\includegraphics[width=0.48\linewidth]{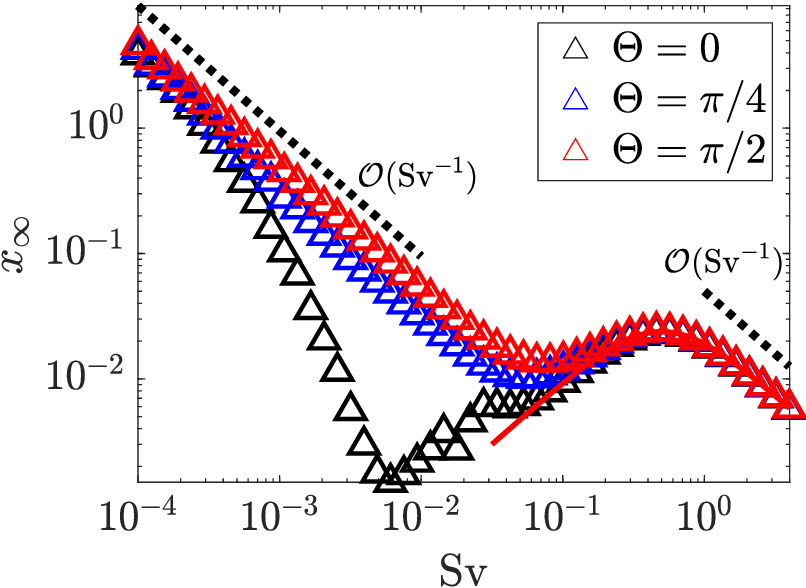}}
\subfigure[]{\includegraphics[width=0.5\linewidth]{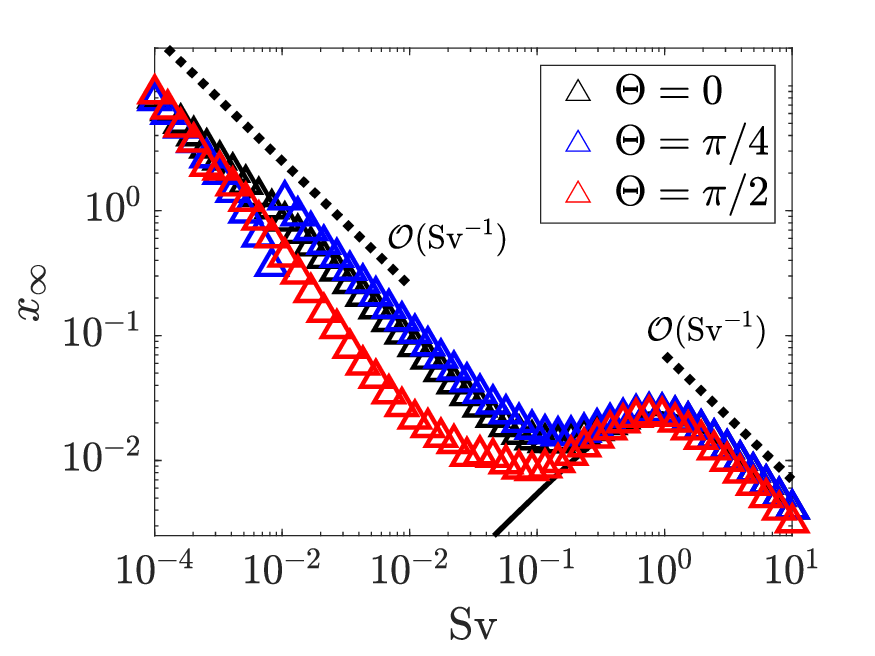}}
\caption{(a) Spreading length $x_{\infty}$ versus $\mathrm{Sv}$ for a prolate spheroid. Here, triangles represent numerical results and the solid red line corresponds to the asymptotic result at $\Theta=\pi/2$. (b) Spreading length $x_{\infty}$ versus $\mathrm{Sv}$ for an oblate spheroid. Here, triangles represent numerical results and the solid black line corresponds to the asymptotic result at $\Theta=0$. Fixed parameters for both plots are $|\mathcal{B}|=0.9$ and $\mathrm{Re}_{w}=3000$. } 
\label{fig:xinf_sv_diffini}
\end{figure}

In figure~\ref{fig:xinf_sv_diffini}, we present the spreading length $x_{\infty}$ of a spheroidal particle as a function of $\mathrm{Sv}$ for different initial orientations $\Theta$. The triangles denote numerical results, while the solid lines correspond to the asymptotic predictions above. For prolate spheroids, spreading is maximal when initialized at $\Theta = \pi/2$, whereas for oblate spheroids, the peak occurs at $\Theta = \pi/4$. In general, oblate spheroids exhibit greater horizontal displacement than prolate ones. In both cases, $x_{\infty}$ varies non-monotonically with $\mathrm{Sv}$ and scales as $\mathcal{O}(\mathrm{Sv}^{-1})$ in the asymptotic limits of small and large $\mathrm{Sv}$.

To interpret this behavior, the $\mathrm{Sv}$ dependence of spreading is divided into three regimes: in Regime I, $x_{\infty}$ is large and decreases with increasing $\mathrm{Sv}$; in Regime II, it increases with $\mathrm{Sv}$; and in Regime III, it again decreases. The final regime has limited practical importance, as the particle’s horizontal displacement becomes small relative to its body size.

\begin{figure}
\centering  
\subfigure[]{\includegraphics[width=0.5\linewidth]{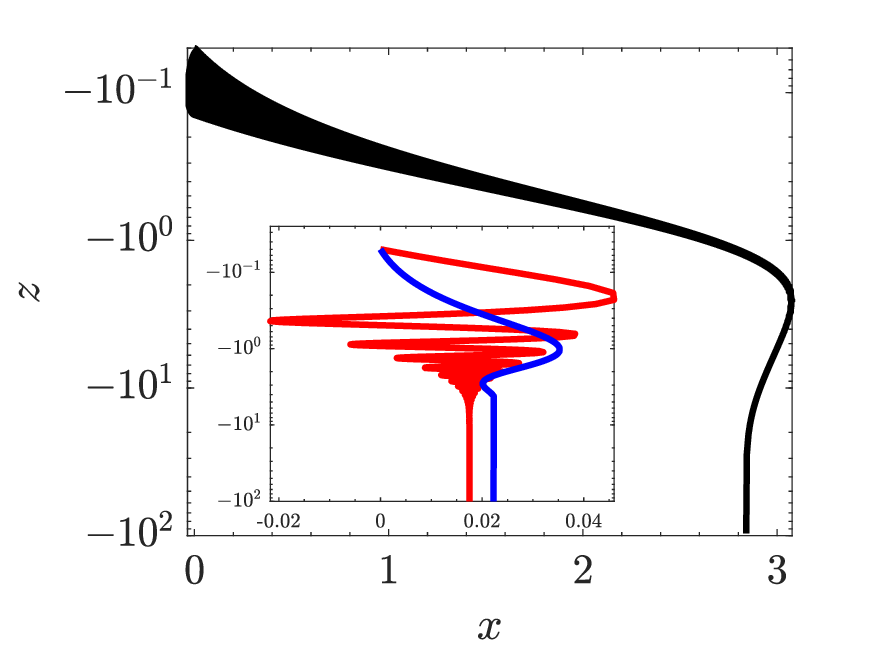}}
\subfigure[]{\includegraphics[width=0.47\linewidth]{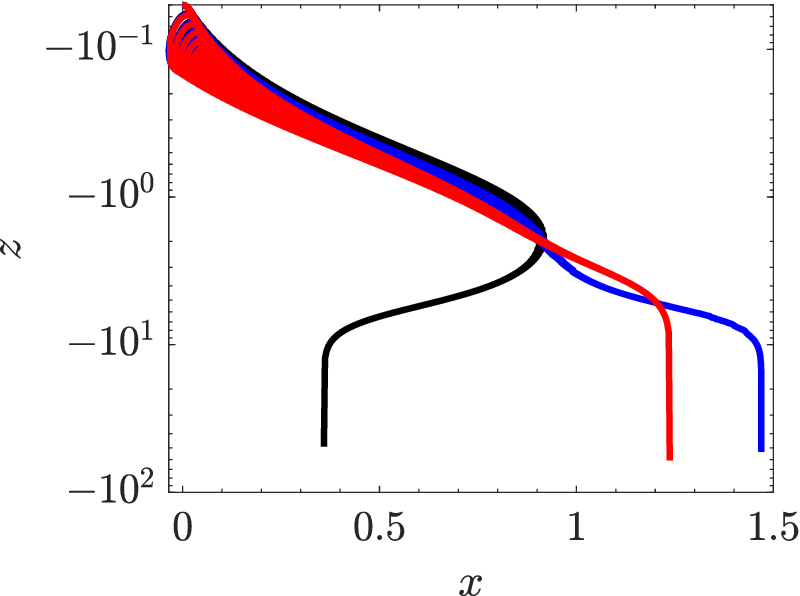}}
\caption{(a) Trajectories of a prolate spheroid in the $x$--$z$ plane. Values of $\mathrm{Sv}$ correspond to figure~\ref{fig:xinf_sv_diffini}a: $\mathrm{Sv}=1.5\times 10^{-4}$ (black), $3.55\times 10^{-2}$ (red), and $2.9\times 10^{-1}$ (blue). Initial orientation: $\Theta=\pi/2$. Inset shows magnified view of the blue trajectory. (b) Same for an oblate spheroid, with $\mathrm{Sv}=8.28\times 10^{-4}$ (black), $9\times 10^{-4}$ (blue), and $1.04\times 10^{-3}$ (red), initialized at $\Theta=\pi/4$. Fixed parameters: $|\mathcal{B}|=0.9$, $\mathrm{Re}_{w}=3000$.} 
\label{fig:traj_fidiffsv}
\end{figure}

In figure~\ref{fig:traj_fidiffsv}, we show the $x$--$z$ trajectories of prolate and oblate spheroids for selected values of $\mathrm{Sv}$, chosen based on the data points in figure~\ref{fig:xinf_sv_diffini}. In panel (a), black, red, and blue lines correspond to $\mathrm{Sv}=1.5\times 10^{-4}, 3.55\times 10^{-2}$, and $2.9\times 10^{-1}$, respectively, with initial orientation $\Theta=\pi/2$. In regime I ($\mathrm{Sv}=1.5\times 10^{-4}$), the prolate particle displays substantial horizontal spreading as it spends a significant time in orbital motion before transitioning to vertical descent. In regime II ($\mathrm{Sv}=2.9\times 10^{-1}$), the trajectory becomes predominantly vertical with reduced oscillations and a slight increase in $x_{\infty}$ compared to the intermediate case ($\mathrm{Sv}=3.55\times 10^{-2}$). Panel (b) shows the trajectories of oblate spheroids initialized at $\Theta=\pi/4$, with $\mathrm{Sv}=8.28\times 10^{-4}, 9\times 10^{-4}$, and $1.04\times 10^{-3}$ (black, blue, and red lines). The abrupt jump in $x_{\infty}$ observed in figure~\ref{fig:xinf_sv_diffini}b is clearly reflected in the corresponding trajectories. These differences are rooted in the orientation dynamics of the particle, as explored next.

\begin{figure}
\centering  
\subfigure[]{\includegraphics[width=0.51\linewidth]{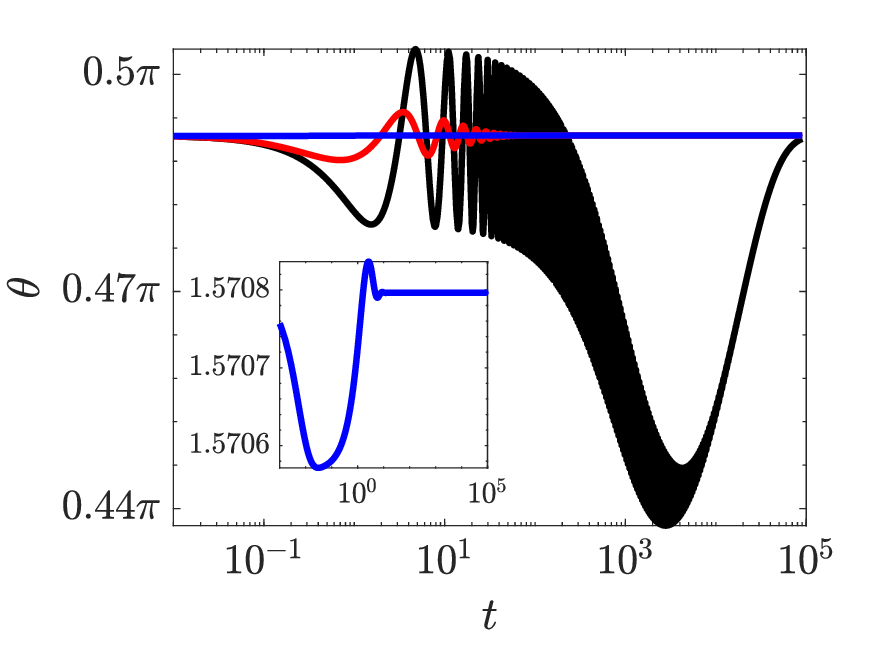}}
\subfigure[]{\includegraphics[width=0.465\linewidth]{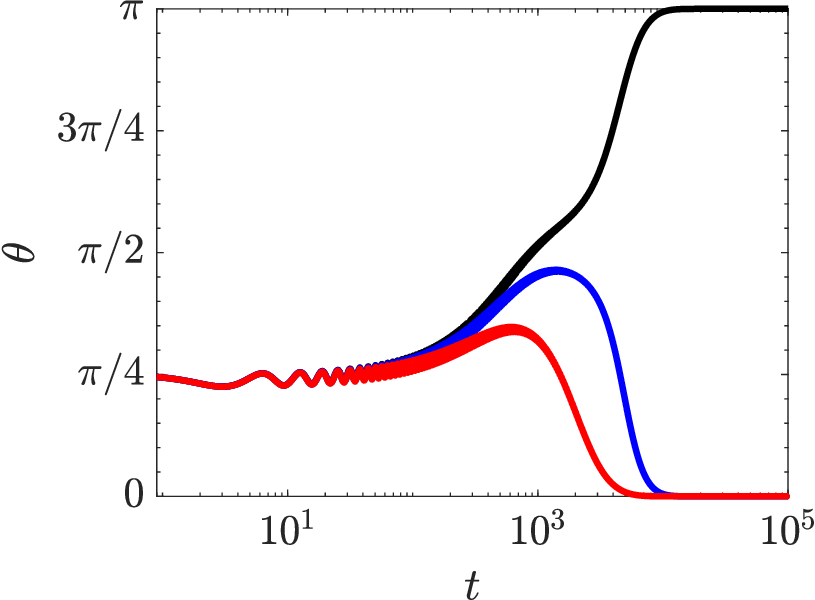}}
\caption{(a) Orientation $\theta(t)$ for a prolate spheroid. $\mathrm{Sv}=1.5\times 10^{-4}$ (black), $3.55\times 10^{-2}$ (red), and $2.9\times 10^{-1}$ (blue). Initial orientation: $\Theta=\pi/2$. Inset magnifies the trajectory for $\mathrm{Sv}=2.9\times 10^{-1}$. (b) Same for an oblate spheroid initialized at $\Theta=\pi/4$, with $\mathrm{Sv}=8.28\times 10^{-4}$ (black), $9\times 10^{-4}$ (blue), and $1.04\times 10^{-3}$ (red). Parameters: $|\mathcal{B}|=0.9$, $\mathrm{Re}_{w}=3000$.}
\label{fig:th_t_fidiffsv}
\end{figure}

Since all particles ultimately attain a broadside-on orientation, their trajectories are primarily governed by the time evolution of their orientation. Figure~\ref{fig:th_t_fidiffsv} displays $\theta(t)$ for both prolate and oblate spheroids. For prolates (panel a), strong oscillations at low $\mathrm{Sv}$ induce time-dependent drag variations, resulting in pronounced horizontal drift. As $\mathrm{Sv}$ increases, these oscillations weaken; at $\mathrm{Sv}=2.9\times 10^{-1}$, the orientation evolves monotonically, corresponding to a modest increase in $x_{\infty}$. Notably, even a mild transient deviation in $\theta(t)$ before reaching the equilibrium state can significantly influence the final horizontal displacement. At larger $\mathrm{Sv}$, orientation rapidly settles to its steady-state, and spreading becomes negligible.

In the oblate case (panel b), the orientation dynamics exhibit a sharp transition around $\mathrm{Sv}=8.2\times 10^{-4}$. Below this value, the spheroid completes a full rotation before settling at $\theta=\pi$, while above it, the particle aligns without undergoing a full turn. This qualitative difference in orientation history explains the discontinuous change in $x_{\infty}$ and the associated trajectory shift. These complete revolutions at low $\mathrm{Sv}$, dependent on initial conditions, are also visible in the phase portraits shown earlier (figure~\ref{fig:phase_sv}).

\begin{figure}
\centering  
\subfigure[]{\includegraphics[width=0.495\linewidth]{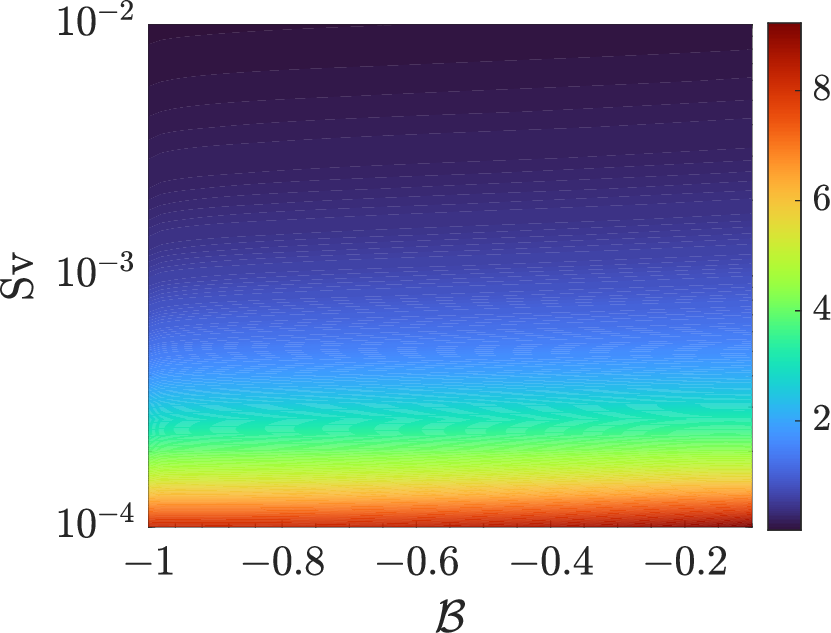}}
\subfigure[]{\includegraphics[width=0.475\linewidth]{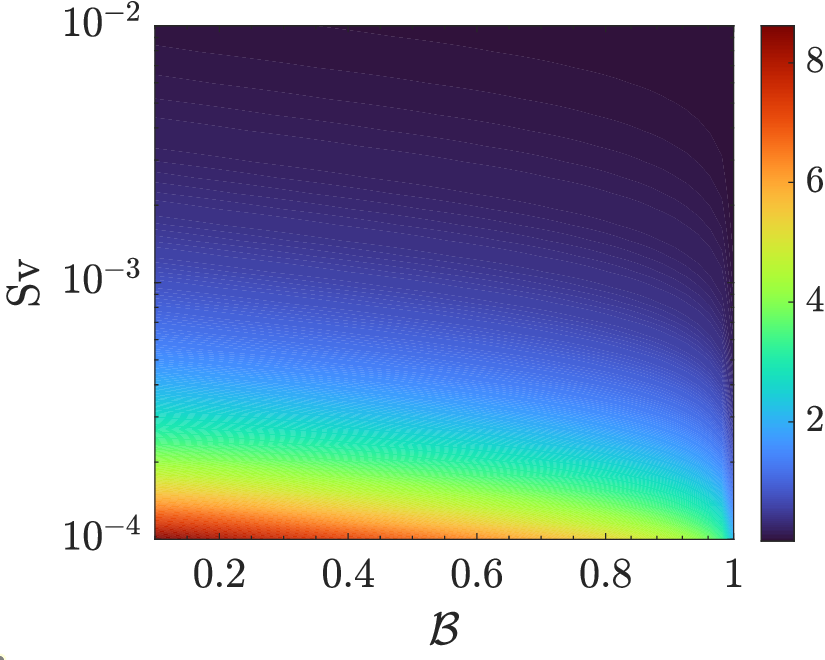}}
\caption{ Contour plot of horizontal spreading $x_{\infty}$ in a $\mathrm{Sv}-\mathcal{B}$ plane.  (a) Oblate spheroid and (b) Prolate spheroid. Parameters are: $\mathrm{Re}_{w}=10^{4}$, $|\mathcal{B}|=0.9$ and $Z=-0.05$. } 
\label{fig:contxfi}
\end{figure}

In figure~\ref{fig:contxfi}, we present contour plots of the horizontal spreading $x_{\infty}$ in the $\mathrm{Sv}$–$\mathcal{B}$ parameter space for both prolate and oblate spheroids, with a fixed wave Reynolds number $\mathrm{Re}_{w}=10^{4}$. The initial orientation is taken as $\Theta=0$ for prolates and $\Theta=\pi/2$ for oblates. Once a particle attains its steady-state alignment, it settles without further horizontal motion, unlike in the buoyancy-driven case, where horizontal translation persists even after alignment. Both prolate and oblate spheroids exhibit significant horizontal drift at low $\mathrm{Sv}$, as they take longer to reach their settling-induced orientation. 

Spreading also depends strongly on particle shape. Rod-like prolates exhibit smaller $x_{\infty}$ compared to disk-like oblates. As $\mathrm{Sv}$ increases, the spreading $x_{\infty}$ decreases markedly. For $\mathrm{Sv}=\mathcal{O}(10^{-2})$, the particle translates only a few wavelengths before aligning vertically, in contrast to the buoyancy-driven case where high $\mathrm{Sv}$ leads to extensive horizontal displacement. Similar to prolates, disk-like oblates show reduced translation relative to their nearly spherical counterparts.

Horizontal displacement arises almost entirely during the transient evolution of orientation. Once the particle reaches its broadside-on alignment, further translation ceases. Therefore, the extent of horizontal motion is governed by parameters in the orientation dynamics, namely $\mathrm{Sv}$, $\mathrm{Re}_{w}$, and $\mathcal{F}_{p}$, as they determine the timescale for achieving steady orientation. If the initial orientation already coincides with the steady-state value, transient effects are minimal and the translation is primarily due to wave-induced drift. At large $\mathrm{Sv}$, particles settle to finite depth rapidly (in under a second), and the influence of the wave field diminishes accordingly.

\begin{figure}
\centering  
\subfigure[]{\includegraphics[width=0.49\linewidth]{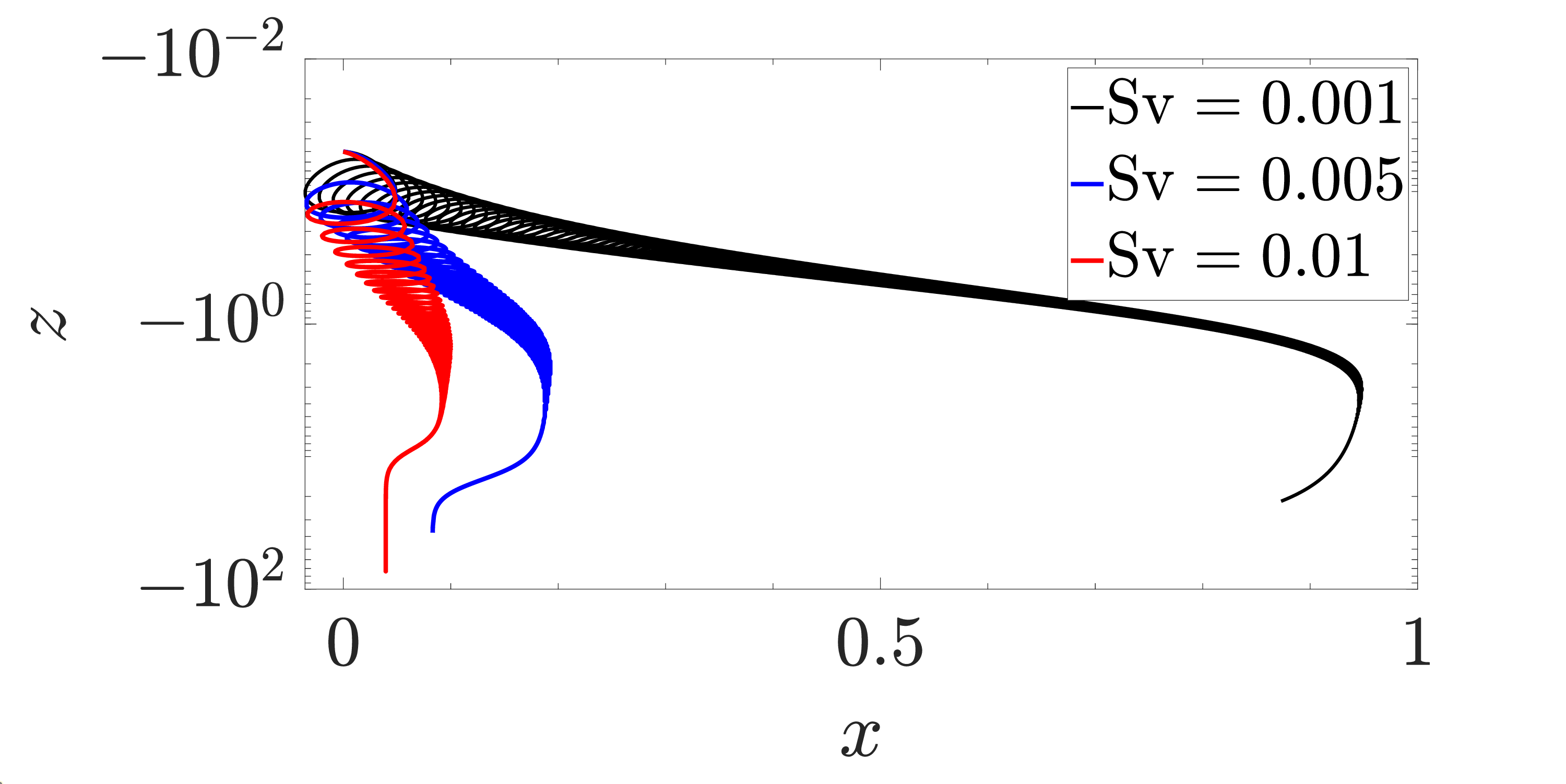}}
\subfigure[]{\includegraphics[width=0.49\linewidth]{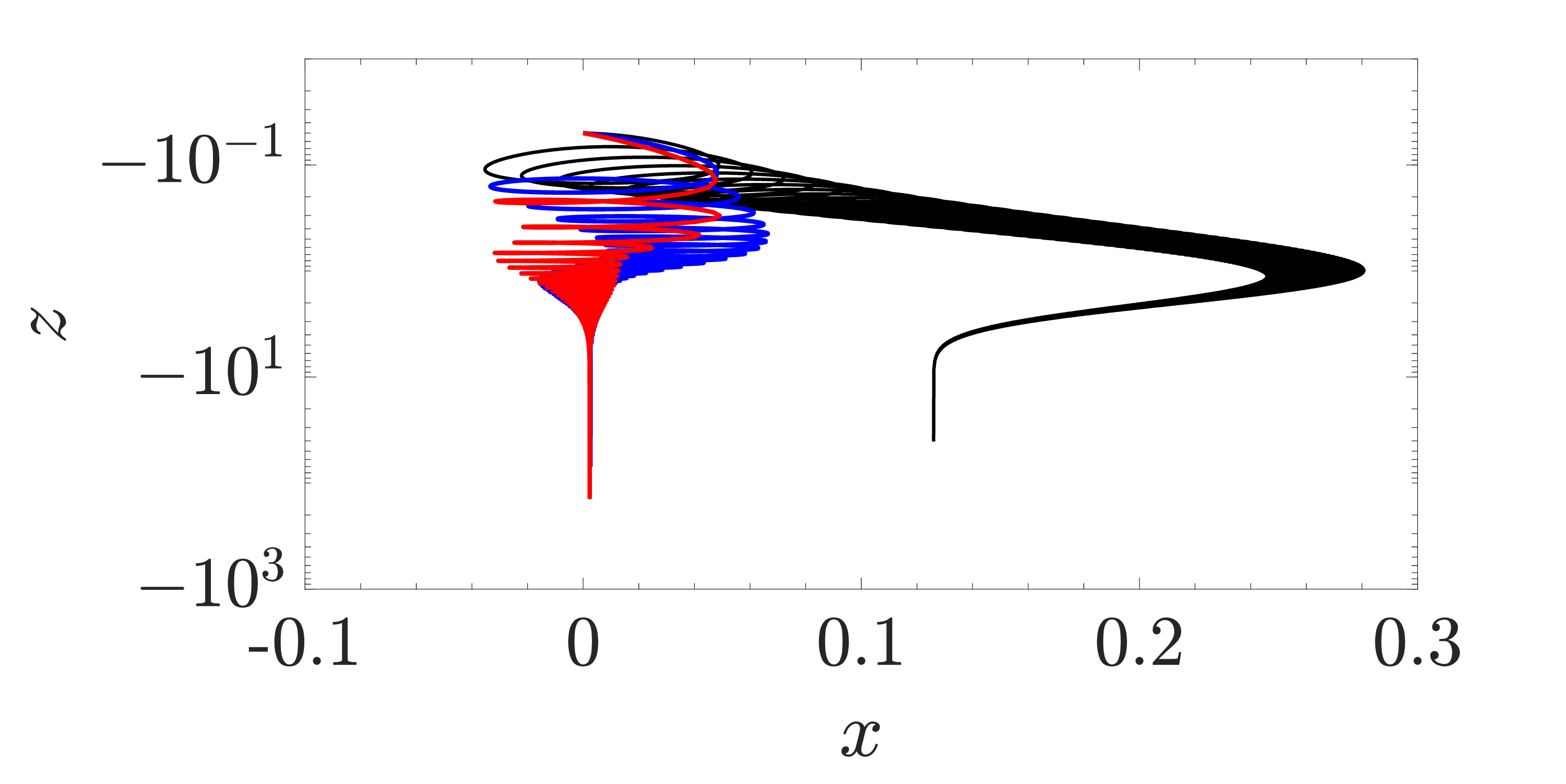}}
\caption{ Trajectories of a prolate spheroid when accounting for the inertial torque at $\Theta=0$.  (a) $\mathcal{B}=0.1$ and (b) $\mathcal{B}=0.9$. Other initial conditions are $X=0$ and $Z=-0.05$. Also, $\mathrm{Re}_{w}=3000$. } 
\label{fig:trjfipro}
\end{figure} 

In figure~\ref{fig:trjfipro}, we present trajectories of a prolate spheroid for varying $\mathrm{Sv}$. All cases are simulated up to $t = 500$, except for $\mathcal{B}=0.1$ (black curve), which is extended to $t = 5000$ to highlight long-time settling behavior. The particle’s motion can be divided into three distinct stages. In the first, wave-induced orbital motion dominates. In the second, as this motion damps out, the particle may translate horizontally, particularly within the previously identified WOM region. During this stage, the orientation evolves gradually and non-monotonically. If inertial torque is neglected, the orientation settles during this phase and horizontal drift persists indefinitely. In the third stage, the particle aligns with its broadside-on orientation, after which horizontal translation ceases.

At low $\mathrm{Sv}$, particles linger in the first two stages, resulting in prolonged horizontal motion and slower vertical descent. At higher $\mathrm{Sv}$, rapid alignment truncates the orbital and drift phases, eliminating the second stage entirely, as seen for $\mathcal{B}=0.9$. In such cases, vertical descent is more pronounced. Although differences in vertical displacement across $\mathrm{Sv}$ values appear modest at a fixed time, the settling behavior is strongly influenced by the duration of the orbital phase. The overall trajectory characteristics are qualitatively similar for oblate spheroids initialized at $\Theta = \pi/2$.

\begin{figure}
    \centering
    \includegraphics[width=0.65\linewidth]{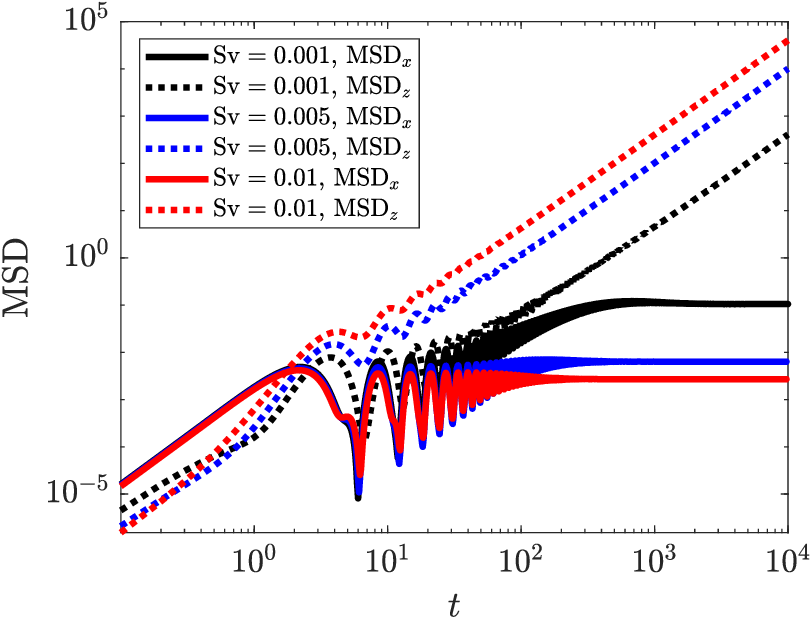}
    \caption{Mean-square displacement (MSD) curve of a prolate spheroid versus time for different values of $\mathrm{Sv}$ when inertial torque is considered. Solid lines indicate MSD of a horizontal displacement and dashed lines correspond to the vertical displacement. Parameters: $\mathcal{B}=0.9$ and $\mathrm{Re}_{w}=3000$.}
    \label{fig:msd_it}
\end{figure}

In figure~\ref{fig:msd_it}, we present the mean-squared displacement (MSD) of a prolate spheroid while accounting for inertial torque. Initially, both horizontal and vertical MSDs scale as $t^2$, indicating ballistic motion. At later times, the horizontal MSD saturates and scales as $t^0$, consistent with the cessation of horizontal translation once the particle aligns with its equilibrium orientation. This contrasts with the buoyancy-only case, where horizontal motion persists. At lower $\mathrm{Sv}$, $\mathrm{MSD}_x$ is larger due to extended transients in orientation, which result in greater variability in lateral displacement. In contrast, $\mathrm{MSD}_z$ shows negligible deviation from the previous case (Fig.~\ref{fig:msd_aniso}), confirming that inertial torque primarily affects horizontal, rather than vertical, motion.

\section{On the role of particle finite-size effects}\label{sec:fssec}

We now examine how finite particle size modifies the dynamics in a wavy flow. In oceanic settings, particles span a broad range of sizes and shapes, and the point-particle approximation need not apply. When the particle dimension becomes comparable to the wavelength, finite-size corrections must be included. The dominant contributions can be incorporated through Fax\'en-type terms, as shown by \cite{homann2010finite}. For spheroids, the quasi-steady viscous force and torque—including higher-order multipoles and Fax\'en corrections—are well established in unbounded Stokes flow \citep{KIM1985}. These provide the appropriate leading-order framework for assessing finite-size effects in the present system.

We first ignore sedimentation and set $\mathrm{Sv}=0$. For a force- and torque-free prolate spheroid, the expressions (\ref{eq:for}) and (\ref{eq:tor}) lead directly to the classical Fax\'en-type relations for a finite spheroid in an arbitrary flow. The dimensional translation and rotation rates, for a prolate spheroid, can be written as
\begin{subequations} \label{eq:spheroid_faxen}
\begin{equation}\label{eq:unof}
    v_{i}
    =\frac{1}{2\alpha}\int_{-\alpha}^{\alpha}
      \left\{
      1+(\alpha^{2}-\xi^{2})\frac{1-e^{2}}{4e^{2}}\nabla^{2}
      \right\}
      u_{i}(\xi)\,d\xi,
\end{equation}
\begin{multline}\label{eq:onot}
    \Omega_{i}
    =\frac{3}{8\alpha^{3}}
      \int_{-\alpha}^{\alpha}(\alpha^{2}-\xi^{2})
      (\nabla\times u(\xi))_{i}\,d\xi \\
    +\frac{e^{2}}{2-e^{2}}
      \frac{3}{4\alpha^{3}}
      \int_{-\alpha}^{\alpha}
      \left\{
      (\alpha^{2}-\xi^{2})
      \left[1+(\alpha^{2}-\xi^{2})\frac{1-e^{2}}{8e^{2}}\nabla^{2}\right]
      \right\}
      \varepsilon_{ijk}p_{j}S_{kl}(\xi)p_{l}\,d\xi,
\end{multline}
\end{subequations}
where $\alpha=a e$, and $\xi$ spans the singularity distribution from $-\alpha$ to $\alpha$.

Since the wave field is incompressible and irrotational, $\nabla^{2}u=0$, and the higher-order Fax\'en correction in (\ref{eq:unof}) vanishes. A direct corollary is that a spherical particle moves identically to a tracer in any irrotational flow, since the finite-size correction is identically zero in that case.

Evaluating~(\ref{eq:spheroid_faxen}) for deep-water waves yields the finite-size particle translational and angular velocities,
\begin{subequations}\label{eq:vel2fs}
\begin{eqnarray}
\dot{x}
&=&\frac{e^{z}\epsilon}{\sigma}\,
\Re\!\left[e^{i(x-t-\theta)}\sinh(\sigma e^{i\theta})\right],
\label{eq:vel2fs_a}
\\[6pt]
\dot{z}
&=&\frac{e^{z}\epsilon}{\sigma}\,
\Im\!\left[e^{i(x-t-\theta)}\sinh(\sigma e^{i\theta})\right],
\label{eq:vel2fs_b}
\\[6pt]
\dot{\theta}&=& -\frac{3 B e^{z} \epsilon}{\sigma^{3}} \Re\left[ e^{i(x - t - \theta)} \left( \sigma e^{i \theta} \cosh(\sigma e^{i \theta}) - \sinh(\sigma e^{i \theta}) \right) \right] 
\end{eqnarray}
\end{subequations}

where $\sigma=k\alpha$. In the limit $\sigma\rightarrow 0$, the finite-size velocities in~(\ref{eq:vel2fs}) reduce to those of a tracer. For $\sigma>0$, however, translation becomes coupled to rotation, modifying both drift and alignment. Earlier studies \citep{dibenedetto_2018,pujara2023wave} have shown that, in the point-particle limit, a spheroid aligns to a shape-controlled, wave-preferred angle determined solely by $\mathcal{B}$. A key question is therefore: how does finite size alter this preferred alignment?

\begin{figure}
\centering  
\subfigure[]{\includegraphics[width=0.49\linewidth]{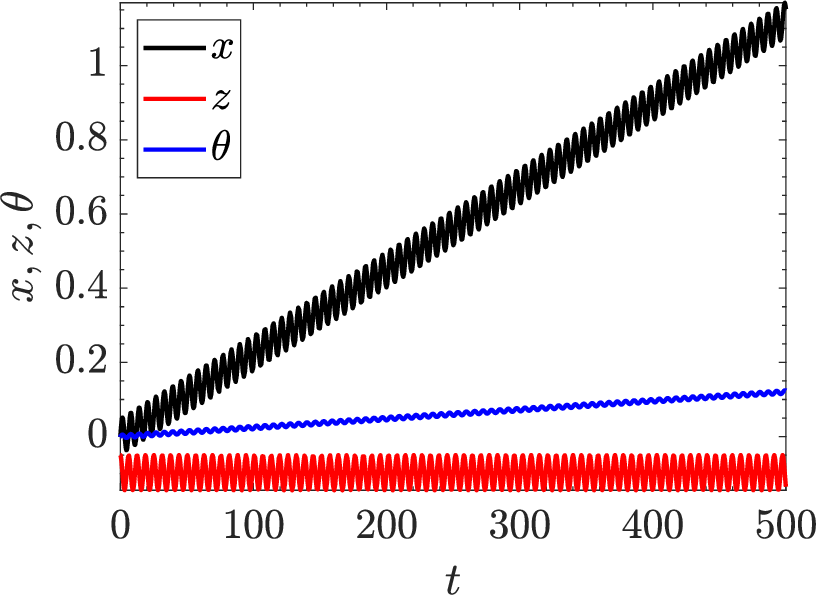}}
\subfigure[]{\includegraphics[width=0.49\linewidth]{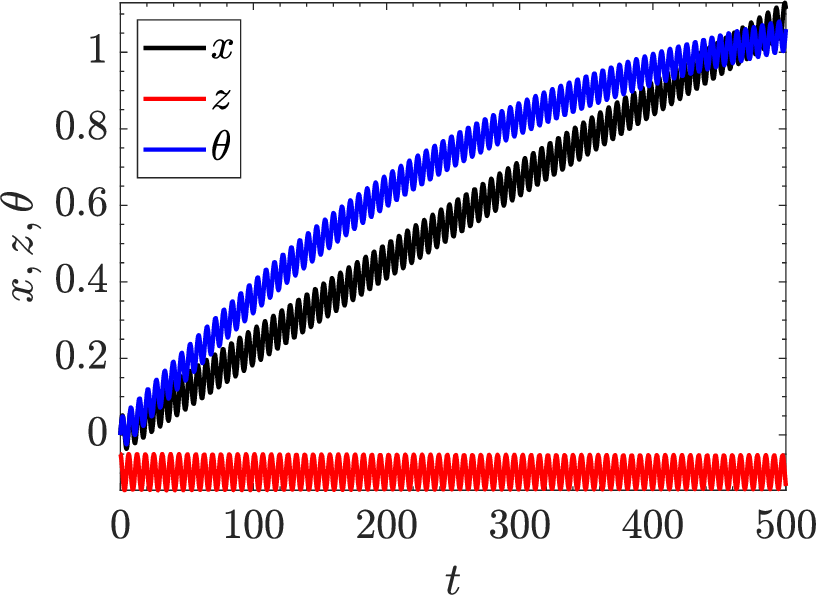}}
\caption{ Numerical solutions of the translation and orientation of a finite-size prolate spheroid.  
(a) $\mathcal{B}=0.1$; (b) $\mathcal{B}=0.9$.  
Initial conditions: $X=0$, $Z=-0.05$, $\phi=0$.  
Also, $\sigma=0.5$.} 
\label{fig:fspro2}
\end{figure}

Numerical integration of the full finite-size system (figure~\ref{fig:fspro2}) illustrates the essential behaviour. For a nearly spherical prolate ($\mathcal{B}=0.1$), the qualitative dynamics resemble those of a point particle: the horizontal motion oscillates, the orientation monotonically approaches its wave-preferred angle, and the particle remains at fixed depth. Increasing $\mathcal{B}$ accelerates alignment. Because the particle is fore--aft symmetric, finite-size corrections must be even in $\sigma$, as reflected directly in~(\ref{eq:vel2fs}). To quantify the leading modification to drift and orientation, we expand~(\ref{eq:vel2fs}) to $O(\sigma^{2})$:

\begin{subequations}\label{eq:perfseq}
 \begin{equation}
     \dot{x}=\epsilon e^{z}\left[\cos(x-t)+\frac{1}{6}\cos(x-t+2\theta)\,\sigma^{2}\right]+O(\sigma^{4}),
 \end{equation}
 \begin{equation}
     \dot{z}=\epsilon e^{z}\left[\sin(x-t)+\frac{1}{6}\sin(x-t+2\theta)\,\sigma^{2}\right]+O(\sigma^{4}),
 \end{equation}
 \begin{equation}
     \dot{\theta}=\mathcal{B}\epsilon e^{z}\left[\cos(x-t+2\theta)+\frac{1}{10}\cos(x-t+4\theta)\,\sigma^{2}\right]+O(\sigma^{4}).
 \end{equation}
\end{subequations}

Using the $O(\sigma^{2})$ expansion of~(\ref{eq:perfseq}), the wave-averaged drift and long-time orientation follow from a standard multiple-scale analysis. At leading order, $x_{0}=\bar{X}(T)$, $z_{0}=\bar{Z}(T)$ and $\theta_{0}=\vartheta(T)$. The wave-averaged dynamics appear at $O(\epsilon^{2})$:

\begin{subequations}
\begin{align}
    \frac{\partial \bar{X}}{\partial T}
    &= e^{2\bar{Z}}\left[1+\frac{\sigma^{2}}{6}\left(\mathcal{B}+2\cos 2\vartheta\right)\right], \\[4pt]
    \frac{\partial \bar{Z}}{\partial T}
    &= 0, \\[4pt]
    \frac{\partial \vartheta}{\partial T}
    &= \mathcal{B}e^{2\bar{Z}}\left[\mathcal{B}+\cos 2\vartheta
       +\frac{\sigma^{2}}{30}\left(5+9\mathcal{B}\cos 2\vartheta+3\cos 4\vartheta\right)\right].
\end{align}
\end{subequations}

These equations show that the wave-averaged drift depends on the mean orientation due to finite-size effects. Focusing on the $O(\sigma^{2})$ corrections, we obtain:

\begin{subequations}
\begin{align}
    \vartheta_\infty
    &\approx\dfrac{1}{2}\cos^{-1}(-\mathcal{B})
      +\frac{2-3\mathcal{B}^{2}}{60\sqrt{1-\mathcal{B}^{2}}}\,\sigma^{2},
      \\[4pt]
    v_x^{fs}
    &\approx \epsilon^{2} e^{2Z}\left(1-\frac{\mathcal{B}}{6}\sigma^{2}\right).
\end{align}
\end{subequations}

Thus, a neutrally buoyant prolate spheroid lags a tracer with respect to its long-time horizontal dispersion.

\begin{figure}
\centering  
\subfigure[]{\includegraphics[width=0.5\linewidth]{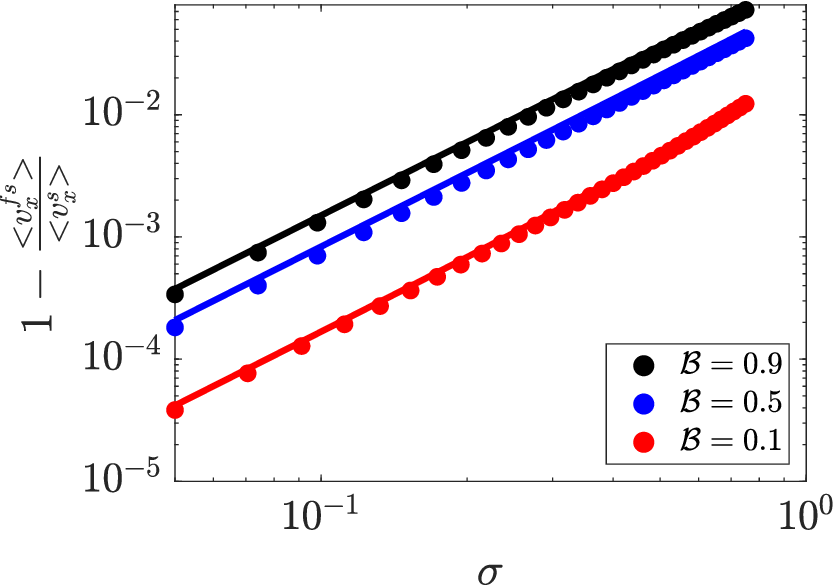}}
\subfigure[]{\includegraphics[width=0.48\linewidth]{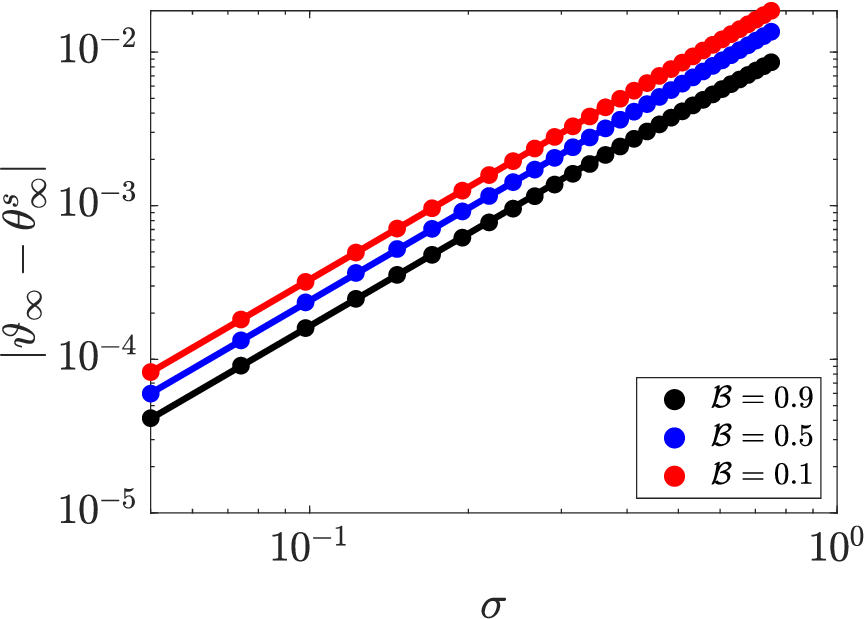}}
\caption{
Validation of finite-size asymptotic calculation against numerical solutions.  
(a) Difference between the mean horizontal drift of a finite-size particle and a point particle.  
(b) Difference in equilibrium orientation (in radians). In both figures, $\epsilon=0.01$ and initial conditions are $X=0$, $Z=-0.05$ and $\Theta=0$.}
\label{fig:fsasmypfig}
\end{figure}

Figure~\ref{fig:fsasmypfig} compares these asymptotic predictions with numerical results. Panel~(a) shows the deviation in mean drift, scaled by the point-particle Stokes drift. Panel~(b) compares the equilibrium orientation: the asymptotics predict $|\vartheta_{\infty}-\theta^{s}_{\infty}|\sim\sigma^{2}$, where $\theta^{s}_{\infty}=\tfrac{1}{2}\cos^{-1}(-\mathcal{B})$. Significant departures arise even for nearly spherical particles ($\mathcal{B}=0.1$) once $\sigma$ is not small.

These results demonstrate that, even in the absence of inertia, finite-size corrections materially affect both drift and alignment. When the particle size is comparable to the wavelength, a finite-size spheroid cannot be treated as a tracer.

\subsection{Effect of inertial torque and finite size on the dynamics}

\begin{figure}
\centering  
\subfigure[]{\includegraphics[width=0.49\linewidth]{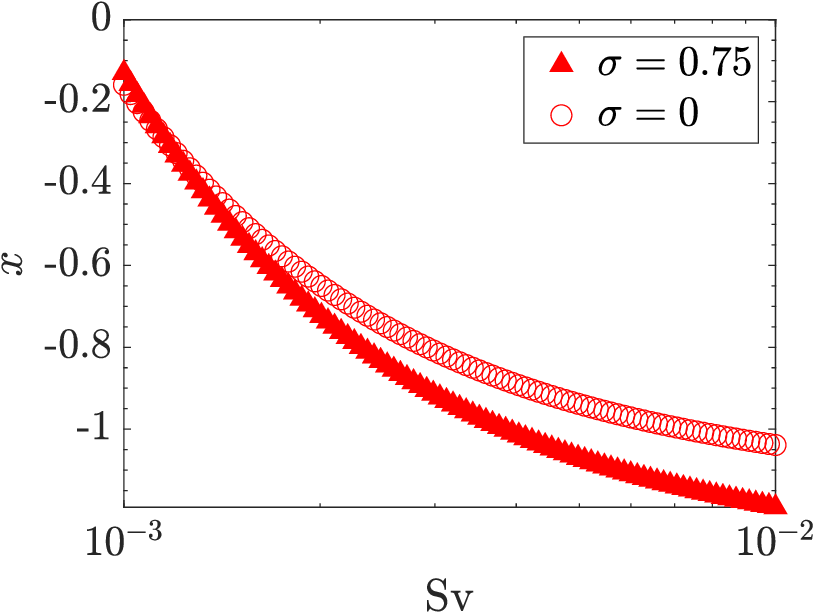}}
\subfigure[]{\includegraphics[width=0.49\linewidth]{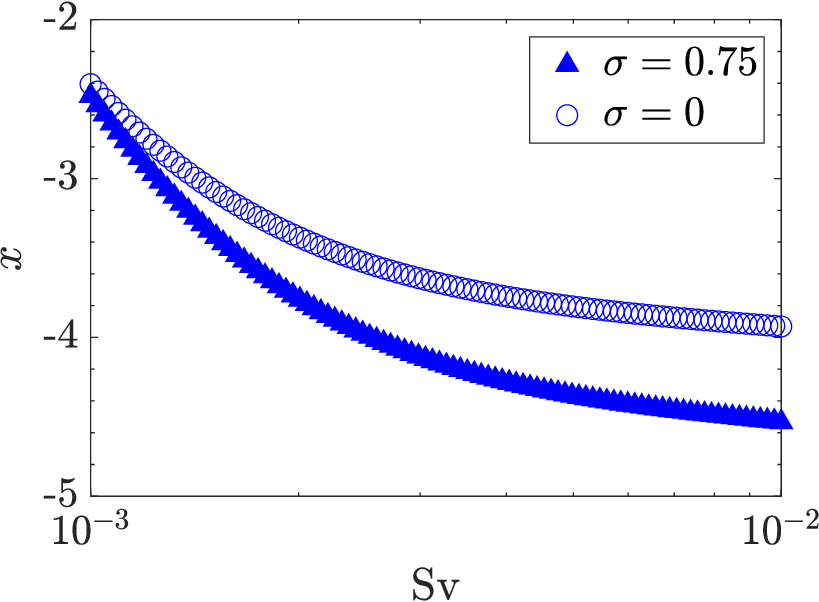}}
\subfigure[]{\includegraphics[width=0.49\linewidth]{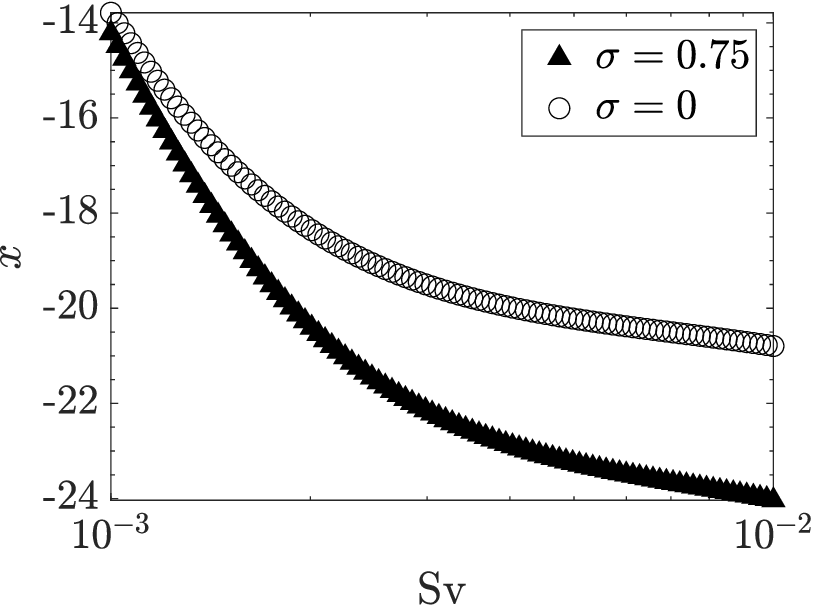}}
\caption{Horizontal translation $x$ at a fixed time as a function of $\mathrm{Sv}$.  
(a) $\mathcal{B}=0.1$, (b) $\mathcal{B}=0.5$, (c) $\mathcal{B}=0.9$.  
Circles: point-particle limit ($\sigma\to0$); triangles: finite-size particle ($\sigma=0.75$).  
Colors denote the three values of $\mathcal{B}$. Initial conditions: $X=0$, $Z=-0.05$, $\Theta=0$.}
\label{fig:setfi}
\end{figure}

We now compare the translation of a finite-size spheroid with a point particle when both buoyancy and inertial torque are present. Earlier, finite-size effects were examined for neutrally buoyant particles; here we focus on negatively buoyant spheroids.

Figure~\ref{fig:setfi} shows the horizontal displacement at a fixed time as a function of $\mathrm{Sv}$. Circles denote the point-particle limit, while triangles correspond to a finite-size particle with $\sigma=0.75$. For all $\mathcal{B}$, the finite-size particle ultimately translates farther against the direction of wave propagation. At small $\mathrm{Sv}$, the correction is weak, but the difference grows with $\mathrm{Sv}$ and becomes more pronounced at longer times. Interestingly, for very small $\mathrm{Sv}$ the finite-size particle initially drifts less than the point particle, then overtakes it at later times, indicating a delayed but stronger horizontal response. Increasing $\mathcal{B}$ amplifies horizontal translation in both point and finite-size cases, consistent with the trends in figure~\ref{fig:anisomobcont}.

\begin{figure}
    \centering
    \includegraphics[scale=0.6]{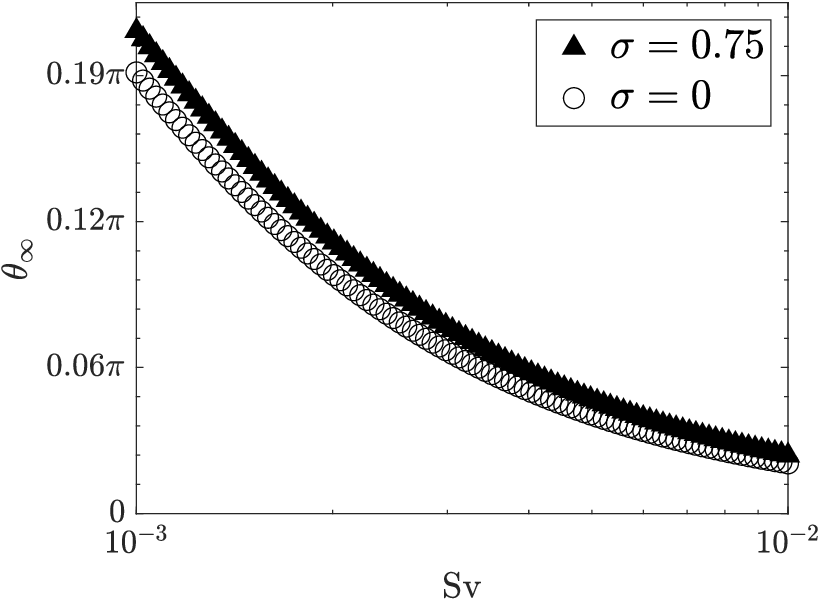}
    \caption{Equilibrium angle $\theta_{\infty}$ versus $\mathrm{Sv}$ for a prolate spheroid with $\mathcal{B}=0.9$, $X=0$, $Z=-0.05$, $\Theta=0$, and $\sigma=0.75$.}
    \label{fig:thtsetfi}
\end{figure}

Figure~\ref{fig:thtsetfi} shows the corresponding long-time orientation. As in the point-particle case, $\theta_{\infty}$ decreases monotonically with $\mathrm{Sv}$. Finite-size effects ($\sigma>0$) shift the equilibrium angle to larger values than in the point-particle limit, but this offset shrinks as $\mathrm{Sv}$ increases, reflecting the weaker influence of Faxén-type corrections at large settling numbers. The same qualitative behavior is observed for other aspect ratios; for $\mathcal{B}=0.1$ the enhancement of alignment is somewhat more pronounced.

\begin{figure}
\centering  
\subfigure[]{\includegraphics[width=0.49\linewidth]{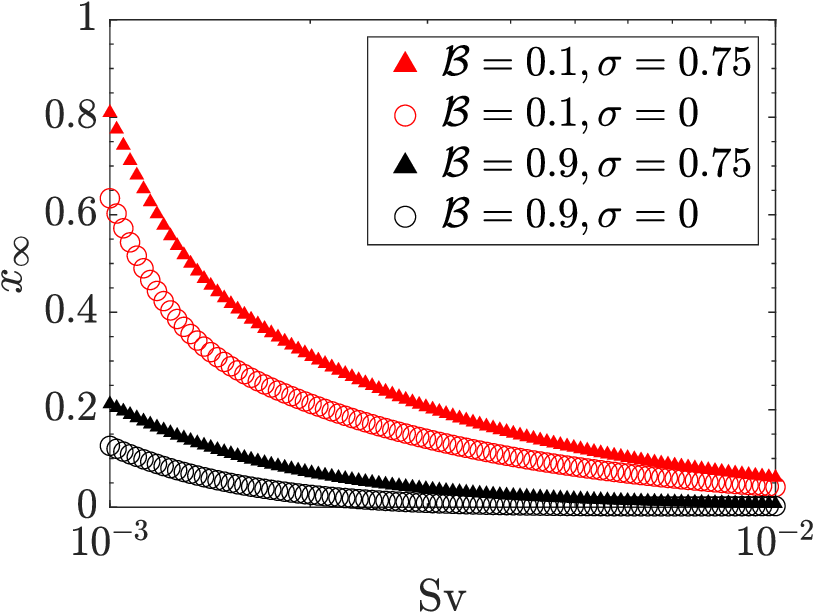}}
\subfigure[]{\includegraphics[width=0.49\linewidth]{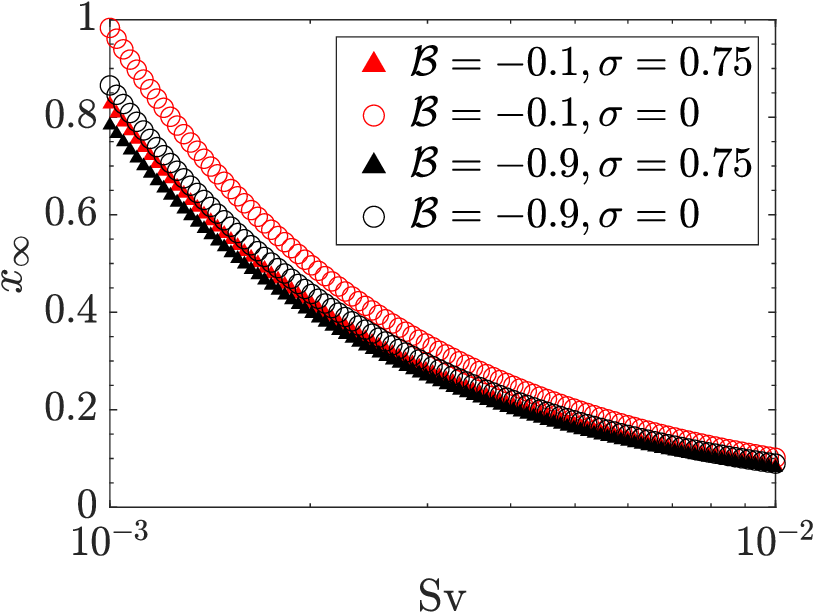}}
\caption{Spreading length $x_{\infty}$ versus $\mathrm{Sv}$ for  
(a) prolate and (b) oblate spheroids.  
Red: $\mathcal{B}=0.1$; black: $\mathcal{B}=0.9$.  
Circles: point particle; triangles: finite-size particle ($\sigma=0.75$).  
$\mathrm{Re}_{w}=3000$.}
\label{fig:fsi1}
\end{figure}

We now include inertial torque in the rotational dynamics and examine the resulting spreading length $x_{\infty}$, shown in figure~\ref{fig:fsi1}. As established earlier, $x_{\infty}$ decreases with increasing $\mathrm{Sv}$ for all aspect ratios. For prolate spheroids (figure~\ref{fig:fsi1}a), finite size enhances horizontal spreading: the finite-size particle travels farther than the point particle, with the largest relative difference at small $\mathrm{Sv}$. At large $\mathrm{Sv}$, inertial torque drives rapid alignment to the broadside orientation, shortening the transient over which horizontal drift occurs and rendering the influence of $\sigma$ negligible.

For oblate spheroids (figure~\ref{fig:fsi1}b), the trend reverses: finite size reduces horizontal spreading relative to the point-particle case. Nonetheless, for fixed $\mathrm{Sv}$ and $\mathcal{B}$, oblates exhibit larger $x_{\infty}$ than prolates, consistent with their broader projected area in the broadside configuration. As in the prolate case, finite-size corrections weaken as $\mathrm{Sv}$ increases and both curves converge once inertial torque enforces rapid orientation equilibration.

In summary, finite size modifies both the magnitude and the $\mathrm{Sv}$-dependence of horizontal translation in the presence of inertial torque. The strongest effects occur when (i) $\mathrm{Sv}$ is small enough that the particle experiences a long-lived misaligned state, and (ii) $\sigma$ is large enough for Faxén corrections to appreciably change the instantaneous wave-induced motion. At large $\mathrm{Sv}$, inertial torque quickly selects the broadside orientation and suppresses finite-size corrections in both prolate and oblate geometries.

The analysis presented so far has focused exclusively on the inertia-less limit, $\mathrm{St}=0$, in which particle motion is instantaneously locked to the local flow through quasi-steady hydrodynamic forces and torques. Having now introduced a reorienting torque arising from fluid inertia, it is important to briefly examine the role that finite particle inertia ($\mathrm{St}>0$) would play in the present problem. For the class of particles relevant to geophysical and environmental applications, the particle-to-fluid density ratio is typically $O(1)$; consequently, a faithful description of the translational dynamics would require accounting for the combined effects of quasi-steady Stokes drag, added mass, and history (Basset) forces. In contrast to the spherical case, an anisotropic particle would additionally experience rotational added-mass effects together with a memory contribution to the hydrodynamic torque.

We do not attempt a quantitative treatment of finite-$\mathrm{St}$ dynamics here, as such an analysis would require several additional caveats and the adoption of semi-empirical forms for the above-mentioned forces, thereby limiting the predictive scope of the results. Instead, we restrict ourselves to a qualitative discussion of the key physical implications of weak but non-zero inertia. For small $\mathrm{St}$, the particle velocity deviates from the local fluid velocity through an inertial slip that depends on both gravitational acceleration and the local fluid acceleration. This slip modifies the translational response to the wave field and, through coupling with the rotational dynamics, can further influence the long-time drift and orientation statistics. The slip induced by gravity is responsible for the inertial torque discussed in detail in the earlier sections. For $\mathrm{St}>0$, however, the slip arises from a competition between the gravitational acceleration and the local fluid acceleration. As a result, the inertial torque tends to align the spheroid along a direction intermediate between the vertical and the instantaneous acceleration vector of the wave field. This leads to a non-monotonic response of the tumbling dynamics: with increasing $\mathrm{St}$, the tumbling rate in the plane of the wave would initially decrease due to the increasing effective heaviness of the particle, and then increase again as inertial torque driven by misalignment with the wave-induced acceleration becomes dominant. These competing effects suggest that a systematic investigation of finite-$\mathrm{St}$ dynamics using fully resolved numerical simulations, for example via an immersed boundary type approach, would be a particularly worthwhile direction for future work.

\section{Conclusion}

In this work, we have investigated the coupled translation and orientation dynamics of spheroidal particles in surface gravity waves, systematically isolating the roles of buoyancy, inertial torque, and finite particle size. By combining asymptotic analysis with numerical calculations, we have established how each of these physical mechanisms controls the long-time horizontal transport, equilibrium orientation, and dispersion properties of anisotropic particles. Our results reveal multiple distinct transport regimes that are inaccessible within the traditional point-particle or neutrally buoyant approximations.

For buoyant anisotropic particles without inertial torque, we showed that the long-time horizontal motion can remain unbounded, with the direction and magnitude of drift governed by the equilibrium orientation and mobility anisotropy. In this regime, particles may translate either along or against the wave-propagation direction, and thin rods and disks exhibit enhanced drift relative to nearly spherical particles. When inertial torque is included, however, the dynamics change qualitatively: particles invariably approach a broadside-on configuration, horizontal translation becomes confined to a finite transient interval, and the total spreading is set by the time required for orientation equilibration. We further demonstrated that this equilibration time depends sensitively on $\mathrm{Sv}$, $\mathrm{Re}_{w}$, and the particle shape factor $\mathcal{F}_{p}$, thereby directly linking orientation dynamics to large-scale particle transport.

When finite-size effects are incorporated through Faxen-type corrections, translation becomes coupled to rotation even in the absence of inertia, including for neutrally buoyant particles. For fore–aft symmetric spheroids, finite size systematically modifies both the drift velocity and the equilibrium orientation, with the leading deviations scaling quadratically with the ratio of particle size to wavelength. As a result, a finite-size spheroid cannot, in general, be treated as a tracer once its size becomes comparable to the wavelength. When combined with buoyancy and inertial torque, finite-size effects further reshape the transport: heavy finite-size prolates exhibit enhanced spreading relative to point particles, while finite-size oblates display reduced spreading. These results demonstrate that particle size, shape, buoyancy, and inertia interact in a strongly non-additive manner.

Microplastics constitute a major and persistent source of contamination in both terrestrial and marine environments. A large fraction of these particles are non-spherical, which leads to complex translation–orientation coupling that strongly influences their transport in geophysical flows. Microplastics span a wide size range, from a few microns to centimetre scales, and this size distribution plays a central role in governing their residence time and dispersion in the deep ocean. Field observations by \cite{isobe2014selective} at coastal stations in Japan demonstrated that large plastic debris gradually fragments into smaller microplastics, which are more likely to penetrate into deeper oceanic layers. Their study further showed that buoyant microplastics are primarily transported by a combination of Stokes drift and turbulent currents, with turbulence represented through stochastic velocity fluctuations. Positively buoyant particles larger than $1$~mm were found predominantly near the surface, whereas particles smaller than $1$~mm exhibited prolonged residence in the deep ocean.

The shape of a microplastic particle is equally important in determining its dispersion characteristics. As discussed earlier, non-spherical particles such as spheroids exhibit strong coupling between translation and orientation. Recent experiments by \cite{tatsii2023shape} investigated the influence of shape on the settling dynamics of atmospheric microplastics by modeling them as spheres and rods. These authors verified the classical settling-induced preferential alignment of rods and showed that rod-shaped particles settle more slowly than spherical ones. However, due to their longer residence times, rods undergo enhanced dispersion. Using a modified Lagrangian dispersion model (FLEXPART v10.4; \citealt{pisso2019lagrangian}), they demonstrated that spherical particles tend to deposit locally near their source, while rod-shaped particles exhibit long-range atmospheric transport with nearly global-scale deposition patterns. 

Motivated by these observations, we investigated the translation and orientation dynamics of a spheroid in surface gravity waves. We began with an idealized model that neglects anisotropy in the translational equations of motion to isolate the influence of orientation dynamics. Using an asymptotic model, we computed the long-time spreading $x_{\infty}$ and the preferred alignment angle, and validated these results against numerical simulations. In this regime, the preferred alignment depends on $\mathrm{Sv}$, aspect ratio, and the initial orientation. Unlike the neutrally buoyant case, the horizontal transport of the particle does not grow linearly in time. Instead, the horizontal displacement becomes finite, and the particle may translate either along or against the wave propagation direction. We observed that at large $\mathcal{B}$, a particle attains a higher steady orientation and reaches its equilibrium angle rapidly. As a consequence, the long-time transport dynamics are entirely governed by the equilibrium orientation.

We then considered a more realistic scenario by incorporating anisotropy into the translational equations. While the qualitative behaviour of the preferred orientation remains similar, the spheroid now undergoes unbounded horizontal translation. With buoyancy present, a particle aligns into a drag-minimizing configuration, and this alignment drives persistent horizontal transport and overall dispersion. The contour plots of horizontal drift velocity reveal that a particle can again migrate against the wave-propagation direction. The direction of translation is controlled by the equilibrium angle and the mobility functions of the spheroid. Thin rods and disks were found to exhibit larger horizontal drift velocities, and spheroids experience stronger translation in both directions at larger $\mathrm{Sv}$. We further observed that thin disks travel farther along the wave-propagation direction but settle more slowly than thin rods at a given instant.

Next, we analysed the dynamics of a spheroid in the presence of inertial torque. A spheroid invariably attains a broadside-on configuration when inertial torque is included, thereby maximizing the drag during settling. Once this inertially preferred orientation is reached, horizontal translation ceases. As a result, the horizontal transport of a spheroid becomes finite in the presence of inertial effects, in sharp contrast with the buoyancy-dominated case. Using asymptotic methods, we derived explicit expressions for the horizontal spreading and validated them against numerical simulations. The spreading depends strongly on $\mathrm{Sv}$ and the initial orientation. Depending on the initial condition, a particle can translate either along the wave or against it. High-aspect-ratio rods and disks translate less in the horizontal direction because they reach the broadside-on alignment rapidly, which suppresses horizontal motion. The rate of orientation evolution depends on $\mathrm{Sv}$, $\mathrm{Re}_{w}$, and $\mathcal{F}_{p}$. For large $\mathrm{Sv}$ and high aspect ratio, this alignment is reached very quickly. We established a direct link between the rate at which a particle reaches its equilibrium angle and the extent of horizontal transport.

The particle trajectories reveal three distinct stages of motion. In the first stage, the particle undergoes oscillatory orbital motion induced by wave effects. In the second stage, after the oscillations decay, the orientation evolves while the particle translates horizontally without orbital motion. In the third and final stage, once the broadside-on orientation is reached, horizontal translation vanishes and the particle settles purely vertically. The occurrence of the second stage is tied to the non-monotonic transient evolution of orientation. When the initial orientation coincides with the final equilibrium orientation, this transient does not occur and the second stage disappears.

Finally, we examined the dynamics of non-inertial finite-size particles. We demonstrated that a finite-size particle cannot be treated as a tracer when the finite-size parameter $\sigma$ is appreciable. Although particle inertia may become important at very large $\sigma$, we focused on intermediate values to isolate and quantify finite-size effects alone. Using multiple-scale analysis, we derived explicit expressions for the mean horizontal drift velocity and equilibrium angle of a finite-size spheroid and validated them against numerical simulations. While the equilibrium orientation is only weakly modified, the long-time drift of a finite-size particle deviates significantly from that of a point particle, with the gap in horizontal displacement continuing to grow in time. Throughout our analysis, we restricted attention to $\sigma < 1$, corresponding to particles smaller than the wavelength.

We further compared the dynamics of heavy point particles and heavy finite-size particles. The heavy finite-size particle exhibits stronger horizontal translation against the wave-propagation direction than its point-particle counterpart. This difference is attributed primarily to the higher equilibrium orientation attained by a finite-size particle. When inertial torque is included, we found enhanced spreading of a finite-size prolate spheroid, whereas a finite-size oblate spheroid exhibits reduced spreading compared with a point oblate particle.

Overall, this study provides a unified theoretical framework for understanding how particle shape, buoyancy, inertial torque, and finite size together govern the transport of spheroidal particles in surface gravity waves. These results offer new physical insight into the dispersion of non-spherical microplastics in the ocean and provide a foundation for incorporating realistic particle geometry into predictive transport models for marine and atmospheric environments.

\backsection[Funding]{H.M. would like to acknowledge financial support from the Prime Minister’s Research Fellow (PMRF) scheme, Ministry of Education, Government of India (Project no. SB22230347AMPMRF008746).}

\backsection[Declaration of interests]{ The authors report no conflict of interest.}

\backsection[Author ORCIDs]{\\ 
Himanshu Mishra https://orcid.org/0000-0001-6255-2124;\\ 
Anubhab Roy https://orcid.org/0000-0002-0049-2653.
}

\appendix

\section{Mobility functions and shape factors for spheroidal particles}\label{appB}

The definition of the resistance function are adapted from \cite{kim2013microhydrodynamics}. The mobility functions are the inverse of the resistance function and can be defined as,

\begin{subequations}
\begin{equation}
    \mathcal{X}_{A}=\dfrac{3\{-2e+(1+e^{2})L\}}{8e^{3}}, \quad \mathcal{Y}_{A}=\dfrac{3\{2e+(3e^{2}-1)L\}}{16e^{3}},
\end{equation}
\begin{equation}
    \mathcal{X}_{C}=\dfrac{6e+3(e^{2}-1)L}{4e^{3}-4e^{5}}, \quad \mathcal{Y}_{C}=-\dfrac{3\{L+e(e L-2)\}}{4e^{3}(e^{2}-2)},
\end{equation}
\begin{equation}
     \mathcal{Y}_{H}=\dfrac{3\{L+e(e L-2)\}}{4e^{5}}.
 \end{equation}
\end{subequations}

Here, $L=\log((1+e)/(1-e))$ and $e=\sqrt{(a^{2}-c^{2})}/a$, while $c$ is a length of semi-minor axis. For an oblate particle, the mobility functions are given as,
 
\begin{subequations}
\begin{equation}
    \mathcal{X}_{A}=\dfrac{3\{e\sqrt{1-e^{2}}+C(2e^{2}-1)\}}{4e^{3}}, \quad \mathcal{Y}_{A}=\dfrac{3\{C+2Ce^{2}-e\sqrt{1-e^{2}}\}}{8e^{3}},
\end{equation}
\begin{equation}
    \mathcal{X}_{C}=\dfrac{3(C-e\sqrt{1-e^{2}})}{2e^{3}}, \quad \mathcal{Y}_{C}=-\dfrac{3\{e\sqrt{1-e^{2}}+C(2e^{2}-1)\}}{2e^{3}(e^{2}-2)},
\end{equation}
\begin{equation}
    \mathcal{Y}_{H}=-\dfrac{3\{e\sqrt{1-e^{2}}+C(2e^{2}-1)\}}{2e^{5}}.
\end{equation}
\end{subequations}

Here, $C=\tan^{-1}(e/\sqrt{1-e^{2}})$. The resistance functions, $X_{A}$, $Y_{A}$, $X_{C}$, $Y_{C}$, $Y_{H}$ are the inverse of the mobility functions. The shape factor for the prolate particle \citep{dabade2015effects,sheikh2020} is given as,

\begin{eqnarray}
    \mathcal{F}& =& \dfrac{-\pi e^{2}(420e+2240e^{3}+4249e^{5}-2152e^{7})}{315\{(e^{2}+1)\tanh^{-1}e-e\}^{2}\{(1-3e^{2})\tanh^{-1}e-e\}} \nonumber\\
    && +\dfrac{\pi e^{2}(420+3360e^{2}+1890e^{4}-1470e^{6})\tanh^{-1}e}{315\{(e^{2}+1)\tanh^{-1}e-e\}^{2}\{(1-3e^{2})\tanh^{-1}e-e\}} \nonumber\\
    && -\dfrac{\pi e^{2}(1260e-1995e^{3}+2730e^{3}-1995e^{7})(\tanh^{-1}e)^{2}}{315\{(e^{2}+1)\tanh^{-1}e-e\}^{2}\{(1-3e^{2})\tanh^{-1}e-e\}}.
\end{eqnarray}
Similarly, for an oblate particle, the expression for $\mathcal{F}$ is given by,

\begin{eqnarray}
    \mathcal{F} &=& \dfrac{\pi e^{3}\sqrt{1-e^{2}}(-420+3500e^{2}-9989e^{4}+4757e^{6})}{315\sqrt{1-e^{2}}(-e\sqrt{1-e^{2}}+(1+2e^{2})\sin^{-1}e)(e\sqrt{1-e^{2}}+(2e^{2}-1)\sin^{-1}e)^{2}}\nonumber\\
    && \dfrac{210\pi e^{2}(2-24e^{2}+69e^{4}-67e^{6}+20e^{8})\sin^{-1}e}{315\sqrt{1-e^{2}}(-e\sqrt{1-e^{2}}+(1+2e^{2})\sin^{-1}e)(e\sqrt{1-e^{2}}+(2e^{2}-1)\sin^{-1}e)^{2}}\nonumber\\
    && \dfrac{105\pi e^{3}(12-17e^{2}+24e^{4})(\sin^{-1} e)^{2}}{315\sqrt{1-e^{2}}(-e\sqrt{1-e^{2}}+(1+2e^{2})\sin^{-1}e)(e\sqrt{1-e^{2}}+(2e^{2}-1)\sin^{-1}e)^{2}}.
\end{eqnarray}

\begin{figure}
    \centering
    \includegraphics[width=1.0\linewidth]{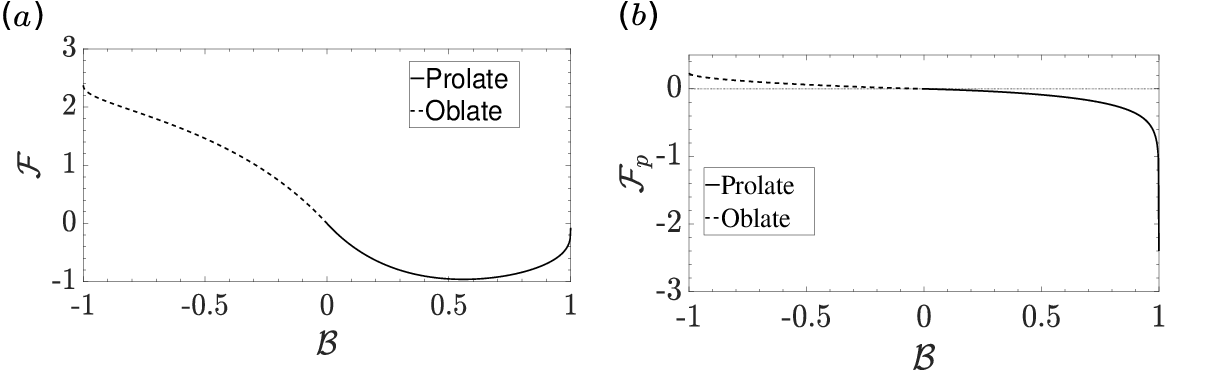}
    \caption{ a) Shape factor $\mathcal{F}$ versus $\mathcal{B}$ plot b) Modified shape factor $\mathcal{F}_{p}$ versus $\mathcal{B}$ plot. The expressions of shape factors are given in Appendix.}
    \label{fig:fb}
\end{figure}

We present the shape factor as a function of $\mathcal{B}$ for both prolate and oblate particles in figure \ref{fig:fb}. The dotted lines indicate the variation of the shape factor in an oblate particle, and the solid line corresponds to a prolate particle. The maximum values of the shape factor for both prolate and oblate particles are obtained when the magnitude of  $\mathcal{B}$ approaches unity. The value of $\mathcal{F}_{p}$ is largest for rod- and disk-shaped particles; therefore, we expect the effect of the inertial torque to be significant for high aspect ratio particles.

\section{Higher order asymptotic relations for long-time orientation }\label{app_asym_orientation}
Asymptotic expression for the long-time orientation of the particle when $\Theta=0$,
\begin{equation}\label{eq:anisothetal}  
\mathcal{J}(\mathcal{B},\mathrm{Sv};0,Z,0)=\left(\dfrac{\mathcal{B}e^{2Z}\left[2X_{1}^{2}+\mathcal{B}\left[2+\mathrm{Sv}^{2}\mathcal{X}_{A}\{\mathcal{Y}_{A}+\mathcal{X}_{A}(7+\mathrm{Sv}^{2}\mathcal{X}_{A}(\mathcal{X}_{A}+5\mathcal{Y}_{A}))\}\right]\right]}{4\mathrm{Sv}\mathcal{X}_{A}X_{1}^{3}}\right)
\end{equation}

In above, $X_{1}=1+\mathrm{Sv}^{2}\mathcal{X}^{2}_{A}$. For $\Theta=\pi/2$,

\begin{equation}
    \mathcal{J}(\mathcal{B},\mathrm{Sv};0,Z,\pi/2)=\left(\dfrac{\mathcal{B}e^{2Z}[-2X_{2}^{2}+\mathcal{B}\{2+\mathrm{Sv}^{4}\mathcal{Y}_{A}^{3}(5\mathcal{X}_{A}+\mathcal{Y}_{A})+\mathrm{Sv}^{2}\mathcal{Y}_{A}(\mathcal{X}_{A}+7\mathcal{Y}_{A})\}]}{4\mathrm{Sv}\mathcal{Y}_{A}X_{2}^{3}}\right).
\end{equation}

Here, $X_{2}=1+\mathrm{Sv}^{2}\mathcal{Y}_{A}^{2}$. Similarly for $\Theta=\pi/4$ we have,

\begin{equation}
    \mathcal{J}(\mathcal{B},\mathrm{Sv};0,Z,\pi/4)=\left(\dfrac{\mathcal{B}e^{2Z}\{\mathcal{B}X_{5}(8+X_{7}+\mathrm{Sv}^{4}(\mathcal{X}_{A}+\mathcal{Y}_{A})^{2}X_{6}+X_{8})-2\mathrm{Sv}(\mathcal{X}_{A}+\mathcal{Y}_{A})X_{4}^{2}\}}{2\mathrm{Sv}(\mathcal{X}_{A}+\mathcal{Y}_{A})X^{3}_{4}}\right)
\end{equation}

Here, $X_{3}=4+2\mathrm{Sv}(\mathcal{X}_{A}-\mathcal{Y}_{A})$, $X_{4}=2+\mathrm{Sv}(2\mathcal{X}_{A}+\mathrm{Sv}\mathcal{X}_{A}^{2}+\mathcal{Y}_{A}(-2+\mathrm{Sv}\mathcal{Y}_{A}))$, $X_{5}=2+\mathrm{Sv}(\mathcal{X}_{A}-\mathcal{Y}_{A})$, $X_{6}=\mathcal{X}_{A}^{2}-4\mathcal{X}_{A}\mathcal{Y}_{A}+\mathcal{Y}_{A}^{2}$, $X_{7}=12\mathrm{Sv}(\mathcal{X}_{A}-\mathcal{Y}_{A})+6\mathrm{Sv}^{2}(\mathcal{X}_{A}-\mathcal{Y}_{A})^{2}$ and $X_{8}=4\mathrm{Sv}^{3}(\mathcal{X}_{A}^{3}-\mathcal{Y}_{A}^{3})$.

\bibliographystyle{jfm}%\bibliography{jfm2esam}
\bibliography{jfm}

\end{document}